\documentclass[12pt]{amsart}
\usepackage{amssymb,multicol,latexsym}
\usepackage[mathscr]{eucal}
\usepackage{graphicx}
\def\mymid{,}

\def\espaceN{\mathsf{N}}

\def\aa{\alpha_1}
\def\ab{\alpha_2}
\def\atr{\alpha_3}
\def\at{\alpha_{\theta}}
\def\am{\alpha^{-}}
\def\amp{\alpha^{'-}}
\def\ap{\alpha^{+}}
\def\ad{\alpha^{*}}
\def\tam{\widetilde{\alpha}^-}
\def\tamp{\widetilde{\alpha}^{'-}}
\def\tap{\widetilde{\alpha}^+}
\def\tad{\widetilde{\alpha}^*}

\def\g{\gamma}

\def\p{p}

\def\mod{\mathop{\mathrm{mod}}\nolimits}
\def\u{\widehat{u}}
\def\Lambdach{\Lambda_{\mathrm{ch}}}
\def\sdim{\mathrm{sdim}}

\def\D{D(2|1;\alpha)}
\def\DD#1{D(2|1;#1)}
\def\hDD#1{\widehat{D}(2|1;#1)}
\def\hD{\widehat{D}(2|1;\alpha)} 
\def\hf{\frac{1}{2}}

\renewcommand{\atopwithdelims}[2]{\genfrac{[}{]}{0pt}{}{#1}{#2}}
\renewcommand{\atop}[2]{\genfrac{}{}{0pt}{}{#1}{#2}}

\def\chr{\mathop{\mathrm{char}}\nolimits}
\def\Re{\mathop{\mathsf{Re}}\nolimits}
\def\Im{\mathop{\mathsf{Im}}\nolimits}

\newcommand{\wt}{\widetilde}

\def\hslc{\widehat{s\ell}(2|1)}

\def\tilde{\widetilde}

\def\zero{{\mathbf{\bar0}}}
\def\one{{\mathbf{\bar1}}}

\def\V{\mathrm{V}} % for the name `Verma'

\def\algebra{\mathfrak}
\def\ac{\algebra{c}}
\def\ag{\algebra{g}}
\def\ww{\algebra{w}}

\def\module{\mathscr}
\def\mA{\module{A}}

\def\mE{\module{E}}
\def\mF{\module{F}}

\def\mL{\module{L}}    %<----- \tSSL21 admissible representations
\def\mV{\module{V}} % |
\def\mm{\module{M}}   % NOT a Verma module!!
\def\Adm{{\module{M}}} %<----- \tSL2 admissible representations
\def\Int{{\module{I}}} %<----- \tSL2 integrable representations
\def\mN{\module{N}}
\def\mP{\module{P}}
\def\mR{\module{R}}
\def\mW{\module{W}}
\def\mX{\module{X}}

\def\qmV{\mathsf{V}}  %quantum group modules

\def\thalf{\tfrac{1}{2}}
\def\half{\frac{1}{2}}
\def\Hplus{H^+}
\def\Hminus{H^-}

\newcommand{\Tr}{\mathop{\mathrm{Tr}}\nolimits}

\newcommand{\UParrow}[1]{\bigm\uparrow{\scriptstyle\!\!#1}\kern-12pt}

\def\ketsl#1#2#3{|{#3}\rangle_{\widehat{s\ell}(#1){#2}}}

\newcommand{\ket}[1]{\mathchoice{%
    {\left|{#1}\right\rangle}}{|{#1}\rangle}{|{#1}\rangle}{|{#1}\rangle}}

\numberwithin{equation}{section}
\makeatletter
 \@addtoreset{equation}{section}
\makeatother

\newcommand{\SL}[1]{s\ell(#1)}
\newcommand{\tSL}[1]{\widehat{s\ell}(#1)}
\newcommand{\tGL}[1]{\widehat{g\ell}(#1)}
\newcommand{\SSL}[2]{s\ell(#1|#2)}
\newcommand{\tSSL}[2]{\widehat{s\ell}(#1|#2)}

\newcommand{\oC}{\mathbb{C}}

\newcommand{\oN}{\mathbb{N}}
\newcommand{\oZplus}{\mathbb{Z}_+}
\newcommand{\oP}{\mathbb{P}}

\newcommand{\oZ}{\mathbb{Z}}

\newtheorem{Thm}{Theorem}[section]
\newtheorem{Lemma}[Thm]{Lemma}

\theoremstyle{definition}

\newtheorem{Rem}[Thm]{Remark}%[section]

{\noindent{\nolinebreak\hfill\mbox{\rule{.5em}{.5em}}\,}\par\medskip}

\def\cL{\mathcal{L}}

\def\cO{\mathcal{O}}

\def\cU{\mathcal{U}}

\def\bar{\overline}

\def\tensor{\otimes}

\renewcommand{\d}{\partial}

\begin{document}
\addtolength{\baselineskip}{4pt}
\addtolength{\parskip}{4pt}
\addtolength{\textfloatsep}{-.5\baselineskip}
\vfuzz=1.3pt
\raggedbottom
\begin{flushright}
DTP/99/39\\
July 1999\\
\end{flushright}
\title[$\smash{\protect\widehat{s\ell}(2|1)}$ and
$\smash{\protect\widehat{D}(2|1;\alpha)}$ Representations]{\hfill{
    \lowercase{\tt hep-th/9907171}}\\[12pt]
 $\tSSL21$ and $\hD$ as Vertex Operator Extensions of Dual
 Affine $\SL2$ Algebras}

\author[Bowcock]{P.~Bowcock $^{\dagger}$} \address{Dept.\ of Mathematical Sciences,
  University of Durham}

\author[Feigin]{B.~L.~Feigin $^*$} \address{Landau Institute for
  Theoretical Physics, Russian Academy of Sciences}

\author[Semikhatov]{A.~M.~Semikhatov $^+$} \address{Tamm Theory Division,
  Lebedev Physics Institute, Russian Academy of Sciences}

\author[Taormina]{A.~Taormina $^{\dagger}$} \address{Dept.\ of Mathematical
  Sciences, University of Durham}

\begin{abstract}
  We discover a realisation of the affine Lie superalgebra $\tSSL21$
  and of the exceptional affine superalgebra $\hD$ as vertex operator
  extensions of two $\tSL2$ algebras with `\textit{dual}' levels (and
  an auxiliary level-1 $\tSL2$ algebra).  The duality relation between
  the levels is $(k_1+1)(k_2+1)=1$.  We construct the representation
  of $\tSSL21_{k_1}$ on a sum of tensor products of $\tSL2_{k_1}$,
  $\tSL2_{k_2}$, and $\tSL2_1$ modules and decompose it into a direct
  sum over the $\tSSL21_{k_1}$ spectral flow orbit.  This
  decomposition gives rise to character identities, which we also
  derive.  The extension of the construction to $\hDD{k_2}_{k_1}$ is
  traced to the properties of $\tSL2\oplus\tSL2\oplus\tSL2$ embeddings
  into $\hD$ and their relation with the dual $\tSL2$ pairs.
  Conversely, we show how the $\tSL2_{k_2}$ representations are
  constructed from~$\tSSL21_{k_1}$ representations.
\end{abstract}

\maketitle
%%%%%%%%%%%%%%%%%%%%%%%%%%%%%%%%%%%%%%%%%%%%%%%%%%%%%%%%%%%%%%%%%%%%%

\thispagestyle{empty} 

\flushcolumns

\setcounter{tocdepth}{2}%3

\vspace*{-36pt}

\begin{center}
  \parbox{.95\textwidth}{
    \begin{multicols}{2}
      {\footnotesize
        \tableofcontents}
    \end{multicols}
    }
\end{center}

\bigskip

\section{\textbf{Introduction}}
In this paper, we address a particular instance of the general problem
of extending infinite-dimensional algebras by vertex operators.  We
show that vertex operator extensions can lead to interesting and
nontrivial constructions of affine Lie (super)algebras and their
representations.  A well-known example of an extension via vertex
operators applies to the sum of two Virasoro algebras with appropriate
central charges, and leads to matter coupled to gravity in the
conformal gauge~\cite{[David],[DisKaw]}.  Another context where vertex
operator extensions of conformal algebras are relevant is that of
coset conformal field theories~$\ac=\ag_1/\ag_2$, where the `inverse'
problem is to reconstruct the $\ag_1$ representations by combining the
representations of~$\ac$ and~$\ag_2$ \hbox{related by the action of
  vertex operators}.

A natural generalisation of the situation encountered in matter plus
gravity is provided by studying extensions of the sum
$\tSL2_{k_1}\oplus\tSL2_{k_2}$ of two affine~$\SL2$ algebras at levels
$k_1$ and~$k_2$.  This is a complicated problem in general; we solve
it in the case where
\begin{equation}\label{kk1}
  (k_1+1)(k_2+1)=1,\qquad k_i\in\oC\setminus\{-2,-1\}.
\end{equation}

The result is that with the help of an additional scalar current,
\textit{spin-$\half$ vertex operators extend
  $\tSL2_{k_1}\oplus\tSL2_{k_2}$ to the affine Lie superalgebra
  $\hDD{k_2}_{k_1}$}.  We study the representations in terms of the
subalgebra of $\hDD{k_2}_{k_1}$ given by $\tSSL21_{k_1}\oplus\u(1)$.
Let $\mW_{n,k}$ be the Weyl module over $\tSL2_k$ induced from the
$n+1$-dimensional representation of $\SL2$.  For generic values of the
level, the vacuum representation $\mL_{0,0,k_1}$ of $\tSSL21_{k_1}$ is
contained in a sum of tensor products of Weyl modules $\mW_{n,k_1}$
and $\mW_{n,k_2}$ of the respective algebras $\tSL2_{k_1}$ and
$\tSL2_{k_2}$.  More precisely,
\begin{equation}\label{decompose-Weyl}
  \bigoplus_{n\geq0}\mW_{n,k_1}\tensor\mW_{n,k_2}\tensor
  \mm_{\frac{1}{2}(n\;\mathrm{mod}\;2),1}=
  \bigoplus_{\theta\in\oZ}\mL_{0,0,k_1;\theta}\tensor\mA_{-\theta},
\end{equation}
where $\mL_{0,0,k_1;\theta}$ denotes the spectral flow transform of
the $\tSSL21$ module, $\mm_{0,1}$ and $\mm_{\half,1}$ are the
irreducible representations of the auxiliary level-1 $\tSL2$ algebra,
and $\mA_{a}$ are Fock modules over a Heisenberg algebra.

\medskip

The basic relation~\eqref{kk1} can be rewritten as~$\frac{1}{k_1+2} +
\frac{1}{k_2+2} = 1$, and can be put into a broader perspective when
viewed as a particular case of the following family of relations
between levels,
\begin{equation}\label{eq:n}
  \tfrac{1}{k_1+2} + \tfrac{1}{k_2+2} =n\in\oN\,.
\end{equation}
We refer to the~$\tSL2$ algebras with the levels~$k_1$ and~$k_2$
satisfying~\eqref{eq:n} as \textit{dual}; Eq.~\eqref{eq:n} may be
viewed as a duality relation between $\tSL2$ levels, which potentially
allows one to study certain classes of integrable and admissible
representations of the extended algebraic structure, starting from the
well-studied $\tSL2$ representation theory at fractional level.  For
simplicity in illustrating the main idea, we can replace
the~$\tSL2_{k_i}$ algebras with the Virasoro
algebras~$\mathrm{Vir}(k_i)$ obtained from them by Hamiltonian
reduction, with central charges $d_i=13-\frac{6}{k_i+2}-6(k_i+2)$.
When extending the sum $\mathrm{Vir}(k_1) \oplus \mathrm{Vir}(k_2)$ by
vertex operators, one must address the problem of their potential
non-locality. For instance, (a component of) the
operator~$\phi^{(i)}_{12}$ has
monodromy~$(-1)^{\frac{1/2}{k_i+2}}=e^{\frac{i\pi/2}{k_i+2}}$ with
itself, which means nonlocality in general; however a bilinear
combination of~$\phi^{(1)}_{12}$ and~$\phi^{(2)}_{12}$ operators for
two dual Virasoro algebras has monodromy
$e^{\frac{i\pi/2}{k_1+2}}\,e^{\frac{i\pi/2}{k_2+2}}=e^{n i\pi/2}$ in
view of relation~\eqref{eq:n}.  It is therefore local (for even~$n$)
or `almost' local.

The precise way to build local or almost local bilinear combinations
of vertex operators of the dual algebras is governed by the
corresponding quantum group~$s\ell(2)_{q_i}$
with~$q_i=e^{\frac{2i\pi}{k_i+2}}$.  The vertex operators, which carry
a representation of the quantum group, have to be contracted `over the
quantum group index,' as $\Tr_q(\phi^{(1)}(z)\,\phi^{(2)}(z))$.
For~$n$ odd in~\eqref{eq:n}, the monodromies~$e^{n\pi i/2}$
of~$\theta(z)=\Tr_q(\phi^{(1)}_{12}(z)\,\phi^{(2)}_{12}(z))$ show that
these operators are not yet local, but they become so when multiplied
with~$e^{\pm\frac{1}{\sqrt{2}}f}$, where~$f$ is a free scalar.  On the
other hand, for~$n$ even these bilinear combinations are local without
the need of a free scalar.  The~$n=0$ case is the well known example
mentioned earlier of matter coupled to gravity in the conformal gauge;
one may view the relation~\eqref{eq:n} with~$n=0$ as the anomaly
cancellation condition between matter ($\mathrm{Vir}(k_1)$), Liouville
($\mathrm{Vir}(k_2)$) and reparametrisation ghosts sectors in the
quantisation of two-dimensional gravity.  For~$n=2$, such a
contraction~$\theta(z)$ is \textit{a fermion} with central
charge~$\hf$.  Subtracting this from the total central charge of the
two Virasoro algebras, we are left with
\begin{equation*}
  [13 - \tfrac{6}{k_1 + 2} - 6(k_1 + 2)] +
  [13 - \tfrac{6}{k_2 + 2} - 6(k_2 + 2)] - \tfrac{1}{2} =
  \tfrac{15}{2} - \tfrac{3}{2 k_1 + 3} - 3(2 k_1 + 3),
\end{equation*}
which is the central charge of an~$N=1$ superVirasoro model.  This
suggests, therefore, that $\mathrm{Vir}(k_1)\oplus\mathrm{Vir}(k_2)$
extends to the~$N=1$ superVirasoro algebra.  It also suggests that the
sum~$\tSL2_{k_1}\oplus\mathrm{Vir}(k_2)$ can be extended, when~$n=2$,
to the affine superalgebra~$\widehat{osp}(1|2)_{k_1}$ by an `inverse'
Hamiltonian reduction process (cf.~\cite{[S-sl21]}).  Actually, it is
known~\cite{[FY931]} that the
coset~$\widehat{osp}(1|2)_{k_1}/\tSL2_{k_1}$ is a Virasoro algebra
with central charge
\begin{equation}
  c_V=\tfrac{2k_1}{2k_1+3}-\tfrac{3k_1}{k_1+2}=
  13 - \tfrac{6}{k_2+2} - 6(k_2+2),
\end{equation}
which can be obtained by Hamiltonian reduction from~$\tSL2_{k_2}$ for
$\tfrac{1}{k_1+2} + \tfrac{1}{k_2+2} = 2$, providing the second, dual
$\tSL2$ algebra in the sum.

\smallskip

In this paper, we study in detail the case of a dual pair
of~$\tSL2_{k_i}$ algebras, where the levels are related by
\eqref{eq:n} with~$n=1$ and, thus, a free scalar is needed to make the
$\tSL2_{k_1}\oplus\tSL2_{k_2}$ vertex operators local after taking the
quantum group trace. If we were to guess what conformal theory the
extended structure might be, a crude hint would be provided by
evaluating its central charge, that is,
\begin{equation}\label{0plus1}
  c_{\mathrm{extended}}
  =c_{s\ell(2)_{k_1}}+ c_{s\ell(2)_{k_2}}+c_{\mathrm{scalar}}
  =\tfrac{3k_1}{k_1 + 2} + \tfrac{3k_2}{k_2 + 2} + 1 
  \stackrel{\eqref{kk1}}{=} 0 + 1\,.
\end{equation}
It turns out that the conformal theory with vanishing central charge
on the right-hand side of the last equation is the~$\tSSL21$ affine
Lie superalgebra (which, remarkably, has the same number of bosonic
and fermionic currents and hence vanishing central charge of the
Sugawara energy-momentum tensor); the remaining~1 corresponds to
$\widehat{u}(1)$ (a free scalar theory). We derive this result in what
follows, but now we give some indirect arguments in favour of the
emergence of the~$\tSSL21_{k_1}$ algebra from
$\tSL2_{k_1}\oplus\tSL2_{k_2}$.  There exists a~$W$ algebra associated
with~$\tSSL21$; it can be arrived at, for example, by taking two
scalar fields and constructing the combinations that commute with
two~$\tSSL21$ screenings (which can be either both fermionic or one
bosonic and one fermionic).  This~$W$ algebra has several different
descriptions, one of which is that of the coset
\begin{equation*}
  \ww=\tSSL21\bigm/\tGL2\,.
\end{equation*}
The~$\tGL2=\tSL2\oplus u(1)$ algebra from the denominator provides the
first of the~$\tSL2$ algebras making the dual pair (this will
eventually become the~$\tSL2$ \textit{sub}algebra of~$\tSSL21$).  One
can, in principle, `reconstruct' the $\tSSL21$ algebra from
$\tGL2\oplus\ww$ (possibly with the help of some additional
constructions, for example choosing a particular bosonisation). On the
other hand, the coset~$W$ algebra can also be described as
\begin{equation*}
  \ww=\tSL2\bigm/u(1)\,.
\end{equation*}
Up to the~$u(1)$ mixing, therefore, the~$W$ algebra in the
`reconstruction'~$\tGL2\oplus\ww\to\tSSL21$ can be replaced with
$\tSL2$; this is the origin of the second (dual)~$\tSL2$ in our
construction~$\tSL2\oplus\tSL2\to\tSSL21$ (this second~$\tSL2$ is
obviously \textit{not} a subalgebra of~$\tSSL21$).  The above argument
is also strongly supported by the detailed knowledge we have of a
class of irreducible representations of~$\tSSL21$ at admissible level
$k_1=\tfrac{1}{u}-1$, $u\in\oN+1$, and their corresponding characters.
Their branching functions into characters of the
subalgebra~$\tSL2_{k_1}$ were shown in~\cite{[HT98],[BHT98]} to
involve characters of a rational torus~$A_{u(u-1)}$ and of the
parafermionic algebra~$\oZ_{u-1}$.  The latter can be obtained as the
coset~$\tSL2_{u-1}/\widehat{u}(1)$, providing us with the dual
$\tSL2_{k_2}$ in the construction of~$\tSSL21_{k_1}$ (note that one
indeed has \eqref{kk1} with~$k_2=u-1$).

But there is an additional remarkable observation: the~$u(1)$ mixings
involved in the above coset argument conspire so as to extend
$\tSSL21_{k_1}\oplus u(1)$ to the exceptional affine Lie superalgebra
$\hDD{k_2}_{k_1}$.
The crucial circumstance here is the relation existing between
conformal embeddings $\tSL2\oplus\tSL2\oplus\tSL2\subset\hD$ and the
dual~$\tSL2$ pairs.  Such an embedding is conformal if and only if two
of the~$\tSL2$ algebras are dual (and the third has level~1).
Extending~$\tSL2_{k_1}\oplus\tSL2_{k_2}\oplus\tSL2_1$ by (the quantum
trace of the bilinears in) the spin-$\hf$ vertex operators thus gives
a realisation of the~$\hDD{k_2}_{k_1}$ algebra.  The representations
of $\hD$ that can be obtained in this way are very special, however,
in that they appear to be exactly the direct sums of $\tSSL21$
representations over the spectral flow orbits.  Anyway, we do not
focus on the $\hD$ representations 
and concentrate instead on the~$\tSSL21$ algebra, whose
representations have been studied in more detail (the extension
to~$\hD$ is however very interesting because of its close relation to
the $N=4$ superconformal algebra, and its study certainly deserves to
be deepened).  Our original interest in the study of the~$\tSSL21$
representation theory, especially at admissible values of the level,
stems from the r\^ole this affine algebra might play in the
quantisation of the~$N=2$ non-critical
strings~\cite{[HY],[Yank],[FY932],[S-sl21]}.

{}From the $\tSSL21$ algebra standpoint, the representations
constructed by taking tensor products of~$\tSL2_{k_1}$, $\tSL2_{k_2}$,
and~$\tSL2_1$ representations are reducible, because all the~$\tSSL21$
generators commute with the third~$\hD$ Cartan current.  The
decomposition of such a representation~$\espaceN$ into
different~$\tSSL21$ representations is governed by the spectral flow:
along with each $\tSSL21$ representation, $\espaceN$~contains all of
its spectral-flow images (we saw this in~\eqref{decompose-Weyl} for
the vacuum representation).

Associated with this construction of~$\tSSL21$ representations are the
character identities in the form of sumrules between
$\tSSL21\oplus\widehat{u}(1)$ and~$\tSL2\oplus\tSL2\oplus\tSL2$
characters, which we give for Verma modules and for a class of
irreducible admissible representations.  Each sumrule is invariant
under a spectral flow whose generator can be added to
the~$\widehat{u}(1) \oplus \tSSL21$ generators to extend the algebra
to~$\hD$. It is therefore expected that each sumrule in a given sector
of the theory describes a~$\hD$ character belonging to a special class
corresponding to conformal embeddings.

As well as building $\tSSL21$ representations from tensor products of
$\tSL2$ representations as described above, we also construct
representations of the dual $\tSL2_{k_2}$ algebra starting with an
$\tSSL21_{k_1}$ representation. A key step in doing so was to
establish a correspondence between Verma modules of $\tSSL21_{k_1}$
and relaxed Verma modules of $\tSL2_{k_2}$. The latter provide one of
the starting points in constructing representations of the vertex
operator extensions of $\tSL2_{k_1}\oplus\tSL2_{k_2}$, and therefore,
the above correspondence is partly what allows one to invert, up to
spectral flow, the vertex operator construction of $\tSSL21$
representations.  We find several indications that the established
correspondences between representations have functorial properties;
however we leave for the future the very interesting separate problem
of constructing the functors between the $\tSSL21_{k_1}$ and
$\tSL2_{k_1}\oplus\tSL2_{k_2}$ representation theories.

\bigskip

This paper is organised as follows.  In Sec.~\ref{sec:basic}, we
describe the basic vertex operator construction allowing us to extend
$\tSL2_{k_1}\oplus\tSL2_{k_2}$ to $\tSSL21_{k_1}$.  In
Sec.~\ref{sec:D}, we study embeddings into the $\hD$ algebra, find the
relation between the conformal embeddings and the dual $\tSL2$ pairs,
and conclude that our vertex operator algebra extends
to~$\hDD{k_2}_{k_1}$. We then concentrate on $\tSSL21_{k_1}$
representations in Sec.~\ref{sec:reps}.  In
Secs.~\ref{sec:reconstructing} and~\ref{sec:reps-inverse}, we show how
the $\tSL2_{k_2}$ representations can be reconstructed starting from
$\tSSL21_{k_1}$ representations.  With the experience gained here, we
then address in Sec.~\ref{sec:construct} the problem of combining
$\tSL2_{k_1}$ and $\tSL2_{k_2}$ representations into a representation
of~$\tSSL21_{k_1}$.  In Sec.~\ref{sec:characters}, we consider the
decomposition of the $\tSSL21$ representations constructed; for Verma
modules, we confirm the decomposition formula by calculating the
characters in Sec.~\ref{sec:Verma-char}, and in Sec.~\ref{sec:char} we
give the character identities relating~$\tSSL21$ characters with those
of the dual~$\tSL2$ pair for an interesting class of
irreducible~$\tSSL21$ representations.  We conclude with remarks on
several related research directions.

Our conventions on the different algebras are summarised in the
appendices.  Appendix~\ref{sec:quantum-group} gives the basic
relations describing the quantum~$\SL2$ algebra and its `spin-$\half$'
representations; these are the representations realised on the vertex
operators involved in our construction.  Appendix~\ref{app:sl21}
summarises our conventions and some useful facts about the
affine~$\tSSL21$ algebra, and Appendix~\ref{app:D-algebra} describes
several basic facts about the exceptional Lie superalgebras~$\D$.  In
Appendix~\ref{app:ope}, we explicitly check that the operator product
expansions of the currents constructed in terms of vertex operators
yield the $\tSSL21$ algebra.

\section{\textbf{The vertex operator construction of
  $\protect\widehat{s\ell}(2|1)$}}\label{sec:basic}
\subsection{Dual $\protect\widehat{s\ell}(2)$ algebras and vertex
  operators}\label{sec:direct-construction} We take two algebras,
$\tSL2_{\p-2}$ and~$\tSL2_{p'-2}$ with
\begin{equation}\label{pp'relation}
  \frac{1}{\p} + \frac{1}{p'} = 1,
\end{equation}
and $\p\in\oC\setminus\{0,1\}$.  In terms of the levels~$k=\p-2$ and
$k'=p'-2$, this condition can also be written as~\eqref{kk1} (with
$k_1=k$ and~$k_2=k'$ from now on).

The~$\tSL2_k$ algebra can be described in terms of the
currents~$J^\pm(z)$ and~$J^0(z)$ with the operator products
\begin{equation}\label{sl2-OPE}
  J^0(z)J^{\pm}(w)=\frac{\pm J^{\pm}(w)}{z-w},\quad
  J^+(z)J^-(w)=\frac{k}{(z-w)^2} + \frac{2 J^{0}(w)}{z-w},\quad
  J^0(z)J^0(w)=\frac{k/2}{(z-w)^2},
\end{equation}
and similarly for the `primed'~$\tSL2_{k'}$ algebra.

In what follows, we let~$\mV_{j,k}$ denote an~$\tSL2$ module with the
highest-weight vector~$\ket{j,k}$ satisfying
\begin{equation}
  J^+_{\geq0}\,\ket{j,k}=0\,,\qquad
  J^0_{\geq1}\,\ket{j,k}=0\,,\qquad
  J^-_{\geq1}\,\ket{j,k}=0\,,\qquad
  J^0_0\,\ket{j,k}=j\,\ket{j,k}\,.
\end{equation}

A fundamental r\^ole in our construction is played by the vertex
operators corresponding to the top level states in the
\textit{irreducible} spin-$\half$~$\tSL2_k$ representation,
\begin{equation}
  J^+_0\,\ket{\thalf,k}=0\,,\qquad
  J^-_0\,\ket{\thalf,k}=\ket{-\thalf,k}\,,\qquad
  J^-_0\,\ket{-\thalf,k}=0\,,
\end{equation}
and similarly in the primed sector.  Let $V_{\half}$ and $V_{-\half}$
be the vertex operators corresponding to $\ket{\thalf,k}$ and
$\ket{-\thalf,k}$, respectively.  So, $V_{\half}(z)$
and~$V_{-\half}(z)$ make up an $\SL2$ doublet, which we express by
writing~$(V_{\half}(z),V_{-\half}(z))\in\oC^2(z)$.  In addition, each
of these operators has a second component and thus is an element of a
two-dimensional space~$\oC_q^2(z)$ associated with the point~$z$,
where the subscript indicates that this space is a representation of
the quantum group $\SL2_q$, with quantum group parameter~$q=e^{2 \pi
  i/p}$~\cite{[BMP-q]}.  The~$\SL2_q$-module is the quotient of a
Verma module over a singular vector; in the notations of
Appendix~\ref{sec:quantum-group}, where we collect some simple facts
about quantum groups, this can be~$\qmV_{\epsilon,1}$
with~$\epsilon=\pm1$.  Thus, the vertex operator associated with the
spin-$\half$ representation is a `two-indexed object' with~$\SL2$
index~$a=1,2$ and quantum group index $\alpha=1,2$, and therefore a
tensor component in the space $\oC^2(z)\tensor\oC_q^2(z)$.  Its action
on~$\tSL2$ modules is given by
\begin{equation}
  \oC^2(z)\tensor\oC^2_q(z):
  \mV_{j,k}\to{}\mV_{j+\half,k}\tensor{}z^{\Delta_{j+\half}-\Delta_j-
    \Delta_{\half}}\,\oC((z))
  \oplus
  \mV_{j-\half,k}\tensor{}z^{\Delta_{j-\half}-\Delta_j-
    \Delta_{\half}}\,\oC((z))\,,
\end{equation}
where~$\Delta_j=\frac{j(j+1)}{p}$ and~$\oC((z))$ is the completion of
the Laurent polynomial ring~$\oC[z,z^{-1}]$ with positive powers
of~$z$.  In what follows, we distinguish one of the quantum group
components from the other with a tilde; in the free-field realisation
that we use they act as
\begin{equation}\label{act-01}
  V_{\pm\half}(z):
  \mV_{j,k}\to{}\mV_{j+\half,k}\tensor{}z^{\Delta_{j+\half}-\Delta_j-
    \Delta_{\half}}\,\oC((z)),
\end{equation}
and 
\begin{equation}\label{act-02}
  \wt{V}_{\pm\half}(z):
  \mV_{j,k}\to{}\mV_{j-\half,k}\tensor{}z^{\Delta_{j-\half}-\Delta_j-
    \Delta_{\half}}\,\oC((z)).
\end{equation}

The monodromies of vertex operators are defined by analytically
continuing from the real line, where they are fixed as
\begin{equation}\label{rearrange}
  V_{j_1,m_1}(w)\,V_{j_2,m_2}(z) = e^{\frac{2\pi i}{p}j_1j_2}\,
  V_{j_2,m_2}(z)\,V_{j_1,m_1}(w)\,,\qquad
  \Re z>\Re w\,.
\end{equation}
To evaluate the operator product of vertex operators~$\prod_i
V_{j_i,m_i}(z_i)$, we use~\eqref{rearrange} to rearrange the
operators~$V_{j_1,m_1}(z_1)\dots V_{j_N,m_N}(z_N)$ such that~$\Re
z_i>\Re z_{i+1}$, and then use the operator products
$V_{j_i,m_i}(z_i)\cdot V_{j_\ell,m_\ell}(z_\ell)\sim(z_i-z_\ell)^{j_i
  j_\ell/(2p)}$ with~$\Re z_i>\Re z_\ell$.  The point of the
subsequent construction is that different factors of the form
$(z_i-z_\ell)^{j_i j_\ell/(2p)}$ combine into \textit{integral} powers,
after which we no longer need to assume~$\Re z_i>\Re z_{i+1}$.

\medskip

Consider now the dual~$\tSL2$ algebra, with the vertex operators
viewed as elements of~$\oC'{}^2(z)\tensor\oC_{q'}^2(z)$, where the
quantum group parameter is~$q'=e^{2\pi i/p'}=q^{-1}$ in view
of~\eqref{pp'relation}.  Similarly to the above, the monodromies in
the primed sector are given by~$e^{\frac{2\pi i}{p'}j_1j_2}=e^{2\pi i
  j_1j_2}\,q^{-j_1j_2}$.

The module~$\qmV'_{\epsilon',1}$ over~$\SL2_{q^{-1}}$ is also a module
over~$\SL2_q$, as we see in~\eqref{q-inverse1}--\eqref{q-inverse2}.
Moreover, depending on whether~$\epsilon'\epsilon$ is~$-1$ or~$+1$,
this is (almost) the dual of~$\qmV_{\epsilon,1}$.  Thus, there exists
the trace~\eqref{q-trace} on the tensor product of the unprimed and
primed vertex operators.  Taking this trace `over the quantum group
index' leaves us with a four-dimensional space:
\begin{equation}\label{take-trace}
  \oC{}^2(z)\tensor\oC_{q}^2(z) \tensor
  \oC'{}^2(z)\tensor\oC_{q'}^2(z)
  \xrightarrow{\langle\cdot,\cdot\rangle}
  \oC^2(z)\tensor\oC'{}^2(z),
\end{equation}
(more precisely, there is an extra factor on the right-hand side given
by~$\qmV_{-1,0}$ for~$\epsilon\epsilon'=-1$). 

Now, the monodromies of the operators from
$\oC^2(z)\tensor\oC'{}^2(z)$ with each other are given by
$q^{j_1j_2}\cdot e^{2\pi i j_1j_2}\,q^{-j_1j_2}= e^{2\pi i
  \half\cdot\half}=e^{\frac{\pi i}{2}}$.  Therefore, the operators
from~$\oC^2(z)\tensor\oC'{}^2(z)$ are `almost' local with respect to
each other.  They can be made local by introducing an auxiliary scalar
field~$f$ with the operator product
\begin{equation}\label{ff}
  f(z)\,f(w)=\log(z-w),
\end{equation}
and multiplying the operators from~$\oC^2(z)\tensor\oC'{}^2(z)$ with
$e^{\pm\frac{1}{\sqrt{2}}f(z)}$.  Note that adding up the dimensions
of the primed and unprimed~$V_{\pm\half}$ operators, we find
\begin{equation}
  \frac{\half(\half+1)}{p} + \frac{\half(\half+1)}{p'}=\frac{3}{4}\,,
\end{equation}
which after multiplying with~$e^{\pm\frac{1}{\sqrt{2}}f(z)}$
becomes~$\frac{3}{4}+\frac{1}{4}=1$.  Their monodromies are
$e^{\frac{\pi i}{2}}\cdot e^{\frac{\pi i}{2}}=-1$.  Denoting
by~$\oC_1^2(z)$ the two-dimensional space generated by
$e^{\pm\frac{1}{\sqrt{2}}f(z)}$, we thus see that the basis elements
of~$\oC^2(z)\tensor\oC'{}^2(z)\tensor\oC_1^2(z)$ represent
\textit{eight dimension-1 fermionic currents}.

It turns out that these eight fermionic currents generate an affine
Lie superalgebra~$\hD$, as will be discussed in Sec.~\ref{sec:D}.
Until then, we concentrate on the algebra generated by \textit{four}
of these eight fermionic currents, which is the affine Lie
superalgebra~$\tSSL21$ (see Appendix~\ref{app:sl21} for notations and
conventions).  We, thus, define
\begin{equation}\label{E0}
  \begin{split}
    E^1 ={}& -\left( \wt{V}_{\half}\,V'_{\half} +
      V_{\half}\,\wt{V}'_{\half}\right)
    \Phi_{\frac{1}{\sqrt{2}}}\,,\\
    E^2 ={}& \tfrac{i}{\pi p'}\cos\tfrac{\pi}{p}\left(
      \wt{V}_{\half}\,V'_{-\half} +
      V_{\half}\,\wt{V}'_{-\half}\right)
    \Phi_{-\frac{1}{\sqrt{2}}}
  \end{split}
\end{equation}
and
\begin{equation}\label{F0}
  \begin{split}
    F^1 ={}& \tfrac{i}{\pi p'}\cos\tfrac{\pi}{p}\left(
      \wt{V}_{-\half}\,V'_{-\half}
      + V_{-\half}\,\wt{V}'_{-\half}\right)
    \Phi_{-\frac{1}{\sqrt{2}}}\,,\\
    F^2 ={}& \left(
      \wt{V}_{-\half}\,V'_{\half}
      + V_{-\half}\,\wt{V}'_{\half}\right)
    \Phi_{\frac{1}{\sqrt{2}}},
  \end{split}
\end{equation}
where
\begin{equation}
  \Phi_{\alpha}(z) = e^{\alpha f(z)}\,.
\end{equation}
\begin{Lemma} \label{Lemma:sl21}
  Let the \;$\tSL2_k$\; and \;$\tSL2_{k'}$\; algebras be generated by
  the currents~$(J^+,\;J^0,\;J^-)$ and~$(J'{}^+,J'{}^0,J'{}^-)$,
  respectively.  Then the fermionic currents~\eqref{E0}--\eqref{F0}
  generate the affine $\tSSL21$ algebra of level~$k=p-2$.  Its~$\tSL2$
  subalgebra coincides with $\tSL2_k$, i.e.,
  \begin{equation}\label{Hminus0}
    E^{12} ={} J^+,\qquad
    H^- = {}J^0,\qquad
    F^{12} = {}J^-,
  \end{equation}
  and its $\u(1)$ subalgebra is generated by the current
  \begin{equation}\label{Hplus0}
     H^+ ={}(k+1)J'{}^0 - \tfrac{k}{\sqrt{2}}\,\d f\,.
  \end{equation}
  The current
  \begin{equation}\label{A-current0}
    A ={}\sqrt{2(k+1)}(J'{}^0 - \tfrac{1}{\sqrt{2}}\d f)
  \end{equation}
  commutes with all the~$\tSSL21$ currents.
\end{Lemma}

We prove the Lemma by evaluating the operator product expansions of
the currents \eqref{E0}--\eqref{Hplus0}.  Some of them are
straightforward. In particular, one has,
\begin{equation}
  H^+(z)\cdot H^+(w)=
  \frac{-k/2}{(z-w)^2}, 
\end{equation}
and since the generators in~\eqref{F0} are obtained from those
in~\eqref{E0} by the action of the~$\tSL2$ subalgebra (the unprimed
$\tSL2$), we also effortlessly obtain the operator product expansions
\begin{gather}
  F^{12}(z)\,E^1(w) = \frac{-F^2}{z-w}\,,
  \qquad
  F^{12}(z)\,E^2(w) = \frac{F^1}{z-w},\\
  E^{12}(z)\,F^1(w) = \frac{E^2}{z-w}\,,
  \qquad
  E^{12}(z)\,F^2(w) = \frac{-E^1}{z-w}\,.
\end{gather}

The remaining relevant operator product expansions are calculated in
Appendix~\ref{app:ope} with the help of the Wakimoto bosonisation
discussed in the following section. The latter explicitly shows the
quantum group action in terms of the screening operator and contour
manipulations (more precisely, screenings (one screening in the
current case of~$\SL2_q$) generate the nilpotent subalgebra
of~$\SL2_q$, while the entire quantum group is represented in terms of
a construction~\cite{[GS],[RRRa]} using more involved contour
operations).

Let us remark at this point that the currents in formulas~\eqref{E0}
and~\eqref{F0} are normalised in accordance to a particular choice of
screening contour in the Wakimoto representation. In general, the
normalisation depends on the precise definition of
the screened vertex operators, cf.~\cite{[GS],[RRRa]}.

\subsection{The Wakimoto representations of
  $\protect\widehat{s\ell}(2)$ and screened operator
  products}\label{sec:waki} We define a first-order bosonic system and
an independent current by the operator products
\begin{equation}
  \beta(z)\,\gamma(w)=\frac{-1}{z-w}\,,\qquad
  \d\varphi(z)\,\d\varphi(w)=\frac{1}{(z-w)^2}\,.
\end{equation}
In terms of these free fields, the~$\tSL2_{\p-2}$ currents can be
written as
\begin{equation}\label{Wakimoto}
  \begin{split}
    J^+ ={}& -\beta\,,\displaybreak[0]\\
    J^0 ={}& \sqrt{\tfrac{\p}{2}}\,\d\varphi  + \beta\gamma\,,\\
    J^- ={}& \beta\gamma\gamma + \sqrt{2\p}\,\gamma\d\varphi +
    (\p-2)\d\gamma\,,
  \end{split}
\end{equation}
and the vertex operators are,
\begin{equation}\label{vo}
  V_{\half}=e^{\frac{1}{\sqrt{2\p}}\varphi},\qquad
  V_{-\half}=\gamma\,e^{\frac{1}{\sqrt{2\p}}\varphi}.
\end{equation}

Similar formulae for the primed~$\tSL2$ algebra are obtained by
consistently replacing the free fields with the primed ones and
replacing~$p$ with~$p'$.

The Wakimoto bosonisation screening acts on a vertex operator as
\begin{equation}
  S:V(z)\mapsto 
  \tfrac{1}{2\pi i}\int_{C_z} dx\, S(x) V(z)\,.
\end{equation}
The integration contour can be chosen in different ways; one of the
possibilities is to take the contour
\begin{equation}
  \unitlength=1pt
  \begin{picture}(480,20)
    \put(100,10){
      \put(50,-.5){\vector(1,0){8}}
      {\thicklines
        \qbezier(0,0)(110,0)(130,-5)
        \qbezier(130,-5)(160,-10)(160,5)
        }
      \put(0,10){
        \put(62,.5){\vector(-1,0){8}}\thicklines
        \qbezier(0,0)(110,0)(130,5)
        \qbezier(130,5)(160,10)(160,-5)
        }
      \put(145,2){$\bullet$}
      \put(0,5){\line(1,0){145}}
      }
  \end{picture}
\end{equation}
where the straight line shows the branch cut due to the nonlocality in
the operator product of~$S(x)$ and~$V(z)$.  As is well known (see,
e.g.,~\cite{[GS],[RRRa],[BMP-f]}), up to a normalisation factor this
contour can be replaced with
the contour running along the cut (in this case the integral is
defined by analytic continuation).  We choose the latter possibility
in what follows;
it turns out that we will have fewer~$(-1)^\alpha$ factors if we
define
\begin{equation}\label{S-act}
  S:V(z)\mapsto\widetilde{V}(z)\equiv
  \int_z^\infty dx\, S(x) V(z)\,,
\end{equation}
where the integral has to be evaluated for those parameter values
where it converges and then analytically continued.  

In order to verify the~$\tSSL21$ operator product expansions between
the currents constructed in the previous section, we need to evaluate
products involving screened and unscreened vertex operators in the
combinations determined by the quantum-group
trace~\eqref{take-trace}, typically,
\begin{equation*}
  (SV_1(z)\,V'_1(z) + V_1(z)\,S'V'_1(z))\cdot
  (SV_2(w)\,V'_2(w) + V_2(w)\,S'V'_2(w)),
\end{equation*}
where~$V$ and~$V'$ some $\tSL2$ and~$\tSL2'$ vertex operators.  In all
of the operator products that we encounter in what follows, a nonzero
contribution comes from the terms
\begin{multline}\label{ope-ww}
  SV_1(z)\,V'_1(z)\,V_2(w)\,S'V'_2(w) +
  V_1(z)\,S'V'_1(z)\,SV_2(w)\,V'_2(w)={}\\
  {}=\int_z^\infty\,du\,S(u)\,V_1(z)\,V'_1(z)\cdot
  \int_w^\infty\,dx\,S'(u)\,V_2(w)\,V'_2(w)+{}\\
  {}+ \int_z^\infty\,du\,S'(u)\,V_1(z)\,V'_1(z)\cdot
  \int_w^\infty\,dx\,S(u)\,V_2(w)\,V'_2(w)\,,
\end{multline}

We first assume~$\Re z>\Re w$, evaluate these operators products, and
analytically continue the result. The above expression~\eqref{ope-ww}
thus becomes,
\begin{multline}
  \int_z^\infty\! du\,S(u)\cdot V_1(z)\cdot V_2(w)\;
  \int_w^z dx\,V'_1(z)\cdot S'(x)\cdot V'_2(w)+{}\\
  {}+\int_w^z du\,V_1(z)\cdot S(u)\cdot V_2(w)\;
  \int_z^\infty\! dx\,S'(x)\cdot V'_1(z)\cdot V'_2(w)+{}\\
  {}+(-1)^{\alpha'_1}\int_z^\infty\!\,du\,S(u)\cdot V_1(z)
  \cdot V_2(w) 
  \int_z^\infty\! dx\,S'(x)\cdot V'_1(z)\cdot V'_2(w) +{}\\
  {}+(-1)^{\alpha_1}\int_z^\infty\!\,du\,S(u)\cdot V_1(z)\cdot
  V_2(w) 
  \int_z^\infty\! dx\,S'(x)\cdot V'_1(z)\cdot V'_2(w)\,,
\end{multline}
where~$(-1)^{\alpha'_1}$ and~$(-1)^{\alpha_1}$ are the monodromies
coming from
\begin{equation}
  V'_1(z)\,S'(x)=(-1)^{\alpha'_1}\,S'(x)\,V'_1(z)\,,\qquad
  V_1(z)\,S(x)=(-1)^{\alpha_1}\,S(x)\,V_1(z)\,,
\end{equation}
(where~$\Re x>\Re z$). When~$V_1$ is either
$e^{\frac{1}{\sqrt{2p}}\varphi}$ or~$\gamma
e^{\frac{1}{\sqrt{2p}}\varphi}$, as prescribed in \eqref{vo}, the
monodromies are~$(-1)^{\alpha_1}=e^{-\pi i/p}$ and
$(-1)^{\alpha'_1}=e^{-\pi i/p'}= -e^{\pi i/p}$, and \eqref{ope-ww}
becomes,
\begin{multline}\label{recipe}
  (SV_1(z)\,V'_1(z) + V_1(z)\,S'V'_1(z))\cdot
  (SV_2(w)\,V'_2(w) + V_2(w)\,S'V'_2(w))={}\displaybreak[0]\\
  {}=-2i\sin\tfrac{\pi}{p}
  \int_z^\infty\!\,du\,S(u)\,V_1(z)\,V_2(w) 
  \int_z^\infty\! dx\,S'(x)\,V'_1(z)\,V'_2(w) +{}\displaybreak[3]\\
  \kern70pt{}+\int_z^\infty\! du\,S(u)\,V_1(z)\,V_2(w)\;
  \int_w^z dx\,V'_1(z)\, S'(x)\,V'_2(w)+{}\\
  {}+\int_w^z du\,V_1(z)\, S(u)\,V_2(w)\;
  \int_z^\infty\! dx\,S'(x)\,V'_1(z)\,V'_2(w)\,.\kern80pt
\end{multline}

Appendix~\ref{app:ope} is devoted to explicit checks of the~$\tSSL21$
operator product expansions in the Wakimoto representation.  
In addition to establishing the operator product expansions, the
Wakimoto bosonisation is a useful tool to verify the relation between
the energy-momentum tensors underlying the central charge balance in
Eq.~\eqref{0plus1}. 
\begin{Lemma}\label{lemma:T}  The combined energy-momentum tensors of
  the~$\tSSL21$ algebra generated by~\eqref{E0}--\eqref{Hplus0} and
  the independent~$u(1)$ current~$A(z)$, Eq.~\eqref{A-current0},
  equals the sum of the energy-momentum tensors for two~$\tSL2$ and
  the free~$u(1)$ current~$\d f(z)$:
  \begin{equation}\label{T-identity}
    T_{\mathrm{Sug}} + \thalf A A =
    \underbrace{\thalf \d\varphi \d\varphi
      - \tfrac{1}{\sqrt{2\p}}\d^2\varphi -
    \beta\d\gamma}_{\tSL2_{\p-2}} 
    {}+{}
    \underbrace{\thalf \d\varphi'\d\varphi'  
      - \tfrac{1}{\sqrt{2p'}}\d^2\varphi' -
      \beta'\d\gamma'}_{\tSL2_{p'-2}}
    + \thalf \d f\d f\,.
  \end{equation}
\end{Lemma}
\noindent{\textsc{Proof.}}
To show this, we directly calculate the Sugawara energy-momentum
tensor given by Eq.~\eqref{Tsug-sl21}.  First, the~$\tSL2$ subalgebra
contribution is given by,
\begin{equation}
  \tfrac{1}{\p-1}\bigl(E^{12}\,F^{12} + \Hminus\, \Hminus \bigr) =
  -\beta\d\gamma + \tfrac{1}{\p - 1}\,\d\beta\gamma +
  \tfrac{\p}{2(\p - 1)}\,\d\varphi\d\varphi,
\end{equation}
(which, obviously, is \textit{not} the~$\tSL2$ energy-momentum
tensor).  Second, the fermionic contribution can be read off
Appendix~\ref{app:ope},
\begin{multline}
  \tfrac{1}{\p-1}\left(E^1F^1 - E^2F^2\right)=
  \tfrac{1}{2(1 - \p)}\d\varphi\d\varphi + 
  \thalf\d\varphi'\d\varphi' + 
  \sqrt{p'}\d\varphi'\d f - 
  \tfrac{\p - 2}{2(\p - 1)}\d f\d f \\
  {}-\beta'\d\gamma' -
  \tfrac{1}{\p - 1}\d\beta\,\gamma +
  \sqrt{2}\beta'\gamma'\d f - \tfrac{1}{\sqrt{2\p}}\d^2\varphi - 
  \tfrac{1}{\sqrt{2p'}}\d^2\varphi',
\end{multline}
and, in addition, we have 
\begin{multline}
  -\tfrac{1}{\p-1}\,H^+\,H^+=
  \sqrt{p'}(\p - 2)\d\varphi' \d f
  -\tfrac{(\p - 2)^2}{2(\p - 1)}\d f \d f 
  - \tfrac{\p}{2}\d\varphi' \d\varphi'\\
  {}+ \sqrt{2}(\p - 2)\beta' \gamma' \d f +
  (\p-1)\beta' \d\gamma' -
  (\p-1)\d\beta' \gamma' -
  \sqrt{2\p(\p-1)}\beta' \gamma' \d\varphi' -
  (\p-1)\beta'\beta' \gamma' \gamma'.
\end{multline}
Finally, the free scalar current~$A$ has the energy-momentum tensor
\begin{multline}
  \thalf A A  =
  \tfrac{\p}{2}\d\varphi' \d\varphi' - 
  \sqrt{\p(\p - 1)}\,\d\varphi' \d f + 
  \thalf (\p - 1) \d f\d f
  +\sqrt{2\p(\p - 1)}\beta'\gamma'\d\varphi'\\
   {} - \sqrt{2}(\p - 1) \beta'\gamma'\d f
     + (\p - 1) \beta' \beta'\gamma'\gamma'
  - (\p - 1) \beta'\d\gamma' +
  (\p - 1) \d\beta'\gamma'.  
\end{multline}
Adding all this together, we arrive at~\eqref{T-identity}.
\hfill\mbox{\rule{.5em}{.5em}}

\section{\textbf{Extending $\protect\widehat{s\ell}(2|1)_k$ to the
    $\protect\widehat{D}(2|1;k')_k$ algebra}}\label{sec:D} So far, we
have seen that the appropriate bilinear combinations of vertex
operators extend the $\tSL2_k\oplus\tSL2_{k'}\oplus\u(1)$ algebra
to~$\tSSL21_k$.  We now show that this is in fact part of a larger
construction, that of the exceptional affine Lie superalgebra~$\hD$.

\subsection{The `auxiliary' level-1~$\protect\widehat{s\ell}(2)$
  algebra} We first use the auxiliary scalar~$f$ to construct
an~$\tSL2$ algebra of level~1,
\begin{equation}\label{aux-sl2}
  j^+(z)=e^{\sqrt{2}f(z)}\,,\qquad
  j^0(z)=\tfrac{1}{\sqrt{2}}\,\d f(z)\,,\qquad
  j^-(z)=e^{-\sqrt{2}f(z)}\,.
\end{equation}
Then the two Cartan currents of~$\tSSL21 _{p-2}$ and the extra
$\widehat{u}(1)$ current become
\begin{align}
  H^-={}&J^0\,,\label{Hminus2}\displaybreak[1]\\
  H^+ ={}&\tfrac{p}{p'}{J^0}' - (p - 2)\,j^0
  \,,\label{Hplus2}\displaybreak[1]\\
  A ={}& \sqrt{2(p-1)}({J^0}' - j^0)\,,\label{A-current2}
\end{align}
where, as before,~${J^0}'$ is the Cartan current of~$\tSL2_{p'-2}$
and~$J^0$ is that of~$\tSL2_{p-2}$. We thus clearly see the interplay
of three~$\tSL2$ algebras, two at dual levels~$k=p-2$, $k'=p'-2$, and
one at level~1, in our construction of~$\tSSL21$.  
  
In the next subsection, we discuss how to embed~$\tSL2 \oplus \tSL2
\oplus \tSL2$ and~$\tSSL21\oplus\u(1)$ in the affine
superalgebra~$\hD$ in such a way that the~$\widehat{u}(1)$
currents~\eqref{Hminus2}--\eqref{A-current2} are recovered.  This will
mean that the \textit{eight} basis elements of
$\oC^2(z)\tensor\oC^{'2}(z)\tensor\oC_1^2(z)$ (four of which were
explicitly given in Eqs.~\eqref{E0}--\eqref{F0} in terms of vertex
operators) represent in fact the~$\hD$ fermions.

\subsection{Conformal embeddings in
  $\protect\widehat{D}(2|1;\alpha)$}\label{sec:D-algebra} Our
conventions on $\D$ are summarised in Appendix~\ref{app:D-algebra}.
Using the notations for the roots introduced there, we write the
bosonic subalgebra of $\D$ as $s\ell(2)^{(\ab)} \oplus
s\ell(2)^{(\atr)}\oplus s\ell(2)^{(\at)}$.  When~$s\ell(2|1)$ is
regularly embedded in~$\D$ (i.e.\ when the roots of the subalgebra are
chosen as a subset of the roots of the embedding algebra), any of
these three~$\SL2$ algebras can be taken as the~$\SL2$ subalgebra
of~$s\ell(2|1)$, for which we introduce the notation (reminding us of
the~$H^-$ current) $s\ell(2)^{(\am)}$, so that~$\am=\ab,\atr$
or~$\at$.  For each choice of~$\am$, there exist two regular
embeddings of~$s\ell(2|1)$ in~$\D$.  The corresponding~$s\ell(2|1)$
root lattices contain~$\am$ and four of the eight odd roots of~$\D$.
{}From now on, we use the parameter~$\gamma =
\frac{\alpha}{1+\alpha}$, $\gamma \in \oC \setminus \{0,1,\infty\}$
instead of~$\alpha$.  This parameter governs the norm of~$\ab$
and~$\atr$, and~$\at$ is always the longest root if $\g \in [\hf,1[$
(see Appendix~\ref{app:D-algebra}).

For the affine algebras, the levels of the algebras involved in the
embedding
\begin{equation}\label{sl21-emb}
  \tSSL21 _{\lambda} \oplus \widehat{u}(1) \subset
  \hDD{\tfrac{\gamma}{1-\gamma}} _{\kappa}
\end{equation}
are related as~$\lambda = -\frac{\kappa}{\gamma},
-\frac{\kappa}{1-\gamma}$ or~$ \kappa$ according to whether~$\am=\ab$,
$\atr$ or~$\at$.  On the other hand, one also has the regular
embedding,
\begin{equation}\label{regular-embedding}
  \tSL2^{(\alpha_2)}_{-\frac{\kappa}{\gamma}} \oplus
  \tSL2^{(\alpha_3)}_{\frac{-\kappa}{1-\gamma }} \oplus
  \tSL2^{(\alpha_{\theta})}_{\kappa} \subset
  \hDD{\tfrac{\gamma}{1-\gamma}}_{\kappa}\,.
\end{equation}

We now note that the Sugawara central charge of
$\hDD{\frac{\gamma}{1-\gamma}}_{\kappa}$ is 1 for {\em any\/} value of
the level~$\kappa$, because the dual Coxeter number
of~$\hDD{\frac{\gamma}{1-\gamma}}_{\kappa}$ is zero and its
superdimension (number of bosonic generators minus number of fermionic
generators) is one; hence, the corresponding central charge is one.
But the central charge of $\tSSL21 _{\lambda} \oplus \widehat{u}(1)$
is also~1 for any level $\lambda$, this time because the
superdimension of~$\SSL21$ is zero, and it turns out that for certain
relations between the level~$\kappa$ and the parameter~$\gamma$, which
we analyse below, the central charge of
the~$\tSL2_{\lambda_1}\oplus\tSL2_{\lambda_2}\oplus\tSL2_{\lambda_3}$
subalgebra of $\hDD{\frac{\gamma}{1-\gamma}} _{\kappa}$ is
\textit{also} one.  It is therefore not surprising that~$\tSL2 \oplus
\tSL2 \oplus \tSL2$ and $\tSSL21 \oplus \widehat{u}(1)$ are intimately
linked, a fact already noticed earlier when the energy-momentum
tensors of the two theories were shown to coincide in a free field
representation~\eqref{T-identity}, and also seen from the
representation decompositions in Sec.~\ref{sec:characters} and the
character identities~\eqref{uu-sumrule}.

The relation between~$\tSL2 \oplus \tSL2 \oplus \tSL2$ and~$\tSSL21
\oplus \widehat{u}(1)$ is based on the fact that the embeddings into
$\hDD{\frac{\gamma}{1-\gamma}} _{\kappa}$ which we consider
\textit{give rise to dual~$\tSL2$ pairs}:
\begin{Lemma}\label{Lemma:cc-1}
  The Sugawara central charge
  of~$\tSL2^{(\alpha_2)}_{-\frac{\kappa}{\gamma}} \oplus
  \tSL2^{(\alpha_3)}_{\frac{-\kappa}{1-\gamma }} \oplus
  \tSL2^{(\alpha_{\theta})}_{\kappa}$ is one if and only if two of
  the~$\tSL2$ subalgebras are dual to each other in the sense of
  Eq.~\eqref{kk1}.  Moreover, when this is so, the third~$\tSL2$
  algebra has level~1.
\end{Lemma}
\noindent\textsc{Proof}.
Indeed, adding up the central charges on the left-hand side
of~\eqref{regular-embedding}, we find the sum is one for~$\kappa
=\gamma -1$ (in which case~$\tSL2^{(\alpha_2)}$ and
$\tSL2^{(\alpha_{\theta})}$ are dual and~$\tSL2^{(\alpha_3)}$ has
level~1), for~$\kappa = -\gamma$ (with~$\tSL2^{(\alpha_3)}$ and
$\tSL2^{(\alpha_\theta)}$ dual), or for~$\kappa = 1$ (with
$\tSL2^{(\alpha_2)}$ and~$\tSL2^{(\alpha_3)}$ dual).  Conversely, we
have to consider three different possibilities to pick out two dual
$\tSL2$ algebras.
\begin{enumerate}

\item Taking the pair~$\tSL2^{(\alpha_2)}$ and
  $\tSL2^{(\alpha_{\theta})}$ leads to~$\kappa =\gamma -1$ and the
  relevant embedding is
  \begin{equation}\label{case-2}
    \tSL2^{(\alpha_2)}_{\frac{1-\gamma}{\gamma}} \oplus
    \tSL2^{(\alpha_3)}_1 \oplus
    \tSL2^{(\alpha_{\theta})}_{\gamma -1}\subset
    \hDD{\tfrac{\gamma}{1-\gamma}}_{\gamma -1}\,.
  \end{equation}  

\item For the pair~$\tSL2^{(\alpha_3)}$ and
  $\tSL2^{(\alpha_{\theta})}$ chosen as dual, $\kappa=-\gamma$ and the
  relevant embedding is
  \begin{equation}\label{case-3}
    \tSL2^{(\alpha_2)}_1 \oplus
    \tSL2^{(\alpha_3)}_{\frac{\gamma}{1-\gamma}} \oplus
    \tSL2^{(\alpha_{\theta})}_{-\gamma} \subset
    \hDD{\tfrac{\gamma}{1-\gamma}}_{-\gamma}.
  \end{equation}
  
\item Taking~$\tSL2^{(\alpha_2)}$ and~$\SL2^{(\alpha_3)}$ as the dual
  pair leads to~$\kappa=1$ and, thus, to the embedding
  \begin{equation}\label{case-1}
    \tSL2^{(\alpha_2)}_{-\frac{1}{\gamma}} \oplus
    \tSL2^{(\alpha_3)}_{\frac{1}{\gamma -1}} \oplus
    \tSL2^{(\alpha_{\theta})}_{1} \subset
    \hDD{\tfrac{\gamma}{1-\gamma}}_{1}\,.
  \end{equation}
\end{enumerate}
This completes the proof of the lemma.  \hfill\mbox{\rule{.5em}{.5em}}

So we have two subalgebras of
$\widehat{g}=\hDD{\tfrac{\gamma}{1-\gamma}}_{\kappa}$, namely
$\widehat{h}_1= \tSSL21 _{\lambda} \oplus \widehat{u}(1)$ and
$\widehat{h}_2=\tSL2 _{\lambda_1}^{(\alpha_2)} \oplus \tSL2
_{\lambda_2}^{(\alpha_3)} \oplus \tSL2 _{\lambda_3}^{(\at)}$, whose
central charges coincide with that of~$\widehat{g}$ as long as two of
the three levels~$k_1=-\tfrac{\kappa}{\g}$,
$k_2=-\tfrac{\kappa}{1-\g}$, and $k_3=\kappa$ obey $(k_a+1)(k_b+1)=1$.

For unitary representations, we would then conclude that embeddings
\eqref{sl21-emb}, \eqref{case-2}, \eqref{case-3}, and \eqref{case-1}
are conformal, since the coset Virasoro algebra
$L={\cL}_{g}-{\cL}_{h}$ has vanishing central charge, and the only
\textit{unitary} representation of the coset algebra in this case is
the trivial one~\cite{[GO]}, leading to the following formal relation
between characters,
\begin{equation}
  \Tr_{\Lambda}q^{\cL_{g}}= \Tr_{\Lambda}q^{\cL_{h}}.
\end{equation}
In the case at hand, though, we are dealing with non unitary
representations.  A sum (of products) of characters corresponding to
$\widehat{h}_1$ or to~$\widehat{h}_2$ representations can only be
interpreted as an irreducible~$\widehat{g}$ character if the
corresponding representation is acted upon trivially by the coset
Virasoro algebra.  We want to argue that, in order for this to be so
in the case of a~$\widehat{g}$ representation with highest vector
$|\Lambda\rangle$, it is sufficient to demand that the state
$L_{-2}|\Lambda\rangle$ is null (i.e.,~singular or a descendant of a
singular state), and that the same condition holds for the vacuum
representation, i.e.~$L_{-2}|{0}\rangle$ is null, where~$|{0}\rangle$
is the usual~$sl(2,C)$ invariant vacuum, which is also invariant under
the action of the finite dimensional algebra~$g$. Indeed, the operator
product expansion of~$L$ with one of the currents $J^{a}(\zeta)$
of~$\widehat{g}$ is of the form,
\begin{equation}
  L(z)J^{a}(\zeta)=
  \frac{X(\zeta)}{(z-\zeta)^{2}}+\frac{Y(\zeta)}{z-\zeta}+
  \text{regular}, 
\end{equation}
and the state associated with the residue of the double pole is,
\begin{equation}
  X_{-1}|{0}\rangle=
  (J^{a})_{1}L_{-2}|{0}\rangle=\lambda 
  (J^{a})_{-1}|{0}\rangle ,
\end{equation}
for some constant~$\lambda$. Since~$(J^{a})_{-1}|\Lambda_{0}\rangle$
is not null and yet our assumption demands that the vector
$(J^{a})_{1}L_{-2}|\Lambda_{0}\rangle$ vanishes or is null, it follows
that~$\lambda=0$, so that~$X(\zeta)$ must vanish identically.  Thus,
\begin{equation}
  \left[J^{a}_{m},L_{n}\right]=M^{ab}Y^b_{m+n},
\end{equation} 
so that~$L_{n}$ and~$Y^a_{n}$ are just two Ka\v c--Moody primary
fields in some representation of~$g$. Moreover this representation is
finite since the number of weight-two states in $\Lambda_{0}$ is
finite, and all the states in the multiplet are null.  We now return
to the representation~$|\Lambda\rangle$ which is not the vacuum
representation. It is easy to show that
\begin{equation}
  (\cL_{g})_{1}L_{-2}|\Lambda\rangle=3L_{-1}|\Lambda\rangle ,
\end{equation}
so that this state must also be null. Since~$L_{-1}$ and~$L_{-2}$
generate the whole Virasoro algebra, it follows that
$L_{-n}|\Lambda\rangle$ must be null for any~$n$, and hence any of the
$Y_{-n}|\Lambda\rangle$ will also be null for a~$Y$ in the null
multiplet of weight-two fields.  As a consequence of the above
remarks, a state of the form
$|W\rangle=L_{-n}\{\ldots\}|\Lambda\rangle$, where~$\ldots$ consists
of some arbitrary collection of~$\widehat{g}$ current modes, is null,
as can be shown inductively by pushing~$L_{-n}$ to the right. The
action of~$L$ on the representation~$\Lambda$ is therefore effectively
trivial as claimed.

For instance, in the case of embedding \eqref{case-1}, where
$\hDD{\tfrac{\gamma}{1-\gamma}}$ has level one, the state
$(E^{\at}_{-1})^{2}|0\rangle$ is a singular vector, where
$E^{\at}_{-1}$ is the usual affine simple root associated with the
step operator corresponding to the highest root~$\at$ of~$g$. This
state is a singular vector at grade two as is~$L_{-2}|0\rangle$, so it
isn't unreasonable to argue that they are in the same multiplet under
$g$. If this were the case, our condition that~$L_{-2}|\Lambda\rangle$
is null is equivalent to the condition that
$({E^{\at}}_{-1})^{2}|\Lambda\rangle$ is null.  The latter is implied
by the vanishing of~$\langle
\Lambda|(E^{-\at}_{1})^{2}(E^{\at}_{-1})^{2}|\Lambda\rangle$, yielding
the familiar~$\tSL2$ result
\begin{equation}
  \at.\Lambda(\at.\Lambda-1)=0.
\end{equation}
This means we only should see the unitary singlet and doublet
representations of the~$s\ell(2)^{(\at)}_1$ appearing in the
$\hDD{\tfrac{\gamma}{1-\gamma}}_{1}$ character decomposition if the
embedding is to be conformal, and indeed inspection of our character
sum rules reveals that only the characters of
these~$s\ell(2)^{(\at)}_1$ representations occur.

We now proceed to identify which of the~$\tSSL21$
embeddings~\eqref{sl21-emb} (with~$\kappa = \gamma
-1,-\gamma~\text{or}~1$) is the one implied
in~\eqref{Hminus2}--\eqref{A-current2}.  To make contact with the
latter expressions, we must fix the level of $\tSSL21$ to be~$k$, so
that its subalgebra~$\tSL2^{(\am)}$ is also at level~$k$.

According to the above discussion, we have the embeddings
\begin{equation}\label{alg}
  \begin{array}{rcccl}
    &&\widehat{D}&&\\
    &\nearrow& & \nwarrow&\\
    \tSSL21 _{\lambda} \oplus \widehat{u}(1)\kern-12pt&&&&
    \kern-12pt\tSL2 _{\lambda_1}^{(\alpha_2)} \oplus \tSL2
    _{\lambda_2}^{(\alpha_3)} \oplus \tSL2 _{\lambda_3}^{(\at)}  
  \end{array}
\end{equation}
in~$\widehat{D}\equiv\hDD{\frac{\g}{1-\g}}_{\g-1}$ for the levels,
\begin{equation}
  (\lambda ; \lambda _1,\lambda_2,\lambda_3)=
  \begin{cases}(k;k',1,k),~
    (k';k',1,k), & \text{for}\quad \g=-\tfrac{k}{k'}=k+1,\\
    (k;k,1,k'),~(k';k,1,k'), & \text{for}\quad
    \g=-\tfrac{k'}{k}=k'+1,
  \end{cases}
\end{equation}
in~$\widehat{D}\equiv\hDD{\frac{\g}{1-\g}}_{-\g}$ for the levels,
\begin{equation}
  (\lambda ; \lambda _1,\lambda_2,\lambda_3)=
  \begin{cases}(k;1,k',k),~
    (k';1,k',k), & \text{for}\quad \g=-k,\\
    (k;1,k,k'),~(k';1,k,k'), & \text{for}\quad
    \g=-k',
  \end{cases}
\end{equation}
and in~$\widehat{D}\equiv\hDD{\frac{\g}{1-\g}}_{1}$ for the levels,
\begin{equation}
  (\lambda ; \lambda _1,\lambda_2,\lambda_3)=
  \begin{cases}(k;k,k',1),~
    (k';k,k',1), & \text{for}\quad \g=-\tfrac{1}{k},\\
    (k;k',k,1),~(k';k',k,1), & \text{for}\quad
    \g=-\tfrac{1}{k'}.
  \end{cases}
\end{equation}
We actually have twelve descriptions of~$\tSSL21 _{k} \oplus
\widehat{u}(1)$ in terms of~$s\ell(2)^{(\ab)} \oplus s\ell(2)^{(\atr)}
\oplus s\ell(2)^{(\at)}$, which we summarize in
Table~\ref{tab:Table1}.  There, we label the even subalgebra of
$\tSSL21 \oplus \widehat{u}(1)$ as
\begin{equation}
  [\tSL2^{(\am)} \oplus \widehat{u}(1)^{(\ap)}]\oplus
  \widehat{u}(1)^{(\ad)}. 
\end{equation}

We further define~$\amp$ and~$\beta$ such that~$\tSL2^{(\amp)}$ is the
algebra dual to~$\tSL2^{(\am)}$, and $\tSL2^{(\beta)}$ is the third,
level 1, algebra in the sense described in Lemma~\ref{Lemma:cc-1}.
The comparison of data in Table~\ref{tab:Table1} with the
currents~\eqref{Hminus2}--\eqref{A-current2} requires proper
normalisation, dictated by the operator product expansions derived
earlier. Since $H^-(z)\,H^-(w)=\frac{k/2}{(z-w)^2}$, the properly
normalised corresponding root is~$\tam =\mu \am$ with~$\mu
=\sqrt{k/2(\am)^2}$ (since~$(\tam)^2=k/2$). Similarly,~$H^+(z)$
corresponds to~$\tap= \nu \ap$ with~$\nu =\sqrt{-k/2(\ap)^2}$,
$J^{0'}(z)$ corresponds to~$\tamp=\rho \amp$ with~$\rho
=\sqrt{k'/2(\amp)^2}$, $j^0(z)$ corresponds to~$\tilde{\beta} = \sigma
\beta$ with~$\sigma =\sqrt{1/2(\beta)^2}$, and finally, $A(z)$
corresponds to~$\tad =\tau \ad$ with~$\tau =\sqrt{1/(\alpha^*)^2}$.

We conclude that the first embedding in Table~\ref{tab:Table1} indeed
corresponds to currents~\eqref{Hminus2}--\eqref{A-current2}, with
\begin{align}
  \tam={}&\tfrac{\sqrt{k}}{2}\at\,,\displaybreak[0]\\
  \tap ={}&\tfrac{-\sqrt{k}}{2}(\alpha_2
  +\alpha_3)={}(k+1)\tamp-k\tilde{\beta}\,,\displaybreak[3]\\
  \tad={}&\tfrac{-1}{\sqrt{2(k+1)}}(k\alpha_2 + (k+1)\alpha_3)={}
  \sqrt{2k(k+1)}(\tamp-\tilde{\beta}).
\end{align}
This therefore indicates that the vertex operators corresponding to
all eight elements from $\oC^2(z)\tensor\oC^{'2}(z)\tensor\oC_1^2(z)$
extend our vertex operator construction of~$\tSSL21_k$ to
$\hDD{\tfrac{1}{k'}}_k=\hDD{k'}_k$.  In Sec.~\ref{sec:characters}, we
establish character sumrules which confirm that the coset
of~$\hDD{k'}_{k}$ by $\tSSL21_{k}\oplus\widehat{u}(1)$ actually
coincides with the coset corresponding to the conformal
embedding~\eqref{case-2}.
\begin{table}[htb]
  \begin{center}\renewcommand{\arraystretch}{.69}
    \tabcolsep=4pt
    \footnotesize
    \begin{tabular}{|c|c|c|c|c|c|c|c|}   \hline
  &  &   &  &
  &  &   &  \\ 
$\am$   & roots generating  &  $\ap$  &  $\ad$     &  
$\amp$  & $\beta$  & value of parameter $\g$&  embedding \\
  & $\tSSL21$ plane   &   &  &
  &  &   & $(\lambda;\lambda_1,\lambda_2,\lambda_3)$  \\
  &  &   &  &
  &  &   &  \\ \hline
  &  &   &  &
  &  &   &  \\ 
$\at$ &$(\aa -\at~,~\aa)$  & $-\ab-\atr$  &$(1-\g)\ab-\g\atr$
& $\ab$   & $\atr$   &$-k/k'$  &$(k;k',1,k)$ \\ 
  &  &   &  &
  &  &   &  \\ 
\cline{5-8}
  &  &   &  &
  &  &   &  \\ 
  &  &   &  &
$\atr$  & $\ab$  & $-k=t_1(-k/k')$   &$(k;1,k',k)$  \\
  &  &   &  &
  &  &   &  \\ 
\cline{2-8}
  &  &   &  &
  &  &   &  \\ 
  &$(\aa +\atr~,~-\aa-\ab)$&$-\ab+\atr$  &$(1-\g)\ab +\g\atr$ &
$\ab$   & $\atr$   &$-k/k'$  &$(k;k',1,k)$ \\ 
  &  &   &  &
  &  &   &  \\ 
\cline{5-8}
  &  &   &  &
  &  &   &  \\ 
  &  &   &  &
$\atr$  & $\ab$  & $-k=t_1(-k/k')$   &$(k;1,k',k)$  \\
  &  &   &  &
  &  &   &  \\ 
\hline
  &  &   &  &
  &  &   &  \\ 
$\ab$   &$(\aa +\atr~,~\at-\aa)$&  $\atr+\at$   & $\atr+(1-\g)
\at$ &   $\at$   & $\atr$   &$-k'/k$  &$(k;k,1,k')$ \\
  &  &   &  &
  &  &   &  \\ 
 \cline{5-8}
  &  &   &  &
  &  &   &  \\ 
  &  &   &  &
$\atr$  & $\at$  & $-1/k=t_2t_1t_2(-k'/k)$  &$(k;k,k',1)$  \\
  &  &   &  &
  &  &   &  \\ 
\cline{2-8}
  &  &   &  &
  &  &   &  \\ 
  &$(\aa~,~\aa+\ab)$  &$-\atr+\at$   &$-\atr+(1-\g)\at$ &
$\at$   & $\atr$   &$-k'/k$  &$(k;k,1,k')$ \\ 
  &  &   &  &
  &  &   &  \\ 
\cline{5-8}
  &  &   &  &
  &  &   &  \\ 
  &  &   &  &
$\atr$  & $\at$  & $-1/k=t_2t_1t_2(-k'/k)$  &$(k;k,k',1)$  \\
  &  &   &  &
  &  &   &  \\ 
\hline  
  &  &   &  &
  &  &   &  \\ 
$\atr$  &$(\aa +\ab~,~\at-\aa)$ &  $\ab+\at$  & $\ab+\g \at$  &
$\at$   & $\ab$  &$-k'$   &$(k;1,k,k')$ \\ 
  &  &   &  &
  &  &   &  \\ 
\cline{5-8}
  &  &   &  &
  &  &   &  \\ 
  &  &   &  &
$\ab$   & $\at$  & $-1/k'=t_2(-k')$   &$(k;k',k,1)$  \\
  &  &   &  &
  &  &   &  \\ 
\cline{2-8}
  &  &   &  &
  &  &   &  \\ 
  &$(\aa~,~\aa+\atr)$  &$-\ab+\at$  &$-\ab +\g\at$ &
$\at$   & $\ab $   &$-k'$   &$(k;1,k,k')$ \\ 
  &  &   &  &
  &  &   &  \\ 
\cline{5-8}
  &  &   &  &
  &  &   &  \\ 
  &  &   &  &
$\ab$   & $\at$  & $-1/k'=t_2(-k')$   &$(k;k',k,1)$  \\
  &  &   &  &
  &  &   &  \\ 
\hline
\end{tabular}
\caption[Embeddings in $\widehat{D}(2|1;k)_k$]{\footnotesize  
  Embeddings of
  $\tSSL21_{\lambda}\oplus\widehat{u}(1)\rule{0pt}{20pt}$ and
  $\tSL2_{\lambda_1}^{(\alpha_2)} \oplus
  \tSL2_{\lambda_2}^{(\alpha_3)}\oplus \tSL2_{\lambda_3}^{(\at)}$ in
  $\hDD{\tfrac{\g}{1-\g}}_{\g-1}$ (embeddings 1,3,5,7),
  $\hDD{\tfrac{\g}{1-\g}}_{-\g}$ (embeddings 2,4,9,11)
  and~$\hDD{\tfrac{\g}{1-\g}}_1$ (embeddings 6,8,10,12).  $\am$ and
  $\ap$ are associated with the isospin and hypercharge in~$\tSSL21$,
  while~$\ad$ is in the~$u(1)$ direction orthogonal to the
  $s\ell(2|1)$ root plane in~$D(2|1;\alpha)$.\,$\tSL2^{(\amp)}$ and
  $\tSL2^{(\am)}$ are dual algebras and~$\tSL2^{(\beta)}$ is the
  third, level 1 algebra. The transformations~$t_1:\g \rightarrow
  1-\g$ and~$t_2:\g \rightarrow \g^{-1}$ are discussed in
  Appendix~\ref{app:D-algebra}.}\label{tab:Table1}
  \end{center}
\end{table}

\section{\textbf{Constructing the representations}}\label{sec:reps}
We now construct relations between representations of~$\tSSL21_k$ and
of the~$\tSL2_k$ and~$\tSL2_{k'}$ algebras with dual~$k$ and~$k'$. The
vertex operator realisation of~$\tSSL21$ in Sec.~\ref{sec:basic}
involves taking `quantum traces' of products of~$\tSL2_k$ and
$\tSL2_{k'}$ vertex operators, suggesting that representations of
$\tSSL21$ could be constructed as sums of products of representations
of these dual~$\tSL2$ algebras (and the auxiliary~$\tSL2_1$
representations).  However, identifying which~$\tSL2$ representations
to choose in order to obtain, e.g., the category $\cO$ of~$\tSSL21$
representations does not follow from the operator construction alone,
and the question is analysed in Sec.~\ref{sec:construct}.
 
On the other hand, one can address the inverse problem of
reconstructing~$\tSL2_k \oplus \tSL2_{k'}$ representations out of a
given~$\tSSL21$ representation.  By simply restricting the latter, one
obviously arrives at representations of~$\tSL2_k$.  For the
dual~$\tSL2_{k'}$ algebra, which is not a subalgebra of~$\tSSL21$, the
construction of its representations starting with an~$\tSSL21_k$
representation is provided in Secs.~\ref{sec:reconstructing}
and~\ref{sec:reps-inverse}.  The `direct' and the `inverse'
problems each consists, first, in constructing a representation and
second, in decomposing it.  The second step, which requires a more
detailed analysis of the representations constructed, will be
addressed in Sec.~\ref{sec:characters}.

We continue using both~$k=p-2$ and~$k'=p'-2$ as the parameters.

\subsection{Reconstructing the dual $\widehat{s\ell}(2)$
  pair}\label{sec:reconstructing} We first address the problem of
reconstructing the dual pair~$\tSL2_k\oplus\tSL2_{k'}$ starting with
the~$\tSSL21_k$ algebra.  One of these is obviously the subalgebra of
$\tSSL21$, and thus the problem is to `complete' the~$u(1)$ subalgebra
to the second (`primed')~$\tSL2$.  The construction goes by picking
out the appropriate terms in the operator products
$(\oC^2(z)\oplus\oC^2(z))\cdot(\oC^2(w)\oplus\oC^2(w))$,
where~$\oC^2(z)\oplus\oC^2(z)$ denotes the~$\tSSL21$ operators
representing~$\SL2$ doublets.  This yields the dual~$\tSL2$ operator
product expansions after a `correction' with an extra scalar field
(recall that we also needed an auxiliary scalar in constructing
$\tSSL21$ out of two~$\tSL2$ algebras).  We, thus, introduce the
scalar field~$\phi$ with the operator product
\begin{equation}
  \d\phi(z)\,\d\phi(w)=-\frac{1}{(z-w)^2}
\end{equation}
(note that the sign is opposite to the one for the~$f$ scalar
in~\eqref{ff}) and define
\begin{equation}\label{allPsi}
  \Psi^+=\tfrac{1}{k+1}\,E^1\,F^2\,e^{\sqrt{2}\phi}\,,\quad
  \Psi^-=\tfrac{1}{k+1}\,F^1\,E^2\,e^{-\sqrt{2}\phi}\,,\quad
  \Psi^0=\tfrac{k'}{\sqrt{2}}\,\d\phi + (k'+1)\Hplus\,.
\end{equation}
The operators~$\Psi^{\pm}$ representing $\bigwedge^2\oC^2(z)$
(`corrected' by the free scalar, and therefore, local with respect
to each other)
commute with the~$\tSL2$ subalgebra of $\tSSL21$.  It also follows
from the~$\tSSL21$ algebra that
\begin{multline}\label{bigOPE}
  (E^1F^2)(z)\;(E^2F^1)(w)={}\\
  {}=\frac{k (k+1)}{(z-w)^4}
  -\frac{2(k+1)\Hplus}{(z-w)^3} - (k+1)\,\frac{T_{\mathrm{Sug}} +
    E^1\,F^1 - E^2\,F^2 + \d\Hminus + \d\Hplus }{ (z-w)^2}{}+ \dots,
\end{multline}
where~$T_{\mathrm{Sug}}$ is the Sugawara energy-momentum
tensor~\eqref{Tsug-sl21}.  Thus,
\begin{equation}\label{PsiplusPsiminus}
  \Psi^+(z)\,\Psi^-(w)=\frac{k'}{(z-w)^2} +
  2\,\frac{\Psi^0}{z-w}\,.
\end{equation}
It is easy to see that~$\Psi^+(z)\,\Psi^+(w)$ and
$\Psi^-(z)\,\Psi^-(w)$ are nonsingular.  We also find the
operator~products
\begin{equation}\label{Psi0Psi+}
  \Psi^0(z)\,\Psi^\pm(w)=
  \frac{\pm\Psi^\pm}{z-w}\,,\qquad
  \Psi^0(z)\,\Psi^0(w)=\frac{k'/2}{(z-w)^2}\,.
\end{equation}
Equations~\eqref{PsiplusPsiminus} and~\eqref{Psi0Psi+} show that
\textit{$\Psi^+$, $\Psi^0$, and~$\Psi^-$ satisfy the `dual' $\tSL2$
  algebra of level~$k'$}.  However, we keep the
notation~$\Psi^{\pm,0}$ instead of~$J'{}^{\pm,0}$ in order to stress
that the former are constructed out of the~$\tSSL21$ currents.
In addition, we have the current
\begin{equation}\label{Xi-current}
  \Xi = \tfrac{1}{\sqrt{2}}\,\d\phi + \Hplus\,,
  \qquad \Xi(z)\,\Xi(w)=-\tfrac{(k+1)/2}{(z-w)^2},
\end{equation}
that commutes with the~$\tSL2_{k'}$ algebra.
\begin{Rem}\label{rem:SF}
  Under the~$\tSSL21$ spectral flow
  \begin{equation*}
    \begin{array}{rclcrcl}
      E^1(z)&\mapsto{}& z^\theta\,E^1(z)\,,&&
      E^2(z)&\mapsto{}& z^{-\theta}\,E^2(z)\,,\\[8pt]
      F^1(z)&\mapsto{}& z^{-\theta}\,F^1(z)\,,&&
      F^2(z)&\mapsto{}& z^\theta\,F^2(z)\,,
    \end{array}\qquad
    \Hplus(z)\mapsto{}\Hplus(z)-\theta\,\tfrac{k}{z}\,,
  \end{equation*}
  the~$\tSL2_{k'}$ currents undergo the spectral flow transform with
  the parameter~$2\theta$:
  \begin{equation*}
    \Psi^{\pm}(z)\mapsto z^{\pm2\theta}\,\Psi^{\pm}(z)\,,\qquad
    \Psi^0\mapsto\Psi^0 + \theta\,\tfrac{k'}{z}\,.
  \end{equation*}
  Note also that if the~$\tSSL21$ spectral flow is accompanied by the
  transformation 
  %\begin{equation*}
    $\phi(z)\mapsto\phi(z) - \sqrt{2}\,\theta\log z$
  %\end{equation*}
  in the auxiliary scalar sector, the~$\tSL2_{k'}$ currents remain
  invariant.
\end{Rem}
\begin{Rem}
  The operators emerging in operator product expansion~\eqref{bigOPE}
  already commute with the~$\tSL2$ subalgebra; they can be further
  reorganised by separating the part that commutes with~$\Hplus$.  The
  resulting operators then generate the coset~$W$
  algebra~$\ww=\tSSL21/\tGL2$ mentioned in the Introduction.
  Its central charge
  \begin{equation*}
    c=-2\,\frac{2k+1}{k+2}=2\,\frac{k'-1}{k'+2}
  \end{equation*}
  satisfies, obviously,~$c+1+\frac{3k}{k+2}=0$.  The lowest spin
  (spin-2 and spin-3) generators of~$\ww$ can be read off
  from~\eqref{bigOPE} (with the first-order pole restored) and are
  expressed through the~$\tSSL21$ currents as
  \begin{equation*}
    W_2 = \frac{E^1 F^1 - E^2 F^2}{k+1} +
    \frac{E^{12}F^{12} + \Hminus\Hminus}{(k+1)(k+2)} +
    \frac{\Hplus\Hplus}{k(k+1)} + \frac{\d\Hminus}{k+2}
  \end{equation*}
  and (choosing it to be a $W_2$-primary)
  \begin{multline}
    W_3 =
    \tfrac{1}{k + 1}
    \sqrt{\tfrac{2k}{(k + 2)(k + 4)(3k+2)}}
    \Bigl(\thalf (k + 4) E^1 \d F^1 + 
    E^{12} F^1 F^2  - 
    \thalf E^{12} \d F^{12} + 
    \thalf (k + 4) E^2 \d F^2-{}\notag\displaybreak[1]\\
    {}- F^{12}  E^1  E^2+ 
    \Hminus  E^1  F^1 + 
    \Hminus  E^2  F^2 - 
    \tfrac{k + 4}{k} \Hplus E^1  F^1
    - \tfrac{2}{k}\Hplus E^{12} F^{12} + 
    \tfrac{k + 4}{k} \Hplus E^2  F^2-{} \displaybreak[3]\\
    {}- \tfrac{2}{k}\Hplus \Hminus  \Hminus 
    - \tfrac{2(k + 4)}{3k^2} \Hplus \Hplus  \Hplus - 
    \tfrac{k + 2}{k}\Hplus \d \Hminus -
    \thalf (k + 4) \d E^1 F^1 + 
    \thalf \d E^{12}  F^{12} \\{}
    - \thalf (k + 4) \d E^2 F^2 - 
    \d \Hplus \Hminus - 
    \thalf\d^2\Hminus -
    \tfrac{1}{6}(k + 4) \d^2\Hplus\Bigr)\,.
  \end{multline}
  No higher-spin currents are generated for~$k=-3$ (i.e.,
  the~$W_3\cdot W_3$ operator product is closed to~$W_2$ and~$W_3$),
  and similarly with more negative integer values: for~$k=-n$, the
  highest spin is~$n$.
\end{Rem}

\subsection{From $\protect\widehat{s\ell}(2|1)_k$ to
  $\protect\widehat{s\ell}(2)_{k'}$
  representations}\label{sec:reps-inverse} We now consider the
$\tSL2_{k'}$ representation furnished by the modes~$\Psi^{\pm,0}_n$ of
the currents
constructed in~\eqref{allPsi}.  We start with an~$\tSSL21_k$ module,
which we take to be a twisted Verma module~$\mP_{h_-,h_+,k;\theta}$
(see Appendix~\ref{app:sl21}) with the twisted highest-weight
vector~$\ket{h_-,h_+,k;\theta}$.  Let also~$\mE_\sigma$ be the Fock
module of the free current~$\d\phi$ with the highest-weight
vector~$\ket{e^{\frac{\sigma}{\sqrt{2}}\,\phi}}$ corresponding to the
operator~$e^{\frac{\sigma}{\sqrt{2}}\,\phi}$.
Then~$\mP_{h_-,h_+,k;\theta}\tensor\bigoplus_{n\in\oZ}\mE_{\sigma+2n}$
carries a representation of~$\tSL2_{k'}$. We now construct
the~$\tSL2_{k'}$ highest-weight vector, which we temporarily denote by
$\ket{h_-,h_+;\theta;\sigma}'$, by tensoring the highest-weight
vectors in these modules.
\begin{Lemma}
  The~$\tSL2_{k'}$ state
  \begin{equation}\label{r-state}
    \ket{h_-,h_+;\theta;\sigma}'=
    \ket{h_-,h_+,k;\theta}\tensor
    \ket{e^{\frac{\sigma}{\sqrt{2}}\,\phi}}
  \end{equation}
  is a twisted relaxed highest-weight state.
\end{Lemma}
We recall~\cite{[FST]} that \textit{a twisted relaxed highest-weight
  state $\ketsl2'{j',\Lambda,k';\theta'}$} satisfies the annihilation
conditions
\begin{gather}
  \Psi^+_{\geq1+\theta'}\ketsl2'{j',\Lambda,k';\theta'}=0\,,\qquad
  \Psi^-_{\geq1-\theta'}\ketsl2'{j',\Lambda,k';\theta'}=0\,,\qquad
  \Psi^0_{\geq1}\ketsl2'{j',\Lambda,k';\theta'}=0\,,
\end{gather}
the eigenvalue condition
\begin{equation}\label{relaxed-J0}
  \Psi^0_0\,\ketsl2'{j',\Lambda,k';\theta'}=
  (j'-\theta'\tfrac{k'}{2})\,\ketsl2'{j',\Lambda,k';\theta'}\,,
\end{equation}
and the relation
\begin{equation}\label{Lambda-cond}
  \Psi^-_{-\theta'}\Psi^+_{\theta'}
  \ketsl2'{j',\Lambda,k';\theta'}=\Lambda\,
  \ketsl2'{j',\Lambda,k';\theta'}\,.
\end{equation}
We will write~$\ketsl2'{j',\Lambda,k'}$ for the `untwisted' state
$\ketsl2'{j',\Lambda,k';0}$.  When there are no more independent
relations (no singular vectors are factored over), the module spanned
by the~$\tSL2_{k'}$ generators acting on $\ket{j',\Lambda,k';\theta'}$
is called the \textit{relaxed Verma
  module}~$\mR'_{j',\Lambda,k';\theta'}$~\cite{[FST]}.  The Sugawara
dimension of the twisted relaxed highest-weight
vector~$\ketsl2'{j',\Lambda,k';\theta'}$ is given by
\begin{equation}\label{relaxed-dim}
  \Delta'=\frac{{j'}^2+j'+\Lambda}{k'+2}-\theta' j' +
  \frac{k'}{4}{\theta'}^2\,.
\end{equation}

\noindent\textsc{Proof} of the Lemma.  Let $U_{h_-,h_+;\theta}$ be the
operator corresponding to the twisted highest-weight vector
$\ket{h_-,h_+,k;\theta}$ (see~\eqref{hw-twisted}).  It follows that
the~$\tSL2_{k'}$ currents~$\Psi^{\pm,0}$ develop the following poles
in the operator product with the operator~$X_{h_-,h_+;\theta;\sigma}=
U_{h_-,h_+;\theta}\,e^{\frac{\sigma}{\sqrt{2}}\,\phi}$ corresponding
to~\eqref{r-state}:
\begin{equation}
  \Psi^+(z)\,X_{h_-,h_+;\theta;\sigma}(0)\sim
  z^{2\theta-1-\sigma},\qquad 
  \Psi^-(z)\,X_{h_-,h_+;\theta;\sigma}(0)\sim z^{-2\theta-1+\sigma}.
\end{equation}
This gives precisely the twisted relaxed highest-weight conditions for
the state in~\eqref{r-state}.  We now choose $\sigma=2\theta$ for
simplicity (taking~$\sigma$ different from $2\theta$ would result in
the spectral flow transform of this state and, thus, in the twist of
the entire module).  We then find that the
$\ket{h_-,h_+;\theta;2\theta}'$ state constructed in~\eqref{r-state}
satisfies condition~\eqref{Lambda-cond} with~$\theta'=0$ and
\begin{equation}\label{Lambda-found}
  \Lambda=-\tfrac{1}{(k+1)^2}\,(k+1+h_+-h_-)(h_++h_-)\,.
\end{equation}
We also see that the eigenvalue of~$\Psi^0_0$ on
$\ket{h_-,h_+;\theta;2\theta}'$ is~$j'=\tfrac{h_+}{k+1}$.  Thus,
$\ket{h_-,h_+;\theta;2\theta}'$ is the relaxed highest-weight state
\begin{equation}\label{relaxed-found}
  \ket{h_-,h_+;\theta;2\theta}'=\ketsl2'{\tfrac{h_+}{k+1},
    -\tfrac{1}{(k+1)^2}\,(k+1+h_+-h_-)(h_++h_-),k'}
\end{equation}
and the freely generated module is the relaxed Verma
module~$\mR'_{\frac{h_+}{k+1},
  \frac{-1}{(k+1)^2}\,(k+1+h_+-h_-)(h_++h_-),k'}$.
\hfill\mbox{\rule{.5em}{.5em}}

\medskip

The dependence of the space $\bigoplus_{n}\mE_{2\theta+2n}$ on
$\theta$ is only~$\mod\oZ$, and for the integer $\theta$ that we
consider in what follows, we have the space
$\mE_{*}=\bigoplus_{n}\mE_{2n}$.  The~$\tSL2_{k'}$ representations
on~$\mP_{h_-,h_+,k;\theta}\tensor\mE_{*}$ are distinguished by the
eigenvalues~$\xi$ of the current in~\eqref{Xi-current}.  Let~$\mX_\xi$
be the Fock module over the Heisenberg algebra of Fourier modes of
the~$\Xi$ current with the highest-weight  vector~$\ket{\xi}$ such
that~$\Xi_0\,\ket{\xi}=\xi\,\ket{\xi}$ and $\Xi_{\geq1}\,\ket{\xi}=0$.
It follows that
\begin{equation}
  \Xi_0\,\ket{h_-,h_+,k;\theta}\tensor
  \ket{e^{\sqrt{2}(\theta+n)\phi}}=
  (h_+ - (k+1)\theta-n)\, 
  \ket{h_-,h_+,k;\theta}\tensor
  \ket{e^{\sqrt{2}(\theta+n)\phi}}\,,
\end{equation}
while as regards the~$\tSL2_{k'}$ algebra, the state
$\ket{h_-,h_+,k;\theta}\tensor\ket{e^{\sqrt{2}(\theta+n)\phi}}$ is a
twisted relaxed highest-weight state with the twist $\theta'=2n$.  We
thus expect the decomposition into a sum of spectral-flow transformed
$\tSL2_{k'}$ representations,
\begin{equation}
  \mP_{h_-,h_+,k;\theta}\tensor\mE_{*}=\bigoplus_{n\in\oZ}
  \mR'_{\frac{h_+}{k+1}, \frac{-1}{(k+1)^2}\,(k+1+h_+ - h_-)(h_+
  +h_-),k';2n} 
  \tensor\mX_{h_+-(k+1)\theta-n}\tensor{}\dots
\end{equation}
(where the dots denote~$\tSL2_k$ representations).  Since the
parameters of the $\tSL2_{k'}$ modules appearing on the right-hand
side are insensitive to the~$\tSSL21$ spectral flow
parameter~$\theta$, this induces the correspondence
\begin{equation}\label{func-0}
  \mP_{h_-,h_+,k;\bullet}\leadsto
  \mR'_{\frac{h_+}{k+1},\frac{-1}{(k+1)^2}\,(k+1+ h_+-h_-)(h_+
  +h_-),k';\bullet}  
\end{equation}
between \textit{spectral-flow orbits} of~$\tSSL21_k$ and~$\tSL2_{k'}$
modules.  

An interesting question is whether this correspondence can be extended
to a functor; answering this involves, in particular, investigating
the correspondence between singular vectors appearing in the modules
related by the correspondence.  We now show that the~${}\leadsto{}$
arrow descends to submodules generated from a class of the so-called
\textit{`charged' singular vectors}.

The relaxed Verma modules may contain singular vectors such that the
corresponding quotients are the usual Verma or twisted Verma modules.
These are the charged singular vectors~\cite{[FST]}, which occur in
the module built on the vector~$\ketsl2'{j',\Lambda,k'}$ whenever
$\Lambda=\Lambdach(n,j')$ for~$n\in\oZ$, where
\begin{equation}\label{Lambdach}
  \Lambdach(n,j')=n(n+1)+2nj'\,,\qquad n\in\oZ\,.
\end{equation}
When~\eqref{Lambdach} holds, the charged singular vector reads
\begin{equation}\label{chargedsl2}
  \ketsl2'{C(n,j',k')}=\left\{
    \begin{array}{ll}
      (\Psi^-_0)^{-n}\,\ketsl2'{j',\Lambdach(n,j'),k'}\,,&
      n\leq-1\,,\\[8pt]
      (\Psi^+_0)^{n+1}\,\ketsl2'{j',\Lambdach(n,j'),k'}\,,&n\geq0\,.
    \end{array}
  \right.  
\end{equation}
This state satisfies the usual Verma-module highest-weight conditions
for $n\leq-1$ and the twisted Verma highest-weight conditions with the
twist parameter~$\theta'=1$ for~$n\geq1$~\cite{[FST]}.

Remarkably, we see from~\eqref{Lambda-found} and~\eqref{Lambdach} that
a charged singular vector occurs in the relaxed Verma module
$\mR'_{\frac{h_+}{k+1}, \frac{-1}{(k+1)^2}\,(k+1+ h_+-h_-)(h_+
  +h_-),k'}$ if and only if
\begin{equation}
  h_+=-h_-- (k + 1) n\quad\text{or}\quad
  h_+=h_--(k+1)(n+1)\,,
\end{equation}
which are the conditions~\eqref{charged-exist} for the existence of
charged singular vectors~\eqref{Echminus} and~\eqref{Echplus} in the
$\tSSL21_k$ Verma modules~$\mP_{h_-,h_+,k;\bullet}$ (in the notations
of~\cite{[BHT97]}, such conditions on the highest weight state lead to
class~IV and class~V representations). Thus,
\begin{Lemma}
  The relaxed Verma module~$\mR'_{\frac{h_+}{k+1},
    \frac{-1}{(k+1)^2}\,(k+1+ h_+-h_-)(h_+ +h_-),k'}$ generated from
  the state constructed in~\eqref{r-state} contains a charged singular
  vector if and only if the~$\tSSL21$ Verma module
  $\mP_{h_-,h_+,k;\theta}$ contains a charged singular vector.
\end{Lemma}

In addition to the `existence' result asserted in the Lemma, the
charged singular vectors~\eqref{chargedsl2} in the~$\tSL2_{k'}$
representation on~$\mP_{h_-,h_+,k;\theta}\tensor\mE_{*}$ actually
\textit{evaluate} as the respective~$\tSSL21_k$ charged singular
vectors~\eqref{Echminus} or~\eqref{Echplus}.  Indeed, consider the
representation on $\mP_{h_-, -h_- - (k+1)n,k;\theta}\tensor\mE_{*}$
and let $n\leq-1$.  Then, up to the factor~$(k+1)^n$,
\begin{multline}
  (\Psi^-_0)^{-n}\,
  \ket{h_-,-h_- - (k+1)n,k;\theta}\tensor
  \ket{e^{\sqrt{2}\theta\,\phi}} ={}\\
  {}=
  {E^2_{\theta+n}\,\ldots\,E^2_{\theta-1}}\cdot
  {F^1_{\theta+n+1}\,\ldots\,F^1_{\theta}}\,
  \ket{h_-,-h_- - (k+1)n,k;\theta}\tensor
  \ket{e^{\sqrt{2}(\theta+n)\,\phi}}\,,
\end{multline}
which is precisely the corresponding charged singular vector
in~\eqref{Echplus}.  Next, let~$n\geq0$.  Note that in this case, the
charged singular vector
$(\Psi^+_0)^{n+1}\,\ketsl2'{j',n(n+1)+2nj',k'}$ satisfies the
highest-weight conditions that are twisted by~1 with respect to the
$\tSL2$ Verma-module highest-weight conditions.  It is immediate to
see that (again, up to a nonvanishing factor)
\begin{multline}
  (\Psi^+_0)^{n+1}\,\ket{h_-,-h_- - (k+1)n,k;\theta}\tensor
  \ket{e^{\sqrt{2}\theta\,\phi}}={}\\
  {}=
  {E^1_{-\theta-n-1}\,\ldots\,E^1_{-\theta-1}}
  \cdot
  {F^2_{-\theta-n}\,\ldots\,F^2_{-\theta}}\,
  \ket{h_-,-h_- - (k+1)n,k;\theta}
  \tensor
  \ket{e^{\sqrt{2}(\theta+n+1)\,\phi}}={}\\
  {}=
  E^1_{-\theta-n-1}\,\ket{C^{(+)}(n, h_-, k;\theta)}\tensor
  \ket{e^{\sqrt{2}(\theta+n+1)\,\phi}}\,.
\end{multline}
Now, unless~$k+1-2h_-=0$, the state
$E^1_{-\theta-n-1}\,\ket{C^{(+)}(n, h_-, k;\theta)}$ generates the
same module as the charged singular vector of which it is a
descendant, since
\begin{equation*}
  F^1_{\theta+n+1}\,E^1_{-\theta-n-1}\,\ket{C^{(+)}(n, h_-,
    k;\theta)}=(k+1-2h_-)\ket{C^{(+)}(n, h_-, k;\theta)}\,.
\end{equation*}

A similar analysis applies to the charged~$\tSL2_{k'}$ singular
vectors in~$\mP_{h_-,h_--(k+1)n,k;\theta}\tensor\mE_{*}$, which again
evaluate as the~$\tSSL21$ charged singular vectors.  Namely,
whenever~$h_+=h_--(k+1)n$, the corresponding~$\;\tSL2_{k'}\;$ module
$\;\mR'_{\frac{h_-}{k+1}-n,(n-1)(\frac{2h_-}{k+1}-n),k'}\;$ contains
the charged singular vector $\ketsl2'{C(n-1,\frac{h_-}{k+1}-n,k')}$.
When $n\leq0$, it is given by the action of~$(\Psi^-_0)^{-n+1}$, which
evaluates as~$E^2_{\theta+n-1}\,\ket{C^{(-)}(n,h_-,k;\theta)}$ times a
primary state in the~$\phi$ sector.  Unless~$2h_-=k+1$, this~$\tSSL21$
vector generates the same module as the charged singular vector
$\ket{C^{(-)}(n,h_-,k;\theta)}$.  When~$n\geq1$, the~$\tSL2_{k'}$
charged singular vector~$\ketsl2'{C(n-1,\frac{h_-}{k+1}-n,k')}$ is
given by the action of~$(\Psi^+_0)^{n}$ on the highest-weight vector
and evaluates precisely as~$\ket{C^{(-)}(n,h_-,k;\theta)}$ (times
a~$\phi$-sector state).

\subsection{From
  $\protect\widehat{s\ell}(2)_k\oplus\protect\widehat{s\ell}(2)_{k'}$
  to $\protect\widehat{s\ell}(2|1)_k$
  representations}\label{sec:construct} We now build~$\tSSL21_k$
representations by combining~$\tSL2_k$ and~$\tSL2_{k'}$ (and the
`auxiliary'~$\tSL2_1$) representations.  The construction is
summarised in Theorem~\ref{thm:space}, while
Sec.~\ref{sec:proof-1}--\ref{sec:proof-2} explain how one arrives at
this result.  As regards the $\tSL2_{k'}$ representations, we build on
the result of the previous subsection, where we saw that
the~$\tSSL21_{k}\leadsto\tSL2_{k'}$ correspondence produces relaxed
Verma modules from the~$\tSSL21_{k}$ Verma modules.  We can therefore
expect that the correspondence acting in the opposite direction should
start with relaxed Verma modules.  This proves to be the case, as we
see in what follows.

\subsubsection{Choosing representation spaces: the
  $\protect\widehat{s\ell}(2)_{k}\oplus\protect\widehat{s\ell}(2)_1$
  sector}\label{sec:proof-1} We recall
from~\eqref{act-01}--\eqref{act-02} that the vertex operators
$\oC^2(z)$ map between~$\tSL2_k$ modules as
\begin{equation}\label{vertex-action-0}
  \oC^2(z)\tensor\oC_q^2(z): \mV_{j,k}\to
  \mV_{j - \half,k}\tensor z^{\frac{-j-1}{p}}\oC((z))
  \oplus
  \mV_{j + \half,k}\tensor z^{\frac{j}{p}}\oC((z))\,.
\end{equation}
This gives rise to a chain of~$\tSL2_k$
modules $\mV_{j+\frac{n}{2},k}$, $n\in\oZ$, with the spins differing
from each other by (half-)integers.  For the future convenience, we
rewrite~\eqref{vertex-action-0} for these modules
\begin{equation}\label{vertex-action}
  \oC^2(z)\tensor\oC_{q}^2(z): \mV_{j+\frac{n}{2},k}\to
  \mV_{j + \frac{n-1}{2},k}
  \tensor z^{\frac{-j-1-\frac{n}{2}}{p}}\oC((z))
  \oplus
  \mV_{j + \frac{n+1}{2},k}
  \tensor z^{\frac{j+\frac{n}{2}}{p}}\oC((z))\,.
\end{equation}

In the auxiliary sector, let $\mF_\mu$ be the Fock space of the
auxiliary current~$\d f(z)$ with the highest-weight vector~$\ket{\mu}$
such that
\begin{equation}\label{ket-mu}
  (\d f)_0\,\ket{\mu}=\sqrt{2}\,\mu\,\ket{\mu}\,.
\end{equation}
We then have the vertex operator action
\begin{equation}\label{aux-action}
  \oC_1^2(z):\mF_\mu\to\mF_{\mu+\half}\tensor z^\mu\oC((z))
  +\mF_{\mu-\half}\tensor z^{-\mu}\oC((z))\,.
\end{equation}
The~$\tSSL21$ fermions constructed in terms of the vertex operators
would change both~$n$ and~$m$ in~$\mV_{j+\frac{n}{2},k}\tensor\mF_{\mu
  + \frac{m}{2}}$. However, $\mu$ is changed by a
\textit{half}-integer simultaneously with the~$\tSL2_k$ spin~$j$
changed by a half-integer.  This involves only an index-2 sublattice
of the~$\oZ\times\oZ\ni(n,m)$ lattice in Fig.~\ref{lattice}
(alternatively, one could choose to work with the other half of the
lattice sites).
\begin{figure}[bt]
  \begin{center}
    \leavevmode\unitlength=.7pt
    \begin{picture}(400,200)
      \thinlines
      \put(200,0){\vector(0,1){200}}
      \put(205,190){${}^{m}$}
      \put(50,90){\vector(1,0){300}}
      \put(345,90){${}^{n}$}
      \multiput(0,-30)(0,80){3}{
        \multiput(120,40)(80,0){3}{\makebox(0,0)[cc]{$\bullet$}}
        }
      \multiput(40,10)(0,80){2}{
        \multiput(40,40)(80,0){4}{\makebox(0,0)[cc]{$\bullet$}}
        }
      \multiput(0,-30)(0,40){5}{
        \multiput(80,40)(40,0){7}{\makebox(0,0)[cc]{\Large$\cdot$}}
        }
    \end{picture}
  \end{center}
  \caption[]{$\rule[-20pt]{0pt}{20pt}$}\label{lattice}
\end{figure}

Our aim is to combine Eqs.~\eqref{vertex-action}
and~\eqref{aux-action} with the action of~$\tSL2_{k'}$ vertex
operators such that the~$\tSSL21$ fermions $E^1$, $E^2$, $F^1$,
and~$F^2$ then act on a space of the form
\begin{equation}\label{pre-space}
  \bigoplus_{n\in\oZ}\mV_{j+\frac{n}{2},k}\tensor R'(n)
  \tensor\bigoplus_{m\in\oZ}\mF_{\mu - \frac{n}{2} + m}
\end{equation}
(where~$R'$ are some~$\tSL2_{k'}$ modules).  The very existence of
such a space endowed with the representations induced from the vertex
operator action is not obvious a priori, because vertex operator
action in each sector gives rise to the spaces~$z^\nu\oC((z))$ with
non-integral~$\nu$ that are different for different terms (recall that
$p=k+2\in\oC\setminus\{0,1\}$ and~$j\in\oC$ in~\eqref{vertex-action});
it does not therefore induce a representation on a space of the
form~\eqref{pre-space} unless the modules~$R'(n)$ are judiciously
chosen.  A crucial point making such a choice possible is the quantum
group trace in the construction of the~$\tSSL21$ fermions.

\subsubsection{Choosing representation spaces: the
  $\protect\widehat{s\ell}(2)_{k'}$ sector}\label{sec:proof-2} As we
saw in Sec.~\ref{sec:reps-inverse} from the construction that is in a
certain sense inverse to the present one, the~$\tSL2_{k'}$
representations have to include the relaxed Verma modules.

As a hint in properly choosing the relaxed Verma module
$R'(0)=\mR'_{j',\Lambda,k';\theta'}$ in~\eqref{pre-space}, let us
\textit{assume} that the product of an~$\tSL2_k$ highest-weight state,
an $\tSL2_{k'}$ relaxed highest-weight state, and an~$\tSL2_1$
highest-weight state,
$\ket{j,k}\tensor\ketsl2'{j',\Lambda,k';\theta'}\tensor \ket{\mu}
\in\mV_{j,k}\tensor \mR'_{j',\Lambda,k';\theta'}\tensor\mF_\mu$, is a
twisted highest-weight state with respect to~$\tSSL21_k$
(see~\eqref{hw-twisted}) tensored with a highest-weight vector in the
Fock module~$\mA_a$ over the free current~$A(z)$ in
Eq.~\eqref{A-current0}; we define the highest-weight
vectors~$\ket{a}\in\mA_a$ such that
\begin{equation}\label{ket-a}
  A_0\ket{a}_A=\sqrt{2(k+1)}\,a\,\ket{a}_A\,.
\end{equation}
Assuming, thus, that
$\ket{j,k}\tensor\ketsl2'{j',\Lambda,k';\theta'}\tensor \ket{\mu}=
\ket{h_-,h_+,k;\theta}\tensor\ket{a}_A$, we evaluate the eigenvalues of
the~$\tSSL21$ Cartan generators (and also of~$A_0$,
see~\eqref{A-current0}) on this vector.  Using~\eqref{Hplus0}
and~\eqref{A-current0}, we obtain the eigenvalues
\begin{align}\label{Hplus-eigen}
  H^+_0\approx{}&(k+1)j'-k(\mu - \tfrac{\theta'}{2})\,,\\
  A_0\approx{}&\sqrt{2(k+1)}\,(j'-\tfrac{k'\theta'}{2} - \mu)\,,
  \label{A-eigen}
\end{align}
whence~$h_+ = (1 + k) j' + k(\theta - \mu + \half\theta')$.  We now
compare the dimensions using Eq.~\eqref{T-identity} that we have
established for our~$\tSSL21_k$ generators.  The balance of dimensions
of the highest-weight vectors reads
\begin{multline}
  \tfrac{j(j+1)}{k+2} + {}
  \underbrace{\tfrac{j'(j'+1) + \Lambda}{k'+2} -
    j'\theta' + \tfrac{k'{\theta'}^2}{4}}_{\eqref{relaxed-dim}}
  {} + {}\mu^2 ={}\\
  {}=  
  \underbrace{\tfrac{j^2 - (j'(k+1))^2}{k+1} +
    2\Bigl(\mu-\tfrac{\theta'}{2}\Bigr) j'(k+1) -
    k\Bigl(\mu-\tfrac{\theta'}{2}\Bigr)^2}_{\eqref{Sug-twisted}}
  {}+{}
  (k+1)\Bigl(j'-\tfrac{k'\theta'}{2} - \mu\Bigr)^2\,.
\end{multline}
This fixes~$\Lambda$ equal to~$\Lambda(j,j')$ given by\footnote{We
  note that~\eqref{Lambda-taken} implies
\begin{equation*}
  \Lambda(j,\tfrac{h_+}{k+1}) =
  -\tfrac{(k+1+ h_+- h_-)(h_+ +j)}{(k+1)^2}\,,
\end{equation*}
which reproduces Eq.~\eqref{Lambda-found} obtained in constructing
$\tSL2_{k'}$ states from a given~$\tSSL21_k$ state and, thus, suggests
that the constructions in Secs.~\ref{sec:construct}
and~\ref{sec:reps-inverse} are parts of the direct and the
inverse~functors.}
\begin{equation}\label{Lambda-taken}
  \Lambda(j,j') =-\left(1 + j' -
    (k'+1)j\right)\left(j' + (k'+1)j\right)\,.
\end{equation}

Generalising this, we now take the modules~$R'(n)$
in~\eqref{pre-space} to be
$R'(n)=\mR'_{j'-\frac{n}{2},\Lambda_n(j,j'),k'}$ with
\begin{equation}\label{Lambda-n}
  \Lambda_n(j,j')= -(1 - (k'+1)j + j')(j' + (k'+1)j - n)\,.
\end{equation}  
In accordance with~\eqref{relaxed-dim}, the Sugawara dimension
$\Delta'_n$ of~$\ket{j',\Lambda_n(j,j'),k'}$ is
\begin{equation}\label{Delta1}
  \Delta'_n=\frac{(1 - (k'+1)j + \frac{n}{2})
    (\frac{n}{2} - (k'+1)j)}{k'+2}\,,
\end{equation}
and therefore, evaluating~$\Delta'_{n+1}-\Delta'_n - \frac{3/4}{k'+2}=
\frac{n/2}{k'+2} - \frac{j}{k+2}$ and~$\Delta'_{n-1}-\Delta'_n -
\frac{3/4}{k'+2}= \frac{-n/2-1}{k'+2} + \frac{j}{k+2}$ (where
$\frac{3/4}{k'+2}$ is the Sugawara dimension of~$\ket{\half,k'}$
corresponding to the~$\oC'{}^2(z)$ vertex operator), we find the
exponents in the~$\tSL2_{k'}$ vertex operator action
\begin{multline}\label{vertex-action'}
  \oC'{}^2(z)\tensor\oC_{q'}^2(z)
  {}:{}\mR'_{j'-\frac{n}{2},\Lambda_n(j,j'),k'}\to{}\\
  {}\to
  \mR'_{j'-\frac{n-1}{2},\Lambda_{n-1}(j,j'),k'}\tensor
  z^{\frac{-\frac{n}{2}-1}{p'} + \frac{j}{p}}\oC((z)) \oplus
  \mR'_{j'-\frac{n+1}{2},\Lambda_{n+1}(j,j'),k'}\tensor
  z^{\frac{\frac{n}{2}}{p'} - \frac{j}{p}}\oC((z))\,.
\end{multline}
Remarkably, the exponents~$\frac{-n/2-1}{p'} + \frac{j}{p}$ and
$\frac{n/2}{p'} - \frac{j}{p}$ appearing here add up to
(half-)integers with the exponents from~\eqref{vertex-action}.

\subsubsection{The vertex operator action} The last observation is a
crucial point that, irrespective of the preceding motivations, allows
us to construct the space carrying an $\tSSL21$ action.
\begin{Thm}\label{thm:space}
  For every pair $(j,j')\in\oC\times\oC$, there is an~$\tSSL21_k$
  representation on the space
  \begin{equation}\label{the-space}
    \espaceN_{j,j'}=\bigoplus_{n\in\oZ}
    \mV_{j + \frac{n}{2},k}\tensor
    \mR'_{j'-\frac{n}{2},
      \Lambda_n(j,j'),k'}
    \tensor\bigoplus_{m\in\oZ} \mF_{m - \frac{n}{2}}\,,
  \end{equation}
  where~$\mV_{j,k}$ is an~$\tSL2_k$ module with the highest-weight
  vector~$\ket{j,k}$, $\mR'_{j',\Lambda,k'}$ is an~$\tSL2_{k'}$
  representation with the relaxed highest-weight
  vector~$\ketsl2'{j',\Lambda,k'}$, and~$\mF_\mu$ is the Fock space of
  the auxiliary current with the highest-weight vector~$\ket{\mu}$
  defined in~\eqref{ket-mu}.
\end{Thm}
\begin{Rem}
  The range of the $n$ summation in~\eqref{the-space} can be
  restricted to a subset of~$\oZ$ depending on the $\tSL2_{k}$ and
  $\tSL2_{k'}$ modules involved, as we will see in what follows.  
\end{Rem}

To show~\eqref{the-space}, it only remains to check that
combining~\eqref{vertex-action}, \eqref{vertex-action'},
and~\eqref{aux-action}, we are left with the Laurent spaces
$z^\nu\oC((z))$ with integer~$\nu$.  Indeed, putting
together~\eqref{vertex-action} and \eqref{vertex-action'} \textit{and
  recalling the trace in~\eqref{take-trace}}, we arrive at
\begin{multline}
  \oC^2(z)\tensor\oC^2(z):
  \mV_{j+\frac{n}{2},k}\tensor
  \mR'_{j'-\frac{n}{2},\Lambda_n(j,j'),k'}  
  \to{}
  \\{}\to
  \mV_{j+\frac{n-1}{2},k}\tensor
  \mR'_{j'-\frac{n-1}{2},\Lambda_{n-1}(j,j'),k'}
  \tensor 
  z^{-\frac{n}{2}-1}  \oC((z)) \oplus
  \mV_{j+\frac{n+1}{2},k}\tensor
  \mR'_{j'-\frac{n+1}{2},\Lambda_{n+1}(j,j'),k'}\tensor
  z^{\frac{n}{2}}\oC((z))\,.
\end{multline}
We next combine this with the auxiliary sector from~\eqref{aux-action}
and, thus, obtain the vertex operator action
\begin{multline}\label{vertex-act}
  \oC^2(z)\tensor\oC^2(z)\tensor\oC^2_1(z):
  \mV_{j+\frac{n}{2},k}\tensor
  \mR'_{j'-\frac{n}{2},\Lambda_n(j,j'),k'}
  \tensor\mF_{\mu - \frac{n}{2} + m}
  \to{}
  \\{}\to
  \Bigl(\mV_{j+\frac{n-1}{2},k}\tensor
  \mR'_{j'-\frac{n-1}{2},\Lambda_{n-1}(j,j'),k'}
  \tensor 
  z^{-\frac{n}{2}-1}  \oC((z)) \oplus
  \mV_{j+\frac{n+1}{2},k}\tensor
  \mR'_{j'-\frac{n+1}{2},\Lambda_{n+1}(j,j'),k'}\tensor
  z^{\frac{n}{2}}\oC((z))\Bigr)\tensor{}\\
  {}\tensor\Bigl(
  z^{\mu - \frac{n}{2} + m}\,\mF_{\mu - \frac{n}{2} + m + \half}
  \oplus
  z^{-\mu + \frac{n}{2} - m}\,\mF_{\mu - \frac{n}{2} + m - \half}
  \Bigr)\,.
\end{multline}
This indeed involves the~$z^\nu\oC((z))$ spaces with all~$\nu$ being
integer mod~$\mu$, and therefore, induces an~$\tSSL21$ representation
on the space~$\bigoplus_{n\in\oZ} \mV_{j + \frac{n}{2},k}\tensor
\mR'_{j'-\frac{n}{2},\Lambda_n(j,j'),k'} \tensor\bigoplus_{m\in\oZ}
\mF_{\mu + m - \frac{n}{2}}$.  This space actually depends only on the
fractional part of~$\mu$, and this dependence is nothing but the
overall (non-integral) spectral flow; we can therefore set~$\mu=0$,
which gives the space~$\espaceN_{j,j'}$ in Eq.~\eqref{the-space}
endowed with the structure of an~$\tSSL21_k$ representation.
(Choosing between integral and half-integral~$\mu$ corresponds to
Ramond and Neveu--Schwarz sectors.)
\begin{Rem}
  In the auxiliary sector modules in~\eqref{the-space}, we can replace
  $m-\frac{n}{2}$ with~$m-\half$ for odd~$n$ and with~$m$ for even
  $n$; therefore, for odd and even~$n$ we have the respective spaces
  $\bigoplus_{m\in\oZ}\mF_{m + \half}=\mm_{\half,1}$ and
  $\bigoplus_{m\in\oZ}\mF_{m}=\mm_{0,1}$ that are the spin-$\half$ and
  spin-0 irreducible representations of the~$\tSL2_1$ algebra
  introduced in~\eqref{aux-sl2}.  Thus, the~$\tSSL21$ representation
  space can be described as a sum of tensor products of three $\tSL2$
  representations:
  \begin{equation}\label{the-space-2}
    \espaceN_{j,j'}=\bigoplus_{n\in\oZ}
    \mV_{j + \frac{n}{2},k}\tensor
    \mR'_{j'-\frac{n}{2},
      \Lambda_n(j,j'),k'}
    \tensor\mm_{\frac{\varepsilon(n)}{2},1}\,,\qquad    
    \varepsilon(n)= n\;\mathrm{mod}\;2\,.
  \end{equation}
  Replacing~$\varepsilon(n)$ with~$1-\varepsilon(n)$ gives the
  Neveu--Schwarz representation space~$\espaceN_{j,j'}^{1/2}$.
\end{Rem}
\begin{Rem}
  It is obvious from the above that $\espaceN_{j,j'}$ is in fact a
  representation of $\hDD{k'}_{k}$.
\end{Rem}

\section{\textbf{Decomposition of representations and  character
    identities}}\label{sec:characters} In this section, we decompose
the $\tSSL21_k$ representations constructed as in~\eqref{the-space-2}
and check the decomposition formulas by using character identities.
In Sec.~\ref{decomp:Verma}, we explain the general strategy of
decomposing the representations and obtaining the corresponding
character identities.  In Sec~\ref{sec:Verma-char}, we calculate the
characters of both sides of the decomposition formula for the
corresponding (relaxed) Verma modules and show these characters to be
identical.  In Sec.~\ref{sec:char}, we outline the steps leading from
the Verma-module case to the respective irreducible representations.
We specialise to the admissible representations and, further, to the
`principal admissible' ones.  The relevant decomposition formula is
given in~\eqref{decompose-int}.  As we do not give a direct proof that
the representations on the right-hand side are indeed the
corresponding irreducible $\tSSL21_{\frac{1}{u}-1}$ representations,
we invoke the character sumrules derived independently.  These are
given in Sec.~\ref{sec:SUMRULES}, where we first have to explain how
the characters of the twisted representations are identified with the
$\tSSL21_{\frac{1}{u}-1}$ characters known from~\cite{[BHT97]}.  As
the result, the sumrules in Eqs.~\eqref{uu-sumrule}--\eqref{B} are
found to be precisely the character identities
for~\eqref{decompose-int}.  We conclude the section with a remark on
the modular properties of the characters involved.

\subsection{Decomposing the $\protect\widehat{s\ell}(2|1)$
  representation}\label{decomp:Verma} The~$\tSSL21_k$ representation
on the space~$\espaceN_{j,j'}$ constructed in the previous section is
by no means irreducible, since all the~$\tSSL21$ generators commute
with the current~$A(z)$ constructed in~\eqref{A-current0}.  Thus,
$\espaceN_{j,j'}$ decomposes as
$\bigoplus_{\lambda}\mP(\lambda)\tensor\mA_{a(\lambda)}$, where
$\mP(\lambda)$ are some~$\tSSL21_k$ modules and~$\mA_{a}$ are the Fock
modules over the free scalar;~$\lambda$ labels different eigenvalues
that~$A_0$ has on the highest-weight vector of each~$\mA$ module.

Evaluating the~$\tSSL21$ highest-weight parameters, we find that the
different~$\tSSL21$ representations~$\mP(\lambda)$ are the spectral
flow transform of each other, with the highest-weight parameters
except the twist being the same for all these modules.  Taking for
definiteness~$\mV_{j + \frac{n}{2},k}$ to be the Verma modules (and
$\mR'_{j'-\frac{n}{2},\Lambda_n(j,j'),k';\theta'}$ the twisted relaxed
Verma modules), we thus arrive at the decomposition
\begin{equation}\label{decompose}
  \boxed{
    \bigoplus_{n\in\oZ}
    \mV_{j + \frac{n}{2},k}\tensor
    \mR'_{j'-\frac{n}{2},\Lambda_n(j,j'),k';\theta'}
    \tensor
    \mm_{\frac{\varepsilon(n)}{2},1} 
    =\bigoplus_{\theta\in\oZ}
    \mP_{j,(k+1)j',k;2\mu - \theta' + \theta}\tensor
    \mA_{j' - \theta}\,.}
\end{equation}
where~$\mA_{j'-\theta}$ are Fock modules (see~\eqref{ket-a}).  A
priori, $\mP_{j,(k+1)j',k;2\mu - \theta' + \theta}$ are some $\tSSL21$
representations with the respective twisted highest-weight vectors
$\ket{j,(k+1)j',k;2\mu - \theta' + \theta}$.  That they are in fact
the corresponding twisted Verma modules is confirmed
Sec.~\ref{sec:Verma-char} by showing that the characters on both sides
of~\eqref{decompose} are identical.\footnote{Strictly speaking, this
  requires the assumption that the modules in questions are generated
  from the respective highest-weight vectors.}  

In calculating the characters, we take the trace
$\Tr\bigl(z^{H^-_0}\,\zeta^{H^+_0}\,q^{L_0^{\mathrm{Sug}}}\,
y^{\frac{1}{\sqrt{2(k+1)}}\,A_0}\,q^{\half(AA)_0}\bigr)$ on the
right-hand side of the decomposition formula and then use
Eqs.~\eqref{Hplus0}, \eqref{A-current0}, and~\eqref{T-identity} to
rewrite this in terms of the~$\tSL2_{k}\oplus\tSL2_{k'}\oplus\tSL2_1$
Cartan generators and the energy-momentum tensors,
\begin{equation}\label{basic-char}
  \Tr\Bigl(z^{H^-_0}\,\zeta^{H^+_0}\,q^{L_0^{\mathrm{Sug}}}
  y^{\frac{1}{\sqrt{2(k+1)}}\,A_0}\,q^{\half(AA)_0}\Bigr)
  {}=
  \Tr\Bigl(z^{J^0_0}\,q^{L_0^{\mathrm{Sug}}}
  \bigl(\zeta^{k+1}\,y\bigr)^{J'{}^0_0}\,
  q^{L'_0{}^{\mathrm{Sug}}}
  \bigl(\zeta^{-k}\,y^{-1}\bigr)^{j^0_0}\,
  q^{l_0^{\mathrm{Sug}}}\Bigr)
\end{equation}
(with the respective Sugawara energy-momentum tensor for each
algebra).  The trace on each side of the last formula is taken over
the space on the \textit{complementary} side of~\eqref{decompose}.
The resulting character identities are therefore of the general form
\begin{equation}\label{sumrule-general}
  \sum\chi^{\tSSL21_k}(q, z,\zeta)\,
  \chi^{\mA}(q,y)
  =
  \sum\chi^{\tSL2_k}(q,z)\,
  \chi^{\tSL2_{k'}}(q,\zeta^{k+1}y)\,
  \chi^{\tSL2_1}(q, \zeta^{-k}y^{-1})
\end{equation}
(where $\chi^{\tSL2_1}$ and $\chi^{\mA}$ are the standard expressions
essentially given by the corresponding theta function; the
`nontrivial' ingredients are the $\tSSL21$ and $\tSL2_k$ and
$\tSL2_{k'}$ characters).  

In what follows, we freely interchange between $(q,z,\zeta,y)$
and~$(\tau,\sigma,\nu,\rho)$ related~by
\begin{equation}\label{qzy}
  q=e^{2i\pi\tau},\quad z=e^{2i\pi \sigma},\quad
  \zeta=e^{2i\pi \nu},\quad y=e^{2i\pi\rho}\,,
\end{equation} 
where~$\tau \in \oC$, $\Im\tau>0\Rightarrow|q|<1$, and $\sigma\in\oC$,
$\nu\in\oC$, $\rho\in\oC$.

\subsection{Verma-module case}\label{sec:Verma-char} Comparing the
characters of both sides of~\eqref{decompose}, we have to deal with
formal objects, since already the relaxed Verma module character
involves the nowhere convergent series
$\delta(z,y)=\sum_{n\in\oZ}\bigl(\frac{z}{y}\bigl)^n$: the characters
of a twisted relaxed Verma module~$\mR'_{j',\Lambda,k';\theta'}$ and
of the~$\tSL2_k$ Verma module $\mV_{j,k}$ are
\begin{multline}
  \chr^{\mR'}_{j',\Lambda,k';\theta'}(u,q)\equiv
  \Tr_{\mR'_{j',\Lambda,k';\theta'}}^{\phantom{y}}
  \Bigl(u^{J'{}^0_0}\,q^{L'_0{}^{\mathrm{Sug}}}\Bigr)={}
  \displaybreak[0]\\
  {}=q^{\frac{k'}{4}{\theta'}^2}\,u^{-\frac{k'}{2}\theta'}\,
  \delta(u\,q^{-\theta'},1)\, \frac{q^{\frac{{j'}^2+j'+\Lambda}{k'+2}}
    (u\,q^{-\theta'})^{j'}}{\prod_{i\geq1}(1-q^i)^3}=
  q^{-\frac{k'}{4}{\theta'}^2}\, \delta(u\,q^{-\theta'},1)\,
  \frac{q^{\frac{{j'}^2+j'+\Lambda}{k'+2}} u^{j'}\,q^{-\theta'j'}}{
    \prod_{i\geq1}(1-q^i)^3}\,,
\end{multline}
\begin{equation}
  \chr^\mV_{j,k}(z,q)\equiv
  \Tr_{\mV_{j,k}}^{\phantom{y}}
  \Bigl(z^{J^0_0}\,q^{L_0^{\mathrm{Sug}}}\Bigr)=
  \frac{q^{\frac{j^2+j}{k+2} %+ \frac{1}{8}
      }
    z^j}{\vartheta_{1,1}(q,z)}\,,
\end{equation}
with the Jacobi theta-function~\eqref{Jacobi11}.  In the auxiliary
sector, it is more convenient to return to the description as
in~\eqref{the-space}; the character of~$\mF_\mu$ is simply
\begin{equation}
  \Tr_{\mF_{\mu}}^{\phantom{y}}
  \Bigl(v^{\frac{1}{\sqrt{2}}(\d f)_0}\,q^{\half(\d f\d f)_0}\Bigr)=
  \frac{v^\mu\,q^{\mu^2}}{\prod_{i\geq1}(1-q^i)}\,.
\end{equation}
For the character of~$\bigoplus_{n\in\oZ} \mV_{j +
  \frac{n}{2},k}\tensor \mR'_{j' -
  \frac{n}{2},\Lambda_n(j,j'),k'}\tensor \bigoplus_{m\in\oZ}\mF_{\mu -
  \frac{n}{2} + m}$, we thus have (recall the dimension in
Eq.~\eqref{Delta1})
\begin{multline}  
  \sum_{m,n\in\oZ} \Tr_{\mV_{j+\frac{n}{2},k}}^{\phantom{y}}
  \Bigl(z^{J^0_0}\,q^{L_0^{\mathrm{Sug}}}\Bigr)
  \Tr_{\mR'_{j'-\frac{n}{2},\Lambda_n(j,j'),k'}}^{\phantom{y}}
  \Bigl(u^{J'{}^0_0}\,q^{L'_0{}^{\mathrm{Sug}}}\Bigr)
  \Tr_{\mF_{\mu-\frac{n}{2}+m}}^{\phantom{y}}
  \Bigl(v^{\frac{1}{\sqrt{2}}(\d f)_0}\,q^{\half(\d f\d f)_0}\Bigr)={}
  \displaybreak[0]\\
  {}= \sum_{n\in\oZ}\sum_{m\in\oZ} \frac{q^{\frac{(j+\frac{n}{2})^2 +
        j + \frac{n}{2}}{k+2}}\, z^{j + \frac{n}{2}}\, \delta(u,1)\,
    q^{\frac{\frac{n^2}{4} + \frac{n}{2}}{k'+2} - \frac{j(n+1)}{k+2} +
      \frac{(k'+1)j^2}{k+2}} u^{j'-\frac{n}{2}}\, q^{(\mu + m -
      \frac{n}{2})^2}\, v^{\mu + m - \frac{n}{2}}
    }{\vartheta_{1,1}(q,z)\prod\limits_{i\geq1}(1-q^i)^4}\,.
\end{multline}
Shifting the summation variable as $n\mapsto n + m$, we continue this
as (with $\vartheta_{1,0}$  defined in~\eqref{Jacobi10})
\begin{equation}
  \quad\quad{}=z^{j}\, u^{j'}\,v^{\mu}\,
  q^{\frac{j^2}{k+1} + \mu^2}\,
  \delta(u,1)\,
  \frac{\vartheta_{1,0}(q, z^{\half}u^{-\half}v^{-\half}\,q^{-\mu})\,
    \vartheta_{1,0}(q, z^{\half}u^{-\half}v^{\half}\,
    q^{\mu})}
  {\vartheta_{1,1}(q,z)
    \prod\limits_{i\geq1}(1-q^i)^4}\,.
\end{equation}

In accordance with~\eqref{basic-char}, we now substitute
$u=\zeta^{k+1}y$ and $v=\zeta^{-k}y^{-1}$. Then $uv=\zeta$, and
therefore, the argument of the first theta function in the numerator
becomes $z^{\half}\zeta^{-\half} q^{-\mu}$.  In the second
theta-function, we have $u^{-\half}v^{\half}= u^{-1}(uv)^{\half}=
u^{-1}\zeta^{\half}$; we now use the formal $\delta$-function property
$z\delta(z,y) = y\delta(z,y)$ to replace $u$ with $1$. Thus,
\begin{multline} %%\label{lhsdecomp}
  \sum_{m,n\in\oZ} \Tr_{\mV_{j+\frac{n}{2},k}}^{\phantom{y}}
  \Bigl(z^{J^0_0}\,q^{L_0^{\mathrm{Sug}}}\Bigr)
  \Tr_{\mR'_{j'-\frac{n}{2},\Lambda_n(j,j'),k'}}^{\phantom{y}}
  \Bigl((\zeta^{k+1}y)^{J'{}^0_0}\,q^{L'_0{}^{\mathrm{Sug}}}\Bigr)
  \Tr_{\mF_{\mu-\frac{n}{2}+m}}^{\phantom{y}}
  \Bigl((\zeta^ky)^{\frac{-1}{\sqrt{2}}(\d f)_0}\,
  q^{\half(\d f\d f)_0}\Bigr)={}\notag\displaybreak[1]\\
  {}=z^{j}\, \zeta^\mu\, q^{\frac{j^2}{k+1} + \mu^2}\,
  \delta(\zeta^{k+1}y,1)\, \frac{\vartheta_{1,0}(q,
    z^{\half}\zeta^{-\half}q^{-\mu})\, \vartheta_{1,0}(q,
    z^{\half}\zeta^{\half}\,q^{\mu})} {\vartheta_{1,1}(q,z)
    \prod\limits_{i\geq1}(1-q^i)^4}\,.
\end{multline}
We set $\mu=0$ for simplicity (this parameter has the meaning of the
overall spectral flow transform).

On the right-hand side of~\eqref{decompose}, we have the
character~\eqref{sl21-Verma-char} of the twisted $\tSSL21$ Verma
module $\mP_{h_-,h_+,k;\theta}$, and the character of~$\mA_a$ is given
by
\begin{equation}
  \Tr_{\mA_a}^{\phantom{y}}
  \Bigl(y^{\frac{1}{\sqrt{2(k+1)}}\,A_0}\,q^{L_0}\Bigr)=
  \frac{y^a\,q^{(k+1)a^2}}{\prod\limits_{i\geq1}(1-q^i)}\,,
  \label{Achar}
\end{equation}
and thus the character of the right-hand side of~\eqref{decompose}
equals
\begin{multline}\label{summations}
  \sum_{\theta\in\oZ} \Tr^{\phantom{y}}_{\mP_{j,(k+1)j',k;\theta}}
  \Bigl(z^{H^-_0}\,\zeta^{H^+_0}\,q^{L_0^{\mathrm{Sug}}}\Bigr)
  \Tr^{\phantom{y}}_{\mA_{j' - \theta}}
  \Bigl(y^{\frac{1}{\sqrt{2(k+1)}}A_0}\,q^{\half(A A)_0}\Bigr)={}\\
  {}=\sum_{\theta\in\oZ} z^j\,\zeta^{(k+1)j' - (k+1)\theta}\,
  q^{\frac{j^2}{k+1} - (k+1)j'{}^2
    + 2\theta (k+1)j' - (k+1)\theta^2 
    } y^{j'-\theta}q^{(k+1)(j' - \theta)^2}\, \frac{\vartheta_{1,0}(q,
    z^{\half}\zeta^{\half})\, \vartheta_{1,0}(q,
    z^{\half}\zeta^{-\half})}{
    \vartheta_{1,1}(q,z)\prod\limits_{i\geq1}(1-q^i)^4}\\
  {}=z^j\,y^{j'}\,\zeta^{(k+1)j'}\,q^{\frac{j^2}{k+1} 
    }\, \delta(\zeta^{k+1}\,y,1)\, \frac{\vartheta_{1,0}(q,
    z^{\half}\zeta^{\half})\, \vartheta_{1,0}(q,
    z^{\half}\zeta^{-\half})}{
    \vartheta_{1,1}(q,z)\prod\limits_{i\geq1}(1-q^i)^4}\,,
\end{multline}
where we again can use the~$\delta$-function to replace
$y^{j'}\,\zeta^{(k+1)j'}$ with~1.  This is identical with the
character obtained for the left-hand side (with~$\mu=0$; a nonzero
value of~$\mu$, as well as of the twist~$\theta'$ of the~$\mR'$
modules, is restored by the overall spectral flow transform).  We thus
conclude that there are the twisted $\tSSL21$ Verma modules on the
right-hand side of~\eqref{decompose}.

\subsection{From Verma modules to irreducible
  representations}\label{sec:char} To obtain the decomposition formula
of form~\eqref{decompose} for other (e.g., irreducible)
representations, one can start with appropriate Verma modules (e.g.,
those with dominant highest-weights, so as to arrive at the admissible
representations eventually) and follow the corresponding BGG
resolution, with~\eqref{decompose} applied again to each term in the
resolution.  If the decomposition formula is of a functorial nature
(i.e.,~the appearance of singular vectors is `synchronised' on both
sides), one then arrives at a similar formula for the irreducible
representations.  Here, we follow this program only as far as the
first step consisting in taking the quotients with respect to the
charged singular vectors.  In the end, we confirm the decomposition
identities for irreducible representations by verifying the
corresponding character identities.

We see from~\eqref{Lambdach} and~\eqref{charged-exist} that
\emph{whenever a charged singular vector occurs in one \emph{(and
    hence in all)} of the relaxed Verma modules
  $\mR'_{j'-\frac{n}{2},\Lambda_n(j,j'),k'}$, $n\in\oZ$, each of the
  twisted $\tSSL21$ Verma modules $\mP_{j,(k+1)j',k;\theta}$,
  $\theta\in\oZ$, contains a charged singular vector}. (Proceeding in
the `inverse' direction, we saw the same correspondence in
Sec.~\ref{sec:reps-inverse}, where we made it even more explicit.)
For Weyl modules, we then obtain the result stated in the Introduction
by taking the appropriate quotients of~\eqref{decompose}.

In what follows, we discuss the form that the decomposition takes for
a class of irreducible representations.  We find which character
identities (`sumrules' of type~\eqref{sumrule-general}) must follow
from this formula, and derive them independently.  This is a strong
argument in favour of the functorial properties of
decomposition~\eqref{decompose}.  We now explain which identity is to
be tested at the level of character sumrules.

We take a rational level~$k$ in the positive zone expressed through
two positive integers~$t$ and~$u$ via
\begin{equation}\label{kplus2}
  k+2=\tfrac{t}{u}\neq1\,.
\end{equation}
We require the relaxed Verma module $\mR'_{j',\Lambda_0(j,j'),k'}$ `at
the origin' in~\eqref{decompose} to have the charged singular vector
$\ketsl2'{C(\ell,j',k')}$; more precisely, we satisfy
Eq.~\eqref{Lambdach} by choosing
\begin{equation*}
  (k'+1)j=1 + j' + \ell\,,\qquad\ell\in\oZ\,.
\end{equation*}
For both $\tSL2_k$ and $\tSL2_{k'}$, we are interested in the case
where the irreducible representations are admissible~\cite{[KW]}.  We
recall here that the quotient of $\mR'_{j',\Lambda,k'}/\widetilde{V}$
with respect to the submodule generated from the charged singular
vector $\ketsl2'{C(\ell,j',k')}$, $\ell\in\oZ$, is a Verma module if
$\ell\geq0$ or a twisted Verma module (with the twist~1) if
$\ell\leq-1$, and the spin of the highest-weight vector of the
quotient module is
\begin{equation*}
  j'_\V = j' + \ell + (k'+2)\tfrac{1-\theta_\ell}{2}\,,\quad
  \text{where}\quad\theta_\ell=
  \begin{cases}
    1,&\ell\geq0\,,\\
    0,&\ell\leq-1\,.
  \end{cases}
\end{equation*}
We now require that the spins of~$\mV_{j,k}$ and
$\mR'_{j',\Lambda_0(j,j'),k'}/\widetilde{\mV}$ take the `admissible'
values, 
\begin{equation}\label{admiss-j}
  j = \tfrac{r-1}{2} - \tfrac{t}{u}\,\tfrac{s-1}{2}\,,
  \qquad
  j' + \ell + \tfrac{t}{t-u}\,\tfrac{1-\theta_\ell}{2} =
  \tfrac{r'-1}{2} - \tfrac{t}{t-u}\,\tfrac{s'-1}{2}
\end{equation}
with
\begin{equation}\label{admiss-spins}
  1\leq r\leq t-1\,,\quad 1\leq s\leq u\,,\qquad
  1\leq r'\leq t-1\,,\quad 1\leq s'\leq t-u
\end{equation}
(which obviously requires $t\geq u+1$).  

Conditions~\eqref{admiss-j} give
\begin{equation}
  (r' + s - s' + \theta_\ell)t = (r+r')u\,,
\end{equation}
implying that $r + r'$ is a multiple of~$t$, which means $r+r'=t$ once
$r$ and $r'$ are in the `admissible range' given above.  We then have
$r' = u - s + s' - \theta_\ell$, and therefore (choosing $\ell\geq0$,
so as to have the `untwisted' Verma modules in the quotients of the
relaxed Verma modules, after which we can set $\ell=0$), we obtain the
representations
\begin{equation}\label{pre-N-amd}
  \bigoplus_n\mV_{\frac{t-u + n +1 - s'}{2}
    -\frac{t-u}{u}\frac{s-1}{2},\frac{t-2u}{u}}
  \tensor\mV'_{\frac{u - n - 1 - s}{2} - \frac{u}{t-u}\frac{s'-1}{2},
    \frac{2u-t}{t-u}}
  \tensor
  \mm_{\frac{\varepsilon(n)}{2},1}\,.
\end{equation}
Here, $\mV$ and $\mV'$ are Verma modules; their quotients are the
respective admissible representations $\Adm$ and $\Adm'$ for~$n$ such
that
\begin{equation}
  -t + s' - s + 1 \leq n - u + 1 \leq s' - s - 1\,.
\end{equation}
Shifting the summation index, we thus arrive at the following
combination of \textit{admissible representations}:
\begin{equation}\label{N-amd}
  \espaceN^{\mathrm{adm}}_{s,s'}=\kern-6pt
  \bigoplus_{n=s' - s + 1 - t}^{s' - s - 1}
  \Adm_{\frac{t + n - s'}{2}
    -\frac{t-u}{u}\frac{s-1}{2},\frac{t-2u}{u}}
  \tensor\Adm'_{\frac{-n - s}{2} - \frac{u}{t-u}\frac{s'-1}{2},
    \frac{2u-t}{t-u}}
  \tensor
  \mm_{\frac{\varepsilon(n + u - 1)}{2},1},~  
  \begin{array}{l}
    1\leq s\leq u,\\
    1\leq s'\leq t - u.
  \end{array}\kern-12pt
\end{equation}
\begin{Rem}
  That the representations outside the `admissible range' decouple can
  also be understood by starting with free-field realisations
  of~$\tSL2$ modules.  The corresponding formula of
  type~\eqref{decompose} then gives some $\tSSL21$ representations
  (whose structure depends on the free-field representations chosen
  for~$\tSL2_k$ and $\tSL2_{k'}$).  Now, admissible representations of
  either $\tSL2_k$ or $\tSL2_{k'}$ are singled out from the free-field
  spaces as the cohomology of an appropriate `BRST' complex (for
  example, a Felder-like complex~\cite{[BF]}, but in general other
  free-field resolutions are needed).  The vertex operators mapping
  between the individual $\tSL2_k$ and $\tSL2_{k'}$ representations
  then extend to the entire complex; this allows one to single out
  \textit{trivial} vertex operators (see, e.g.,~\cite{[BMP-f]}) that
  induce zero mapping on the cohomology (as a rule, nontrivial
  mappings via vertex operator exist if and only if the corresponding
  fusion rule coefficient is nonzero).  In the admissible case, the
  vertex operators become trivial as soon as the spin $j+\frac{n}{2}$
  or $j' - \frac{n}{2}$ goes outside the corresponding Ka\v c table
  (\textit{which occurs simultaneously for the unprimed and primed
    modules in~\eqref{pre-N-amd}--\eqref{N-amd}}).  This allows us
  to consistently restrict to only a finite number (exactly $t-1$) of
  terms on the left-hand side of~\eqref{decompose}.
\end{Rem}

For each pair $(s,s')$ in the corresponding range, the space
$\espaceN^{\mathrm{adm}}_{s,s'}$ decomposes into a direct sum over
the~$\tSSL21$ spectral flow orbit.  The $\tSSL21$ representations
involved carry the labels
$\mL_{h_-,h_+,k;\theta}=\mL_{\frac{t-u+1-s'}{2} -
  \frac{t-u}{u}\frac{s-1}{2}, \frac{t-u+1-s'}{2} -
  \frac{t-u}{u}\frac{s+1}{2},\frac{t}{u}-2;m}$, $m\in\oZ$.  One
expects these to be the corresponding \textit{irreducible} $\tSSL21$
representations.  As we have noted, proving this by studying the BGG
resolution is a separate problem which we do not address here, however
this is confirmed by the character identities derived independently.

Specialising to the integrable (unitary) $\tSL2_{k'}$ representations,
we take the level $k'=\frac{2u-t}{t-u}$ to be an integer, whence
$t=u+1$ which implies $s'=1$ and $k'=u-1$.  It follows that the level
$k$ is fractional of the form
\begin{equation}\label{kfrac}
  k=\frac{1}{u}-1,
\end{equation} 
and therefore, belongs to a subclass of $\tSL2$ `admissible' levels
called `principal admissible,' which are parametrised as
\begin{equation}
  u(k+h^{\vee})=k^0+h^{\vee},\qquad k^0 \in \oZplus,
\end{equation}
where $h^{\vee}=2$ is the dual Coxeter number of
$s\ell(2)$~\cite{[KW],[Mathieu]}.  In the present instance, we
have~$k^0=u-1$.  Note that \eqref{kfrac} is also principal admissible
with respect to $\tSSL21$, with $k^0=0$, since
$h^{\vee}_{s\ell(2|1)}=1$.  For each $s\in[1,u]$, we now have the
space
\begin{multline}
  \espaceN^{\mathrm{int}}_{s}= \bigoplus_{n = 1 - s - u}^{-s}
  \Adm_{\frac{u + n}{2} -\frac{1}{u}\frac{s-1}{2},\frac{1}{u}-1}
  \tensor\Adm_{\frac{-n - s}{2}, u-1} \tensor
  \mm_{\frac{\varepsilon(n + u - 1)}{2},1}={}\\
  {}=\bigoplus_{n = 1 - s}^{u-s} \Adm_{\frac{n}{2}
    -\frac{1}{u}\frac{s-1}{2},\frac{1}{u}-1} \tensor\Adm_{\frac{u - s
      - n}{2}, u-1} \tensor \mm_{\frac{\varepsilon(n - 1)}{2},1}\,.
\end{multline}

We remind the reader that the notation $\mX_{j,k}$ for each module
indicates the spin~$j$ and the level~$k$.  When we are dealing with
the characters in what follows, it will be useful to label each
admissible representation of $\tSL2_{\frac{t}{u}-2}$ with the spin
\begin{equation}
  j=\frac{r-1}{2} - \frac{t}{u}\frac{s-1}{2}\,,
  \qquad
  \begin{array}{ll}
    1\leq r\leq t-1\,,\\
    1\leq s\leq u\,,
  \end{array}
\end{equation}
by~$r$ and~$s$ in addition to~$t$ and~$u$, as~$\Adm_{t,u\mymid r,s}$.
In this notation, the integrable representations are~$\Adm_{t,1\mymid
  r,1}=\Int_{t\mymid r}$.  In fact, we encounter the~$\tSL2_{k'}$
integrable representations $\Adm'_{u+1,1\mymid r',1}=\Int'_{u+1\mymid
  r'}$.  Then (after a convenient redefinition of the summation index)
the space $\espaceN^{\mathrm{int}}_{s}$ rewrites
as
\begin{equation}
  \espaceN^{\mathrm{int}}_{s}=
  \bigoplus_{r = 1}^{u}
  \Adm_{u+1,u\mymid r,s}
  \tensor
  \Int'_{u+1\mymid u+1-r}
  \tensor
  \Int_{3 \mymid \varepsilon (r-s-1)+1}\,.
\end{equation}
This is the space to be decomposed into $\tSSL21_{\frac{1}{u}-1}$
representations. 

In addition to only a finite number of \textit{irreducible}
representations appearing on the left-hand side of the decomposition
formula as discussed above, another effect occurring for the
representations under consideration is the \textit{periodicity under
  the spectral flow} on the~$\tSSL21$ side, where the spectral
flow~\eqref{sl21-sf} with $\theta=u$ produces an isomorphic
representation.  In the analogue of~\eqref{decompose} for such
representations~$\mL_{h_-,h_+,k}$, we split the summation over the
twists on the right-hand side by writing $\theta=u \alpha + \beta$
with $\alpha\in\oZ$ and $\beta=0,\dots,u-1$.  Since $\mL_{h_-,h_+,k;
  u\alpha + \beta}\simeq\mL_{h_-,h_+,k; \beta}$, the decomposition
formula for $\espaceN^{\mathrm{int}}_{s}$ takes a remarkable form
involving \hbox{precisely $u$ non-isomorphic $\tSSL21$
  representations}:
\begin{equation}\label{decompose-int}
  \boxed{
    \bigoplus_{r = 1}^{u}
    \Adm_{u+1,u\mymid r,s}
    \tensor
    \Int'_{u+1\mymid u+1-r}
    \tensor
    \Int_{3 \mymid \varepsilon(r - s - 1)+1}
    = \bigoplus_{\beta=0}^{u-1}
    \mL_{\frac{u-s+1}{2u},
      \frac{u-s-1
        }{2u},\frac{1}{u}-1;\beta} 
    \tensor
    \bigoplus_{\alpha\in\oZ}
    \mA_{\frac{u-1-s}{2}
      - u \alpha - \beta}\,.
    }
\end{equation}
In what follows, we test this formula by verifying the corresponding
character identities.

\subsection{Sumrules for irreducible representation
  characters}\label{sec:SUMRULES} We now study the character sumrules
corresponding to Eq.~\eqref{decompose-int}.

\subsubsection{Free-scalar and $\tSL2$ characters} In order to write
down the character identities corresponding to~\eqref{decompose-int},
we first of all note that the sum of the Fock module characters over
$\alpha$ on the right-hand side gives rise to a level $u$ generalised
theta function, namely,
\begin{equation}
  \sum_{\alpha\in\oZ}
  q^{u~[\alpha-\frac{2\beta-(u-1-s)}{2u}]^2}\,
  y^{u~[\alpha-\frac{2\beta-(u-1-s)}{2u}]}\,=
  \theta_{-2\beta+(u-1-s),u}(q,y)= 
  \theta_{2\beta-(u-1-s),u}(q,y^{-1}),
\end{equation}
where we used~\eqref{Achar} and some standard properties of the
generalised theta functions defined by
\begin{equation}\label{future-theta}
  \theta_{\mu,\kappa}(q, \zeta) = 
  \sum_{n\in\oZ} q^{\kappa(n+\frac{\mu}{2\kappa})^2}
  \zeta^{\kappa(n+\frac{\mu}{2\kappa})}
  =\zeta^{\frac{\mu}{2}}\,q^{\frac{m^2}{4\kappa}}\,
  \vartheta_{1,0}(q^{2\kappa},\zeta^\kappa q^{\mu-\kappa})
\end{equation}
(with the Jacobi theta function from~\eqref{Jacobi10}).  This accounts
for the level-$u$ theta function in
Eqs.~\eqref{uu-sumrule}--\eqref{B}.

We also recall~\cite{[KW]} the character formula corresponding to the
admissible $\tSL2_k$ module $\Adm_{t,u\mymid r,s}$ with the level as
in~\eqref{kplus2}, namely,
\begin{equation}
  \chi_{t,u\mymid r,s}^{\tSL2}(\tau, \sigma)=
  \frac{\theta_{b_+,ut}(\tau, \frac{\sigma}{u})
    -\theta_{b_-,ut}(\tau, \frac{\sigma}{u})}
  {\theta_{1,2}(\tau, \sigma)
    -\theta_{-1,2}(\tau, \sigma)},
\end{equation}
where $b_{\pm}=\pm ur - (s-1)t$ (and, as before, $1 \le r \le t-1$, $1
\le s \le u$).  Under the spectral flow, the~$\tSL2$ characters
transform as follows (see also~\cite{[FSST]} for the~$\tSL2$ spectral
flow properties):
\begin{equation}\label{sl2-SF}
  \chi^{\tSL2}_{u+1, u\mymid r,s}(\tau , \sigma - \lambda \tau)=
  (-1)^{\lambda}\,q^{\frac{\lambda^2(u-1)}{4u}}
  z^{-\frac{\lambda(u-1)}{2u}} 
  \chi^{\tSL2}_{u+1, u\mymid r,s+\lambda}(\tau, \sigma ),
\end{equation}
with $\chi^{\tSL2}_{u+1,u,r,s+\lambda}$ understood as follows: writing
$s+\lambda =u\alpha+\beta$ with $\alpha\in\oZ$, $\beta=0,\dots,u-1$,
we~have
\begin{equation}
    \chi^{\tSL2}_{u+1, u\mymid r,s+\lambda}(\tau, \sigma )=
    \begin{cases}
      \chi^{\tSL2}_{u+1, u\mymid r,(s+\lambda)_u}(\tau ,
      \sigma)\, &\text{for}\quad \alpha\in 2\oZ ,\\      
      -\chi^{\tSL2}_{u+1, u\mymid u+1-r,(s+\lambda)_u}    
    (\tau,\sigma)\,&\text{for}\quad \alpha\in 2\oZ +1\,,
  \end{cases}
\end{equation}
where $(x)_u$ denotes the residue mod~$u$.  The integrable
$\tSL2_{u-1}$ representations are periodic under the spectral flow
with period~2, while the spectral flow transform by~1 maps the
representations into one another according to $r\mapsto u+1-r$; we
then have the character transformation formula
\begin{equation}  
  \chi^{\tSL2}_{u+1, 1\mymid r,1}(\tau, \sigma + \tau) =
  q^{-\frac{u-1}{4}}\,z^{-\frac{u-1}{2}}\,
  \chi^{\tSL2}_{u+1, 1\mymid u+1-r,1}(\tau, \sigma)\,.
\end{equation}

\subsubsection{Identification of the $\tSSL21$ characters and the
  spectral flow} Turning to the right-hand side
of~\eqref{decompose-int}, we next identify which of the
$\tSSL21_{\frac{1}{u}-1}$ characters known from our previous work on
the subject~\cite{[BHT97]} correspond to the module
$\mL_{h_-,h_+,k;\beta}$ with twist $\beta$ and
\begin{equation}\label{quantum-numbers}
  h_-=\frac{u-s+1}{2u}, \qquad h_+=\frac{u-s-1}{2u}.
\end{equation}
We recall an extensive study of $\tSSL21_{\frac{1}{u}-1}$ characters
provided in~\cite{[BT96],[BHT97],[BHT98]}; these characters are
organised according to the eigenvalues that $H^-_0$ and $H^+_0$ have
on the \textit{twist-zero} state (see~\eqref{hw-twisted})
\begin{equation}
  \ket{H_-,H_+ = H_- - \tfrac{\nu}{u},k}\equiv
  \ket{H_-,H_- - \tfrac{\nu}{u},k;0}.
  \label{untwisted-hws}
\end{equation}
To see how these are related to~\eqref{quantum-numbers}, we observe
that the same irreducible module can be generated from the twisted
highest-weight state
\begin{equation}
  X^-(\nu,H_-,k)=\underbrace{E^1_{-\nu+1} \,\ldots E^1_{-1}}_{\nu
    -1}\cdot 
  \underbrace{F^2_{1-\nu}\,\ldots
    \,F^2_0}_{\nu}\ket{H_-, H_- - \tfrac{\nu}{u},k;0}, 
\end{equation}
which is singled out by the fact that it satisfies annihilation
conditions of type~\eqref{hwtop1}, namely $E^1_{-\nu}\approx0$,
$E^2_{\nu}\approx0$, $F^1_{\nu}\approx0$, and $F^{12}_{1}\approx0$ (in
the Verma module, the action of~$E^1_{-\nu}$ on $X^-$ produces the
charged singular vector, and hence, $E^1_{-\nu}\ket{X^-}=0$ in the
irreducible representation; in this sense, therefore, $X^-$ is a
pre-singular vector to~\eqref{Echminus}).  We find the eigenvalues
\begin{equation}\label{corner2}
  \begin{split}
    H^-_0(X^-)={}&H_- - \thalf,\\
    H^+_0(X^-)={}&H_+ + \nu -\thalf
  \end{split}
\end{equation}
that the Cartan generators have on this state.  

On the other hand, we see from~\eqref{quantum-numbers} that 
\begin{equation}
  h_+-h_-=-(k+1)=-\tfrac{1}{u}\,.
  \label{isohyprel}
\end{equation}
The $X^-$ state is then expressed through the twisted highest-weight
state as
\begin{equation}
  X^-= F^2_{-\beta}\ket{h_-,h_- - \tfrac{1}{u},k;\beta},
  \label{corner}
\end{equation}
and therefore, 
\begin{equation}
  \begin{split}
    H^-_0(X^-)={}&h_- - \thalf=\tfrac{-s+1}{2u}\,,\\
    H^+_0(X^-)={}&h_+ + 1-(\tfrac{1}{u}-1)\beta-\thalf
    = \tfrac{-s-1}{2u}+1-(\tfrac{1}{u}-1)\beta.
  \end{split}
  \label{corner1}
\end{equation}

By comparing~\eqref{corner1} and~\eqref{corner2}, we obtain 
\begin{equation}
  \begin{split}
    H_-={}&h_- =\tfrac{u-s+1}{2u}\,,\\
    H_+={}&h_+ - \tfrac{\beta}{u} = \tfrac{u - s - 1 - 2\beta}{2u}.
  \end{split}
  \label{corner-corr}
\end{equation}

This gives the parameters of the twist-zero highest-weight vectors in
the $\tSSL21$ modules entering the decomposition formula.  All of
these representations are obtained by taking the quotient of the Verma
modules with \textit{two} charged singular vectors, one of
type~\eqref{Echminus} and the other of type~\eqref{Echplus}, with the
integers labelling these singular vectors being of the
\textit{opposite} signs (obviously, there exist other singular vectors
that have to be factored over to obtain the irreducible
representation).  According to the values of $H_-$ and $H_+$, the
irreducible representation characters in any chosen `sector' of the
theory (Ramond, Neveu--Schwarz, etc.) at level $k=\frac{1}{u}-1$ are
those of class~IV ($\hf u(u+1)$ characters) and of class~V ($\hf
u(u-1)$ characters)~\cite{[BT96],[BHT97],[BHT98]}.  Choosing the
Ramond sector for definiteness, we have that
\begin{itemize}

\item[\textbf{IV}.]The class IV constraints on the highest weight state
  isospin and hypercharge are given by
  \begin{equation}\label{h-iv}
    2H_-+(k+1)m=0\qquad\text{and}\qquad H_+-H_-=m'(k+1),
  \end{equation}
  where $0 \le m' \le m \le u-1$ and $m, m' \in \oZplus$.  For the
  corresponding Verma module, this implies the existence of the
  charged singular vectors $\ket{C^{(-)}(-m',H_-,k)}$ and
  $\ket{C^{(+)}(m-m',H_-,k)}$.  The corresponding $\hf u(u+1)$
  irreducible representation characters are given by,
  \begin{equation}\label{ram4-prod}
    \chi^{\mathrm{R,IV}}_{m,m'}(q,z,\zeta)=
    q^{\frac{H_-^2-H_+^2}{k+1}}  
    z^{H_{-}}                    
    \zeta^{H_{+}}
    F^{\mathrm{R}}(q,z,\zeta)\,
    \frac{q^{-\frac{u}{8}}\,\eta(q^u)^3\,
      \vartheta_{1,1}(q^u,z q^{-m})}{
      \vartheta_{1,0}(q^u,z^\half\zeta^\half q^{-m'})\,
      \vartheta_{1,0}(q^u,z^\half\zeta^{-\half}q^{m'-m})}\,,
  \end{equation}
  with
  \begin{equation}\label{factorR}
    F^{\mathrm{R}}(q,z,\zeta) =
    \frac{\vartheta_{1,0}(q,z^\half\zeta^{-\half})\,
      \vartheta_{1,0}(q,z^\half\zeta^\half)}{
      \vartheta_{1,1}(q,z)
      \prod_{n\geq1}(1 - q^n)^3}
  \end{equation}
  being essentially the Verma module
  character~\eqref{sl21-Verma-char}.  Note that one of these
  characters only (the vacuum representation $m'=m=0$) is regular in
  the limit where~$z\to1$. %$\sigma \rightarrow 0$.

\item[\textbf{V}.]The class V constraints on the highest weight state
  isospin and hypercharge are given by,
  \begin{equation}
    2H_- - (k+1)(M+M'+2)=0\qquad \text{and} \qquad
    H_+ - H_-=-(M'+1)(k+1), 
    \label{MM'}
  \end{equation}
  where $0 \le M+M' \le u-2$ and $M,M' \in \oZplus$.  The charged
  singular vectors in the corresponding Verma module are
  $\ket{C^{(-)}(M'+1,H_-,k)}$ and $\ket{C^{(+)}(-M-1,H_-,k)}$; the
  corresponding $\hf u(u-1)$ irreducible representation characters are
  \begin{equation}\label{ram5}
    \chi^{\mathrm{R,V}}_{M,M'}(q,z,\zeta)=
    q^{{\frac{H_-^2 - H_+^2}{k+1}}}
    z^{H_{-}}       
    \zeta^{H_{+}}   
    F^{\mathrm{R}}(q,z,\zeta)\,
    \frac{-q^{-\frac{u}{8}}\,\eta(q^u)^3\, \vartheta_{1,1}(q^u,z
      q^{M+M'+2})}{ \vartheta_{1,0}(q^u,z^\half\zeta^\half q^{M'+1})\,
      \vartheta_{1,0}(q^u,z^\half\zeta^{-\half}q^{M+1}) }\,.
  \end{equation}
  Note that $u-1$ of them are regular in the limit as~$z\to1$
  %$\sigma \rightarrow 0$
  (they correspond to $M+M'=u-2$).
\end{itemize}

As we will see momentarily, a given orbit of the spectral
flow~\eqref{sl21-sf} involves the~$\chi^{\mathrm{R,V}}$ as well as
the~$\chi^{\mathrm{R,IV}}$ type characters.
\begin{Lemma}\label{lemma:transparent}
  The character of
  $\mL_{\frac{u-s+1}{2u},\frac{u-s-1}{2u},\frac{1}{u}-1;\beta}$ is
  given by
  \begin{multline}
    \kern-6pt
    \chi^{\mL}_{\frac{u-s+1}{2u},\frac{u-s-1}{2u},
      \frac{1}{u}-1;\beta}(q,z,\zeta)={}\\
    = \begin{cases}
      \chi^{\mathrm{R,IV}}_{s-1,u-1-\beta}(q,z,\zeta)&
      \text{for}\quad u-1 < s+\beta  
      \\[4pt]
      \chi^{\mathrm{R,V}}_{u-s-1-\beta,\beta}(q,z,\zeta)&
      \text{for}\quad u-1 \ge s+\beta    
    \end{cases}
    \kern44pt{}\\
    {}=q^{-(\beta+1)\frac{\beta+s}{u}}
    z^{\frac{1 - s}{2u}}
    \zeta^{-\frac{1 + 2\beta + s}{2u}}
    F^{\mathrm{R}}(q,z,\zeta)\,
    \frac{q^{-\frac{u}{8}}\,\eta(q^u)^3\,
      \vartheta_{1,1}(q^u,z q^{1-s})}{
      \vartheta_{1,0}(q^u,z^\half\zeta^\half q^{\beta+1})\,
      \vartheta_{1,0}(q^u,z^\half\zeta^{-\half}q^{-s-\beta})}\,.
    \kern-12pt
  \end{multline}
\end{Lemma}
Indeed, according to \eqref{untwisted-hws}, the relation between $H_+$
and $H_-$ involves a strictly positive integer~$\nu$, and therefore,
one is naturally in a class V context.  The labels $M$ and $M'$ are
given by~\eqref{MM'} with~$H_{\pm}$ from~\eqref{corner-corr},
\begin{equation}
  \begin{split}    
    M={}& u(H_-+H_+)-1=-s-1-\beta + u
    \\
    M'={}&u(H_--H_+)-1=\beta.
  \end{split}
\end{equation}
These can be translated into class-IV labels in the appropriate range
in accordance with the spectral flow properties of the class~IV and~V
characters, which we now consider.  The spectral flow~\eqref{sl21-sf}
acts on the~$\chi^{\mathrm{R,IV}}_{m,m'}$ characters by shifting the
labels as
\begin{equation}\label{map-IV}
  m\mapsto m\,,\qquad m'\mapsto m'-\theta\,,
\end{equation}
and for some values of $\theta$, the parameter $m'$ may flow to a
value outside the fundamental range $0\le m'\le m\le u-1$.  This gives
the class-V characters.  Indeed, first note that it is sufficient to
consider the spectral flow parameter $\theta$ in the range
$0\le\theta\le u-1$, given the (quasi)-periodicity property of the
class IV characters,
\begin{equation}
  \chi^{\mathrm{R,IV}}_{m+nu,m'+n'u}=(-1)^n\chi^{\mathrm{R,IV}}_{m,m'}.
\end{equation}
Then we have,
\begin{alignat*}{2}
  \chi^{\mathrm{R,IV}}_{m,m'}(\tau, \sigma,\nu +2\theta \tau)={}&
  q^{\frac{\theta^2(1-u)}{u}}\zeta^{\frac{\theta (1-u)}{u}}
  \chi^{\mathrm{R,IV}}_{m,m'-\theta}(\tau, \sigma,\nu )
  &&\quad\text{for}\quad0 \le m'-\theta \le m,\displaybreak[1]\\
  \chi^{\mathrm{R,IV}}_{m,m'}(\tau, \sigma,\nu +2\theta \tau)={}&
  q^{\frac{\theta^2(1-u)}{u}}\zeta^{\frac{\theta (1-u)}{u}}
  \chi^{\mathrm{R,V}}_{u-1-(m-m'+\theta),\theta -m'-1}(\tau,
  \sigma,\nu)\kern-70pt
  &&\\
  &&&\quad\text{for}\quad m'+1 \le \theta \le u-1-(m-m'),
  \\
  \chi^{\mathrm{R,IV}}_{m,m'}(\tau, \sigma,\nu +2\theta \tau)={}&
  q^{\frac{\theta^2(1-u)}{u}}\zeta^{\frac{\theta (1-u)}{u}}
  \chi^{\mathrm{R,IV}}_{m,m'-\theta +u}(\tau, \sigma,\nu )
  &&\quad\text{for}\quad u-(m-m') \le \theta \le u-1.
\end{alignat*}

On the other hand, the spectral flow transform~\eqref{sl21-sf} acts on
class V characters by shifting the representation labels as
\begin{equation}\label{map-V}
  M\mapsto M - \theta\,,\qquad M'\mapsto M' + \theta,
\end{equation}
and one has,
\begin{alignat*}{2}
  &\chi^{\mathrm{R,V}}_{M,M'}(\tau, \sigma,\nu +2\theta\tau)=
  q^{\frac{\theta^2(1-u)}{u}}\zeta^{\frac{\theta (1-u)}{u}}
  \chi^{\mathrm{R,V}}_{M-\theta,M'+\theta}(\tau, \sigma,\nu)
  &&
  \quad\text{for}\quad M-(u-2) \le \theta \le u-2-M',\displaybreak[0]\\
  &\chi^{\mathrm{R,V}}_{M,M'}(\tau, \sigma,\nu +2(u-1-M')\tau)=
  q^{\frac{(M'+1-u)^2(u-1)}{u}}\zeta^{\frac{(M'+1-u)(u-1)}{u}}
  \chi^{\mathrm{R,IV}}_{u-2-M-M',0}(\tau, \sigma,\nu)\kern-120pt
  &&\\
  &&&\quad\text{for}\quad \theta =u-1-M',\displaybreak[2]\\
  &\chi^{\mathrm{R,V}}_{M,M'}(\tau, \sigma,\nu +2\theta\tau)=
  q^{\frac{\theta^2(1-u)}{u}}\zeta^{\frac{\theta (1-u)}{u}}
  \chi^{\mathrm{R,V}}_{M-\theta+u,M'+\theta-u}(\tau,
  \sigma,\nu) 
  &&\quad\text{for}\quad u-M' \le \theta \le u-1.
\end{alignat*}

\subsubsection{Sumrules for~$\protect\widehat{s\ell}(2|1)$
  characters}\label{sec:sumrules} A key ingredient in the derivation
of the character identities is provided by the decomposition formulas
of the $u^2$ characters given above into $\tSL2$ characters at the
same level $k=\frac{1}{u}-1$~\cite{[HT98]}, namely,
\begin{align}\label{decomp1}
  \chi^{\mathrm{R,IV}}_{m,m'}(\tau, \sigma,\nu)={}& \sum_{i=0}^{u-1}
  \chi^{\tSL2}_{u+1,u\mymid u-i,m+1}(\tau, \sigma)\,
  F^{\mathrm{IV}}(\tau, \nu),\displaybreak[0]\\
  \chi^{\mathrm{R,V}}_{M,M'}(\tau, \sigma,\nu)={}& \sum_{i=0}^{u-1}
  \chi^{\tSL2}_{u+1,u\mymid u-i,u-1-M-M'}(\tau, \sigma)\,
  F^{\mathrm{V}}(\tau, \nu),\label{decomp2}
\end{align}
where
\begin{align}
  F^{\mathrm{IV}}(\tau,\nu)={}& \sum_{a=0}^{u-2}c_{i,
    X(i,a)}^{(u-1)}(\tau) \theta_{(u-1)(m-2m'+u)+uX(i,a),u(u-1)}(\tau,
  \tfrac{\nu}{u}),\displaybreak[1]\\
  F^{\mathrm{V}}(\tau,\nu)={}& \sum_{a=0}^{u-2}c_{i,
    X(i,a)}^{(u-1)}(\tau) \theta_{(u-1)(M'-M)+uX(i,a),u(u-1)}(\tau,
  \tfrac{\nu}{u}).
\end{align}
In the above, $c^{(u-1)}_{i,j}(\tau)$ are the~$\tSL2$ string functions
at level $u-1$~\cite{[KP]} and
$X(i,a)=(u-1)i+2i(\frac{u}{2}-[\frac{u}{2}])-2a$, where
$[\frac{u}{2}]$ is the integer part of~$\frac{u}{2}$.  The $\tSL2$
string functions at the level $u-1$ were first introduced by Ka\v c
and Peterson to rewrite $\tSL2_{u-1}$ characters in terms of
generalised theta functions at level $u-1$, namely,
\begin{equation}
  \chi^{\tSL2}_{u+1,1\mymid r,1}(\tau, \nu)= \sum_{m=-u+2}^{u-1}
  c^{(u-1)}_{r-1,m}(\tau) \theta_{m,u-1}(\tau,\nu),
\end{equation} 
with the properties
\begin{equation}
  c_{\ell,
    m}^{(k)}(\tau)=c_{\ell,-m}^{(k)}(\tau)=c_{k-\ell,k-m}^{(k)}(\tau)= 
  c_{\ell,m+2nk}^{(k)}(\tau)
\end{equation}
where $n \in \oZ$ and $\ell\equiv m~\text{mod}~2$.

As is clear from~\eqref{decomp1} and~\eqref{decomp2}, each $\tSSL21$
character is decomposed in exactly $u$ characters
$\chi^{\tSL2}_{u+1,u\mymid r,s} (\tau, \sigma)$ with $s$ fixed.  More
precisely, at a {\em fixed\/} value of~$s$ in the range $1 \le s\le
u$, the~$u$ $\hslc$ characters which can be decomposed into the~$u$
characters $\chi^{\tSL2}_{u+1,u\mymid r,s}(\tau,\sigma)$ are,
\begin{itemize}
  
\item[\bf IV.] $\chi^{\mathrm{R,IV}}_{s-1,m'}$ with $0 \le m'\le
  s-1$~\qquad\quad($s$ characters),
  
\item[\bf V.] $\chi^{\mathrm{R,V}}_{M,M'}$ \ with
  $M+M'=u-1-s$\quad($u-s$ characters).
\end{itemize}

Decomposition formulas~\eqref{decomp1} and~\eqref{decomp2} for
the~$\tSSL21_{\frac{1}{u}-1}$ characters are not only explicitly given
in terms of characters of the subalgebra $\tSL2_{\frac{1}{u}-1}$, but
they also encode some information on a dual algebra $\tSL2_{u-1}$
through the~$\tSL2$ string functions at the level $u-1$.  We were able
to derive character identities for the values $u=2$ and $u=3$ by using
standard, albeit somewhat tedious, manipulations on generalised theta
functions.  On the basis of these two cases, we were led to conjecture
identities for arbitrary~$u\in\oN$.  One may write $2u$ sumrules for
Ramond $\hslc$ characters at the level~$k=\frac{1}{u}-1$ corresponding
to the decomposition~\eqref{decompose-int} for irreducible
representations.  As before, we use~$(\lambda)_u$ to denote the
residue modulo~$u$ of~$\lambda$, and~$[\frac{\lambda}{u}]$ to denote
the integer part of~$\frac{\lambda}{u}$.
%%\begin{Lemma}
The  $2u$ sumrules read
\begin{equation}\label{uu-sumrule}
  S^{\mathrm{R}}_{\lambda}(\tau,\rho,\sigma,\nu)~:~\qquad 
  \boxed{A^{\mathrm{R}}_{\lambda}(\tau,\rho,\sigma,\nu)=
    B^{\mathrm{R}}_{\lambda}(\tau,\rho,\sigma,\nu)}
  \qquad\qquad\lambda = 0,\dots,2u-1,
\end{equation}
where
\begin{multline}\label{A} 
  A^{\mathrm{R}}_{\lambda}(\tau,\rho,\sigma,\nu)={}\\
  {}= \sum_{i=0}^{(\lambda)_u}
  \theta_{u-2i+\lambda,u}(\tau,-\rho)             
  \chi^{\mathrm{R,IV}}_{(\lambda)_u,i}(\tau,\sigma,\nu)
  +\sum_{i=(\lambda)_u+1}^{u-1}
  \theta_{u-2i+\lambda,u}(\tau,-\rho)             
  \chi^{\mathrm{R,V}}_{i-1-(\lambda)_u,u-1-i}(\tau,\sigma,\nu)
\end{multline}
and
\begin{multline}\label{B}
  B^{\mathrm{R}}_{\lambda}(\tau,\rho,\sigma,\nu)=\\
  {}=\theta_{1+\lambda,1}(\tau,\tfrac{u-1}{u}\nu - \rho)  
  \sum_{i=0}^N \chi^{\tSL2}_{u+1,u\mymid
    2i+1+(u-1-4i)[\frac{\lambda}{u}], (\lambda)_u + 1} (\tau,\sigma)\,
  \chi^{\tSL2}_{u+1,1\mymid u-2i,1}
  (\tau,\tfrac{\nu}{u}+\rho)             
  \\
  {}+\theta_{\lambda,1}(\tau,\tfrac{u-1}{u}\nu - \rho)  
  \sum_{i=0}^{N'} \chi^{\tSL2}_{u+1,u\mymid
    2i+2+(u-3-4i)[\frac{\lambda}{u}], (\lambda)_u + 1} (\tau,\sigma)\,
  \chi^{\tSL2}_{u+1,1\mymid u-(2i+1),1}
  (\tau,\tfrac{\nu}{u}+\rho),           
\end{multline}
with $u-2 \le 2N \le u-1$ and $u-3 \le 2N' \le u-2$.
%%\end{Lemma}
%%\noindent
Note that the second sum in~\eqref{A} is zero for~$(\lambda)_u=u-1$.

There is complete agreement between the sumrules~\eqref{uu-sumrule}
and the decomposition formula~\eqref{decompose-int}.  The appearance
of the level-$u$ theta functions was explained in~\eqref{future-theta}.
The theta functions at the level $\kappa=1$ represent the level-one
$\tSL2$ integrable characters, since one has
\begin{equation}
  \frac{1}{\eta(\tau)}\,\theta_{r-1,1}(\tau, \nu)=
  \chi^{\tSL2}_{3,1\mymid r,1}(\tau, \nu),\qquad r=1,2.
\end{equation}
Strictly speaking, Eq.~\eqref{decompose-int} provides the first $u$
sumrules, but the other set is obtained by transforming each of them
under the flow,
\begin{align}
  \sigma\mapsto{}&\sigma - u \tau,\notag\\
  \nu\mapsto{}&\nu + u \tau,\\
  \rho\mapsto{}&\rho - \tau.\notag
\end{align}
To make the correspondence between \eqref{uu-sumrule} and
\eqref{decompose-int} transparent, it remains to recall
Lemma~\ref{lemma:transparent}.

We end up this section by observing that the above sumrules behave in
a remarkable way under two particular flows of the variables $\sigma$,
$\nu$, and~$\rho$.  We first note that the class IV and V characters
transform~as
\begin{equation}\label{flow3-IV}
  \chi^{\mathrm{R,IV}}_{m,m'}(\tau, \sigma + u \tau, \nu + (u-2)\tau)=
  \begin{cases}
    P\,\chi^{\mathrm{R,IV}}_{m,m'+1}(\tau, \sigma, \nu)&\text{for}~m'
    \le m-1,\\   
    P\,\chi^{\mathrm{R,IV}}_{u-1,0}(\tau, \sigma,
    \nu)&\text{for}~m=m'=u-1,\\ 
    P\,\chi^{\mathrm{R,V}}_{0,u-2-m'}(\tau, \sigma,
    \nu)&\text{for}~m'=m\le u-2,
  \end{cases}
\end{equation}
\begin{multline}\label{flow3-V}
  \chi^{\mathrm{R,V}}_{M,M'}(\tau, \sigma + u \tau, \nu + (u-2)\tau)\\
  {}=
  \begin{cases}
    P\,\chi^{\mathrm{R,V}}_{M+1,M'-1}(\tau, \sigma, \nu)&\text{for}~
    0 \le M \le u-3~\text{and}~1 \le M' \le u-2,\\
    P\,\chi^{\mathrm{R,IV}}_{u-2-M,0}(\tau, \sigma, \nu)&\text{for}~
    0 \le M \le u-2~\text{and}~M'=0,
  \end{cases}
\end{multline}
where
\begin{equation}\label{P}
  P=(-1)^{u+1}q^{\frac{(u-1)^2}{u}}z^{\frac{u-1}{2}}
  \zeta^{-\frac{(u-1)(u-2)}{2u}},
\end{equation}

\begin{Rem}
  The~$u$ terms in~$A^R_{\lambda}$ (resp.\ $B^R_{\lambda}$) form an
  orbit under the flow
  \begin{equation}\label{SF1}
    \begin{split}      
      \sigma\mapsto{}& \sigma +u \tau,\\
      \nu\mapsto{}& \nu +(u-2) \tau,\\
      \rho\mapsto{}& \rho + 2 \tfrac{\tau}{u},
    \end{split}
  \end{equation}
  as can be readily checked with the help of the spectral flow
  formulas~\eqref{flow3-IV}, \eqref{flow3-V}, and~\eqref{sl2-SF}.  The
  invariance under \eqref{SF1} indicates that each expression
  $A^R_{\lambda}$ (equivalently, $B^R_{\lambda}$) potentially
  describes a character corresponding to a representation of the
  bigger affine Lie superalgebra $\hD$.  Indeed, the spectral flow
  generator has isospin~$H_-=1/2$, hypercharge $H_+=\frac{2-u}{2u}$, a
  $u(1)$ charge proportional to $-\frac{1}{u}$ in the direction
  orthogonal to $s\ell(2|1)$ and conformal weight~$1$.  (This follows
  by comparing the quantum numbers of the vacuum representation
  of~$\tSSL21\oplus\widehat{u}(1)$ with those of the representation
  with character $\theta_{2u-2,u}(\tau,
  \frac{\rho}{u})\chi^{\mathrm{R,V}}_{0,u-2}(\tau,\sigma,\nu)$, which
  are in the same spectral flow orbit and appear in the~$\lambda =u$
  sumrule).  If one extends the algebra $\tSSL21\oplus\widehat{u}(1)$
  by this spectral flow generator, one generates $\hD$, as can be most
  quickly understood by looking at the~$\D$ root diagram in
  Appendix~\ref{app:D-algebra}.  Indeed, consider for example the
  first embedding of~$s\ell(2|1)$ in~$\D$ in Table~\ref{tab:Table1};
  the spectral flow generator can be identified with the current
  corresponding to the root $\alpha_1+\alpha_2$, which is needed in
  order to extend the $s\ell(2|1)$ root diagram to $\D$.
\end{Rem}

We next note that for $\alpha \in \oZ$ chosen in such a way that
$0 \le m+\lambda -u\alpha \le u-1$, we have
\begin{multline}\label{flow2-IV}
  \chi^{\mathrm{R,IV}}_{m,m'}(\tau, \sigma -\lambda \tau, \nu +
  \lambda\tau)\\ 
  {}=
  \begin{cases}
    (-1)^{\lambda +\alpha} z^{-\frac{\lambda(u-1)}{2u}}
    \zeta^{-\frac{\lambda(u-1)}{2u}}
    \chi^{\mathrm{R,IV}}_{m+\lambda -u\alpha,m'}(\tau, \sigma, \nu )
  &\text{for}~\,m'\le m+\lambda-u\alpha\,,\\
    (-1)^{\lambda +\alpha} z^{-\frac{\lambda(u-1)}{2u}}
    \zeta^{-\frac{\lambda(u-1)}{2u}}
    \chi^{\mathrm{R,V}}_{m'-m-\lambda +u\alpha-1,u-1-m'}(\tau, \sigma,
    \nu  
    )\
  &\text{for}~\,m'> m+\lambda-u\alpha\,.
  \end{cases}
\end{multline}
For class V characters, with $\alpha \in \oZ$ chosen such that $0\le
u-2-M-M'+\lambda-u\alpha \le u-1$, similarly,
\begin{multline}\label{flow2-V}
  \chi^{\mathrm{R,V}}_{M,M'}(\tau, \sigma -\lambda \tau, \nu +
  \lambda\tau)\\ 
  ={}
  \begin{cases}
    (-1)^{\lambda +\alpha} z^{-\frac{\lambda(u-1)}{2u}}
    \zeta^{-\frac{\lambda(u-1)}{2u}}
    \chi^{\mathrm{R,V}}_{M-\lambda +u\alpha,M'}(\tau, \sigma,\nu)
    &\text{for}~M-\lambda +u\alpha \ge 0\,,\\    
    (-1)^{\lambda +\alpha } z^{-\frac{\lambda(u-1)}{2u}}
    \zeta^{-\frac{\lambda(u-1)}{2u}}
    \chi^{\mathrm{R,IV}}_{u-2-(M-\lambda+u\alpha)-M',u-1-M'}(\tau,
    \sigma, 
    \nu)
    &\text{for}~M-\lambda +u\alpha < 0.
  \end{cases}
\end{multline}
\begin{Rem}  
  The~${\lambda'}^{th}$ sumrule may be obtained from
  the~${\lambda}^{th}$ sumrule by the transformation
  \begin{equation}\label{SF2}
    \begin{split}
      \sigma \mapsto{}& \sigma - (\lambda '-\lambda) \tau,\\
      \nu  \mapsto{}& \nu + (\lambda '-\lambda) \tau,\\
      \rho \mapsto{}&
      \rho - (\lambda '-\lambda) \tfrac{\tau}{u}\,.         
    \end{split}
  \end{equation}
  In particular, because of the quasi-periodicity of the level-$u$
  theta functions and because $\tSSL21_{\frac{1}{u}-1}$ characters are
  periodic under the simultaneous shifts $\sigma \mapsto \sigma
  -u\tau$ and $\nu \mapsto \nu+u\tau$, one has the quasi-periodicity
  properties
  \begin{equation}\label{periodicity}
    \begin{split}    
      A^{\mathrm{R}}_{\lambda}(\tau, \rho - 2\tau,    
      \sigma - 2u\tau, \nu + 2u\tau)={}&
      q^{-u}y^{u}                  
      z^{1-u}\zeta^{1-u}
      A^{\mathrm{R}}_{\lambda}(\tau,\rho,\sigma,\nu),\\
      B^{\mathrm{R}}_{\lambda}(\tau, \rho - 2\tau,    
      \sigma -2u\tau, \nu +2u\tau)={}&
      q^{-u}y^{u}                   
      z^{1-u}\zeta^{1-u}
      B^{\mathrm{R}}_{\lambda}(\tau,\rho,\sigma,\nu), 
    \end{split}
  \end{equation}
  and the set of~$2u$ sumrules \eqref{uu-sumrule} is closed under the
  above spectral flow.  Moreover, this set carries a unitary
  representation of the modular group, as can be explicitly checked by
  using the modular transformations of~$\tSL2$ characters at
  integrable and fractional levels. A complete treatment of the
  modular properties must include the Neveu--Schwarz, the `super'
  Ramond and `super' Neveu--Schwarz sectors, which may be obtained as
  follows~\cite{[HT98]}.  The Neveu--Schwarz sumrules are derived by
  flowing the Ramond sector sumrules according to,
  \begin{equation}
    S^{\mathrm{R}}_{\lambda}(\tau, \rho,-\sigma -\tau, \nu)
    =q^{-\frac{1-u}{4u}}z^{-\frac{1-u}{2u}}
    S^{NS}_{\lambda}(\tau, \rho, \sigma , \nu),
  \end{equation}
  while the `super' Ramond and Neveu--Schwarz sectors are given by,
  \begin{equation}
    S^{R,NS}_{\lambda}(\tau, \rho,\sigma +1, \nu)
    =
    \tilde{S}^{R,NS}_{\lambda}(\tau, \rho, \sigma , \nu).
  \end{equation}
  It should be noted that the group of modular transformations and the
  group of spectral flow transformations on characters are combined
  into an `extended modular group' via the semi-direct product, in
  which the spectral flow transformations are the invariant subgroup;
  thus, the `extended modular group' representation can be induced
  from the spectral flow representation, and these are the
  transformations that close on the chosen set of
  $\tSSL21_{\frac{1}{u}-1}$ representations.  The relevance of these
  facts to the representation theory of~$\hD$ are beyond the present
  scope and are left aside for future work.
\end{Rem}

\section{\textbf{Conclusions}}\label{sec:conclusions}
We have seen that a vertex operator extension of two $\tSL2$ algebras
at levels $k$ and $k'$ satisfying the duality relation $(k+1)(k'+1)=1$
yields an interesting structure, the exceptional affine Lie
superalgebra $\hD$.  This novel construction should provide the
setting needed to build some classes of $\hD$ representations whose
characters are given by either side of the
sumrules~\eqref{uu-sumrule}.

In this paper however, we made use of the above vertex operator
extension to construct representations of the $\tSSL21$ subalgebra of
$\hD$, and saw that they give sums of representations `twisted' by the
$\tSSL21$ spectral flow~\eqref{decompose}.  We also derived the
corresponding character identities relating $\tSSL21$ characters to
the constituent $\tSL2_k$ and $\tSL2_{k'}$ characters, both for Verma
modules (Sec.~\ref{sec:Verma-char}) and irreducible representations
(Eq.~\eqref{uu-sumrule}).  The latter identities involve $\tSSL21$
representations at admissible (non integrable) level
$k=\frac{1}{u}-1,\, u \in \oN$ and relate them to representations of
$\tSL2_{\frac{1}{u}-1}$ and $\tSL2_{u-1}$.  Interestingly enough, the
admissible $\tSSL21_{\frac{1}{u}-1}$ representations are related to
the admissible $\tSL2_{\frac{1}{u}-1}$ representations, whose
characters are periodic under the spectral flow with period~$2u$, and
to the integrable $\tSL2_{u-1}$ representations, which are themselves
periodic under the spectral flow with period~2~\cite{[FSST]}.  The
interplay of admissible and integrable representations within a bigger
algebraic structure is quite remarkable and should be further
exploited in the way any duality is: in this context for instance, it
should relate the representation theory of $\hD$ at integer and
fractional levels.  We hope to return to these issues elsewhere.

Another important point raised by the relation between representations
and by the corresponding character identities is that of the closure
under modular transformations.  The $\tSSL21$ characters appearing in
these identities do carry a unitary representation of the modular
group~\cite{[J99]}, and using this information, one can explicitly
check that the $2u$ functions $A(\lambda),\, \lambda =0, \ldots 2u-1$
on the left-hand side of the sumrules \eqref{uu-sumrule} also carry a
unitary representation of the modular group. It is therefore tempting
to identify these functions with the characters corresponding to a
particular class of $\hD$ representations satisfying the requirements
for $\tSSL21 \oplus \hat{u}(1)$ to be conformally embedded in $\hD$,
as discussed in Sec.~\ref{sec:D}.

A very interesting general problem is to verify the functorial
properties of the correspondence established between the
$\tSL2_{k}\oplus\tSL2_{k'}$ and $\tSSL21_k$ representations; in the
simpler case of the relation between $\tSL2$ and $N=2$ superconformal
representations~\cite{[FST]}, a similar relation \textit{is} the
equivalence of categories modulo the spectral flow (i.e., the
equivalence of representation theories of two algebras obtained by
extending the universal enveloping of $\tSL2$ and $N=2$, respectively,
by the spectral flow operator).  We have seen that in the present
case, the spectral flow plays a very similar r\^ole, and thus an
interesting problem is whether we again can construct a similar
functor; the two cases are actually related by the Hamiltonian
reduction functor
\begin{equation*}
  \begin{array}{ccc}
    \tSL2_{k}\oplus\tSL2_{k'} &\atop{
      \displaystyle\xleftarrow{\phantom{\text{ mod spectral flow}}}}{
      \displaystyle\xrightarrow[\phantom{\text{ mod spectral
          flow}}]{}}& 
    \tSSL21_k \\[-6pt]
    \biggm\downarrow &   & \biggm\downarrow
    \kern-6pt\begin{array}{l}
      {}_{\text{Hamiltonian Reduction}}\\
      {}^{\text{\cite{[BO],[BLNW],[IK]}}}
    \end{array}\kern-100pt\\[-2pt]
    \tSL2_{k'-1} & \atop{
      \displaystyle\xleftarrow{\simeq\text{ mod spectral flow}}}{
      \displaystyle\xrightarrow[\phantom{\text{ mod spectral
          flow}}]{}}& N=2 
  \end{array}
\end{equation*}
which may be extended to an argument demonstrating the functorial
properties of the correspondence found in this paper.  We note that
the character identity corresponding to~\eqref{decompose-Weyl} turns
out to be \textit{equivalent} to the character identity derived from
the correspondence between the $\tSL2$ and $N=2$ Verma
modules~\cite{[FSST]}; this identity has also been known in a
different representation-theory context~\cite{[KW-number]}.

As another future research direction, we note the possibility to use
various free-field (and other) realisations
of~$\tSSL21$~\cite{[S-sl21],[BKT]} in the construction of
Secs.~\ref{sec:reconstructing}--\ref{sec:reps-inverse}; it would be
interesting, for example, to relate the corresponding screening
operators and interpret this relation in terms of the respective
quantum groups.

Finally, it is worth exploring the \textit{geometric interpretation}
of the construction for $\tSSL21$ found in this paper.  Despite the
complication generated by the presence of the odd integer $n=1$ on the
right-hand side of~\eqref{eq:n}, we expect to be able to proceed
similarly to the $n=0$ case, relevant to the coupling of matter to
gravity, where no auxiliary scalar is needed and the basic geometric
setting involves the \textit{loop group}
$\widetilde{SL}(2,\oC)=\{\ell:S^1\to SL(2,\oC)\}$.  In that case, the
analogue of $\tSSL21$ is an extended algebra consisting of the
semi-direct product of two commuting $\tSL2$ algebras (corresponding
to the left and the right actions on the loop group) with levels $k_1$
and $k_2$ constrained by Eq.~\eqref{eq:n} with $n=0$, and the
contracted vertex operators $\oC^2(z)\tensor\oC^2(z)$, which are
functions on the group and can be viewed as matrix elements of the
evaluation representation (in contrast with the $\tSSL21$ case,
$\oC^2(z)\tensor\oC^2(z)$ do commute). The constraint on $k_1$ and
$k_2$ actually follows from imposing the Knizhnik--Zamolodchikov
equation on the $\oC^2(z)\tensor\oC^2(z)$ vertex operators.  The
vacuum representation of the extended algebra can be described in
terms of distributions on $G_+$ -- the subgroup of the loop group
consisting of mappings that extend to the origin from $S^1=\{z \in \oC
: |z|=1\}$ -- and is given by a sum of Weyl modules of the two $\tSL2$
algebras.  A different piece of this picture is provided by
distributions on the double quotient $G_-\backslash
\widetilde{SL}(2,\oC)/G_+$ where $G_-$ is the subgroup of
$\widetilde{SL}(2,\oC)$ consisting of loops $\ell$ which are the
boundary values of holomorphic maps $\ell: \{z \in \oC \cup \infty:
|z| > 1 \rightarrow SL(2,\oC)\}$ \cite{[PS]}. The equivalence classes
$X(n)$ are labelled by a positive integer $n\in\oZ_+$.  The
distributions living on a given $X(n)$ carry the left and the right
$\tSL2$ actions; this time, however, the Weyl modules are combined
differently since $\mW_\ell\tensor\mW_m$ enters as many times as there
are $n$-dimensional $\SL2$ representations in the tensor product of
the $\ell$- and $m$-dimensional ones.  It is interesting to investigate 
how much of this description can be carried over to the `noncommutative' 
case of the $\tSSL21$ algebra constructed in this paper.

\bigskip

\noindent\textbf{Acknowledgments} We thank A.~Belavin, H.~Kausch, W. ~Oxbury,
I.~Shchepochkina, I.~Tipunin, and G.~Watts for discussions. This work
was supported by the EPSRC grant GR/M12544, by the RFBR Grant 99-01-01169, INTAS-OPEN-97-1312, and partly by the RFBR
Grant~98-01-01155 and the Russian Federation President
Grant~99-15-96037.  A.M.S.~gratefully acknowledges kind hospitality
extended to him at the Department of Mathematical Sciences, University
of Durham.  A.T.~acknowledges The Leverhulme Trust for a fellowship.

\appendix
\section{\textbf{Some $s\ell(2)$ quantum group
  relations}}\label{sec:quantum-group} In describing the~$\SL2_q$
quantum group, we follow the conventions of~\cite{[Kass]}.  The
quantum group relations are
\begin{equation}\label{sl2q}
  \begin{split}
  &K\,K^{-1} = K^{-1}\,K =1\,,\\
  &K\,E\,K^{-1} = q^2\,E\,,\qquad K\,F\,K^{-1} = q^{-2}\,F\,,\\
  &[E, F] = \frac{K - K^{-1}}{q - q^{-1}}\,.
  \end{split}
\end{equation}
The antipode acts on these generators as follows:
\begin{equation}\label{antipode}
  S(E) = -E\,K^{-1}\,,\qquad S(F)=- K\,F\,,\qquad
  S(K)=K^{-1}\,,\qquad S(K^{-1}) = K\,,
\end{equation}
and the comultiplication is given by
\begin{gather}
  \Delta(E)=1\tensor E + E\tensor K\,,\qquad
  \Delta(F)=K^{-1}\tensor F + F\tensor 1\,,\displaybreak[1]\\
  \Delta(K)=K\tensor K\,,\qquad \Delta(K^{-1})=K^{-1}\tensor K^{-1}\,.
\end{gather}
Together with the counit $\varepsilon$ given by
$\varepsilon(E)=\varepsilon(F)=0$,
$\varepsilon(K)=\varepsilon(K^{-1})=1$, these relations endow $\SL2_q$
with a Hopf algebra structure.

For a module $\qmV$ over a Hopf algebra $A$, the~$A$ action on the dual
module $\qmV^*$ is defined by
\begin{equation}\label{dual}
  (a\,f)(v)=f(S(a)\,v)\,,\qquad a\in A\,,\quad f\in \qmV^*\,,\quad
  v\in \qmV\,.
\end{equation}

Let $\qmV_{\epsilon, n}$ be the~$\SL2_q$ module with the
highest-weight vector $v_0$ such that
\begin{equation}
  E\,v_0=0\,,\qquad K\,v_0=\epsilon q^{n}\,v_0\,,
\end{equation}
where $\epsilon^2=1$ and $n$ is a positive integer.  We then define
\begin{equation}
  F\,v_{i-1}=[i]\,v_i\,,
\end{equation}
whence
\begin{equation}
  E\,v_i=\epsilon [n-i+1]\,v_{i-1}\,,
  \qquad
  K\,v_{i}=\epsilon\,q^{n-2i}\,v_i\,.
\end{equation}
We use the standard notation
\begin{equation}
  [i]=\frac{q^i - q^{-i}}{q-q^{-1}}\,,
  \qquad
  \atopwithdelims{n}{i} = \frac{[n]!}{[i]!\,[n-i]!}\,,\qquad
  [i]!=[1]\,[2]\dots[i]\,.
\end{equation}
There is an invariant scalar product
\begin{equation}
  (v_i,\,v_j)=\delta_{ij}\,q^{-i(n-i-1)}\atopwithdelims{n}{i}\,,
\end{equation}
In the~$\qmV_{\epsilon,n}$ modules with $n=1$, in particular, the
$\SL2_q$ action on the basis vectors $v_0$ and $v_1$ is given by
\begin{alignat}{3}
  E\,v_0={}&0\,,&\qquad K\,v_0={}&\epsilon q\,v_0\,,&\qquad
  F\,v_0={}&v_1\,,\\
  E\,v_1={}&\epsilon v_0\,,& K\,v_1={}&\epsilon q^{-1}\,v_1\,,&
  F\,v_1={}&0\,.
\end{alignat}
In the dual module with the dual basis $v^i$, we then find
from~\eqref{dual} and~\eqref{antipode}:
\begin{alignat}{3}
  E\,v^0={}&-q\,v^1\,,&\qquad K\,v^0={}&\epsilon q^{-1}\,v^0\,,&\qquad
  F\,v^0={}&0\,,\displaybreak[0]\\
  E\,v^1={}&0\,,& K\,v^1={}&\epsilon q\,v^1\,,&
  F\,v^1={}&-\epsilon\,q^{-1}v^0\,.
\end{alignat}

Let $\qmV'_{\epsilon',1}$ be a similar module over $\SL2_{q^{-1}}$,
with the basis $v'_0$ and $v'_1$.  As is easy to check, it is also a
module over $\SL2_q$, the~$\SL2_q$ action being given by
\begin{alignat}{3}\label{q-inverse1}
  E\,v'_0={}&v'_1\,,&\qquad K\,v'_0={}&\epsilon' q^{-1}\,v'_0\,,&\qquad
  F\,v'_0={}&0\,,\\
  E\,v'_1={}&0\,,& K\,v'_1={}&\epsilon' q\,v'_1\,,&
  F\,v'_1={}&\epsilon' v'_0\,.\label{q-inverse2}
\end{alignat}
The tensor product of~$\SL2_q$ modules $\qmV'_{\epsilon',1}\tensor
\qmV_{\epsilon,1}$ is decomposed as
\begin{equation}
  \qmV'_{\epsilon',1}\tensor \qmV_{\epsilon,1} =
  \qmV_{\epsilon\epsilon',0} \oplus \qmV_{\epsilon\epsilon',2}\,,
\end{equation}
where $\qmV_{\epsilon\epsilon',2}$ is generated from $v'_1\tensor v_0$,
and $\qmV_{\epsilon\epsilon',0}$ from $v'_0\tensor v_0 - q v'_1\tensor
v_1$.  The projection $\qmV'_{\epsilon',1}\tensor \qmV_{\epsilon,1}\to
\qmV_{\epsilon\epsilon',0}$ can be defined in terms of the trace
$\langle\cdot,\cdot\rangle$ such that
\begin{equation}\label{q-trace}
  \langle v'_0,v_0\rangle=1\,,\qquad \langle v'_1,v_1\rangle=-q\,,
  \qquad
  \langle v'_0,v_1\rangle=0\,,\qquad\langle v'_1,v_0\rangle=0\,.
\end{equation}
The vertex operator construction involves this trace operation in the
case where $\epsilon=1$ and~$\epsilon'=-1$: from the quantum group
representation standpoint, each $\tSSL21_k$ (or $\hD_k$) fermion is
constructed as
\begin{equation}
  -q v'_1\tensor v_1 + v'_0\tensor v_0 =
  q v'_1\tensor F v_0 - \epsilon'F\,v'_1\tensor v_0\,.
\end{equation}
With the basis of the quantum group module represented by the vertex
operators as in Sec.~\ref{sec:basic}, we identify $qF$ in the first
term with the screening operator $S$ acting in the unprimed sector,
and $-\epsilon'F$ in the second terms with $S'$ acting in the primed
sector; we then denote the action of~$S$ or $S'$ on the respective
vertex operator with a tilde.  This gives~\eqref{E0}--\eqref{F0},
where we omit the tensor product sign and \textit{write} the primed
and the unprimed multipliers in the reversed order compared to the
formula with the tensor-product notation; we hope that this minor
notational discrepancy does not lead to confusion.

\section{\textbf{$\protect\widehat{s\ell}(2|1)$ algebra, spectral
    flow, and charged singular vectors}}\label{app:sl21} 
\begin{figure}[tb]
  \rotatebox{270}{\includegraphics{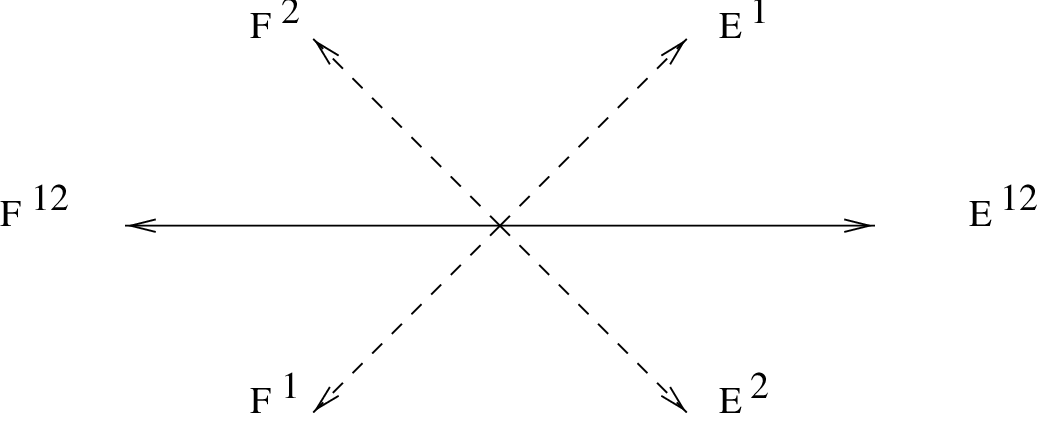}}
  \caption{\textsl{The $s\ell(2|1)$ root diagram.}}
  \label{fig:sl21-roots}
\end{figure}

The affine 
Lie superalgebra $\tSSL21$ consists of four bosonic currents $E^{12}$,
$\Hminus$, $F^{12}$, and $\Hplus$ and four fermionic ones, $E^1$,
$E^2$, $F^1$, and $F^2$.  The $\tSL2$ subalgebra is generated by
$E^{12}$, $\Hminus$, and $F^{12}$, and it commutes with the~$\u(1)$
subalgebra generated by $\Hplus$.  For reference, we give in
Fig.~\ref{fig:sl21-roots} the two-dimensional root diagram of the
finite dimensional Lie superalgebra $s\ell(2|1)$, represented in
Minkowski space with the fermionic roots along the light cone
directions.  We sometimes refer to the eigenvalue corresponding to a
given eigenstate of the Cartan generator $H_0^-$ (resp.\ $H_0^+$) as
the {\em isospin\/} (resp.\ the {\em hypercharge\/}) of that state.  At level~$k$, the
nonvanishing commutation relations are given by
\begin{equation}\label{sl21}
  \begin{array}{rclrcl}\!
    {[}\Hminus_m, E^{12}_n] &=& E^{12}_{m+n}\,,&
    {[}\Hminus_m, F^{12}_n] &=& -F^{12}_{m+n}\,,\\[4pt]
    {[}E^{12}_m, F^{12}_n] &=& m \delta_{m+n, 0} k + 2
    \Hminus_{m+n}\,,& 
    {[}H^\pm_m, H^\pm_n] &=& \mp\half m \delta_{m+n, 0} k\,,\\[4pt]
    {[}F^{12}_m, E^2_n] &=& F^1_{m+n}\,,&
    {[}E^{12}_m, F^2_n] &=& -E^1_{m+n}\,,\\[4pt]
    {[}F^{12}_m, E^1_n] &=& -F^2_{m+n}\,,&
    {[}E^{12}_m, F^1_n] &=& E^2_{m+n}\,,\\[4pt]
    {[}H^\pm_m, E^1_n] &=& \half  E^1_{m+n}\,,&
    {[}H^\pm_m, F^1_n] &=& -\half  F^1_{m+n}\,,\\[4pt]
    {[}H^\pm_m, E^2_n] &=& \mp\half  E^2_{m+n}\,,&
    {[}H^\pm_m, F^2_n] &=& \pm\half  F^2_{m+n}\,,\\[4pt]
    {[}E^1_m, F^1_n]_+ &=& \multicolumn{4}{l}{-m \delta_{m+n, 0} k +
      \Hplus_{m+n} -
      \Hminus_{m+n}\,,}\\[4pt]
    {[}E^2_m, F^2_n]_+ &=& \multicolumn{4}{l}{m \delta_{m+n, 0} k + 
      \Hplus_{m+n} +
      \Hminus_{m+n}\,,}\\[4pt]
    {[}E^1_m, E^2_n]_+ &=& E^{12}_{m+n}\,,&
    {[}F^1_m, F^2_n]_+ &=& F^{12}_{m+n}\,.
  \end{array}
\end{equation}
  
One of the~$\tSSL21_{k}$ spectral flows is given by
\begin{equation}\label{sl21-spectral}
  \cU_\theta:{}
  \begin{array}{rclcrcl}
    E^1_n&\mapsto& E^1_{n-\theta}\,,&&E^2_n&\mapsto&E^2_{n+\theta}\,,
    \\[6pt]
    F^1_n&\mapsto& F^1_{n+\theta}\,,&&F^2_n&\mapsto&F^2_{n-\theta}\,,
  \end{array}
  \quad
  \Hplus_n\mapsto\Hplus_n + k\theta\delta_{n,0}
\end{equation}
(with the~$\tSL2$ subalgebra remaining invariant).  For
$\theta\in\oZ$, this is an automorphism of~$\tSSL21$.  Applying the
spectral flow to modules gives \textit{twisted} modules.  A twisted
module with a vacuum vector is, thus, generated from the state
$\ket{h_-,h_+,k;\theta}$ (which we call the \textit{twisted
  highest-weight vector}) satisfying the \textit{twisted
  highest-weight conditions}
\begin{equation}\label{hw-twisted}
 E^1_{-\theta}\ket{h_-,h_+,k;\theta}=0\qquad
 E^2_{\theta}\ket{h_-,h_+,k;\theta}=0\qquad
 F^{12}_{1}\ket{h_-,h_+,k;\theta}=0,
\end{equation}
and whose quantum numbers of hypercharge and isospin are given by,
\begin{equation}
 \Hplus_0\ket{h_-,h_+,k;\theta}~{}={}~(h_+ - k\theta)\,
      \ket{h_-,h_+,k;\theta}\,,
      \qquad
      \Hminus_0\ket{h_-,h_+,k;\theta}~{}=
     {}~h_-\,\ket{h_-,h_+,k;\theta}\,,
\end{equation} 
where $k$ is the level and $\theta$ is the twist. %\footnote{
The eigenvalue of~$H^+_0$ is parametrised as $h_+ - k\theta$ so as to
have the same value of~$h_+$ for all the modules differing from each
  other by a spectral flow transform. 
We assume $\theta\in\oZ$ in most of our formulas, with the necessary
modifications for~$\theta\in\oZ+\half$ to be done in accordance with
the spectral flow transform.  The dimension of~$\ket{h_-,h_+,k;\theta}$
with respect to the  Sugawara energy-momentum tensor
\begin{equation}\label{Tsug-sl21}
  T_{\mathrm{Sug}} =
  \tfrac{1}{k +1}\Bigl(\Hminus\, \Hminus - \Hplus \Hplus +
  E^{12}\, F^{12} + E^1\, F^1 - E^2\, F^2 \Bigr)
\end{equation}
is given by
\begin{equation}\label{Sug-twisted}
  \Delta_{h_-,h_+,k;\theta} =
  \frac{h_-^2 - h_+^2}{k+1} + 2\theta h_+ -k\theta^2.
\end{equation}

The character of a twisted module ${\mN}_{;\theta}$ is expressed
through the `untwisted' character~$\chi_{\mN}$ as
\begin{equation}\label{sl21-sf}  
  \chi^{\mN}_{;\theta}(q,z,\zeta)=
  \zeta^{-k\theta}\,q^{-k\theta^2}\,\chi^{\mN}(q,z,\zeta\,q^{2\theta}).
\end{equation}

The \textit{twisted Verma module} $\mP_{h_-,h_+,k;\theta}$ is freely
generated from $\ket{h_-,h_+,k;\theta}$ by $E^1_{\leq-\theta-1}$,
$E^2_{\leq\theta-1}$, $F^1_{\leq\theta}$, $F^2_{\leq-\theta}$,
$E^{12}_{\leq-1}$, $F^{12}_{\leq0}$, $H^+_{\leq-1}$, and
$H^-_{\leq-1}$.  For an integral $\theta$, the character of
$\mP_{h_-,h_+,k;\theta}$ is
\begin{equation}\label{sl21-Verma-char}
  \Tr\Bigl(z^{H^-_0}\,\zeta^{H^+_0}\,q^{\cL^{\text{Sug}}_0}\Bigr)=
  z^{h_-}\,\zeta^{h_+-(k+1)\theta}\,
  q^{\frac{h_-^2 - h_+^2}{k+1} + 2\theta h_+ -(k+1)\theta^2
    }\,
  \frac{\vartheta_{1,0}(q, z^{\half}\zeta^{\half})\,
    \vartheta_{1,0}(q, z^{\half}\zeta^{-\half})}{
    \vartheta_{1,1}(q,z)\prod_{m\geq1}(1-q^m)^3}\,,
\end{equation}
where the Jacobi theta functions are defined by
\begin{align}\label{Jacobi11}
  \vartheta_{1,1}(q, z) ={}&
  \sum_{m\in\oZ}(-1)^m q^{\half(m^2 -
    m)} z^{-m} {}= 
  \prod_{m\geq0}(1 - z^{-1} q^m)
  \prod_{m\geq1}(1 - z q^m)\prod_{m\geq1}(1 - q^m)\,,\\
  \label{Jacobi10}
  \vartheta_{1,0}(q,z) ={}&
  \sum_{m\in\oZ}^{}q^{\half(m^2 - m)} z^{-m}
  = \prod_{m\geq0}(1+z^{-1}q^m)\prod_{m\geq1}(1+z q^m)
  \prod_{m\geq1}(1-q^m)\,.
\end{align}

Among singular vectors that can exist in $\mP_{h_-,h_+,k;\theta}$, we
note the so-called \textit{charged singular vectors}.  
They occur whenever
\begin{equation}\label{charged-exist}
  h_+ = \pm h_- - (k+1)n\,,\qquad n\in\oZ
\end{equation}
and are given by an explicit construction as follows~\cite{[S]}.  For
$h_+ -h_- = -(k+1)n$, $n\in\oZ$, the charged singular vector in the
twisted Verma module $\mP_{h_-,h_+,k;\theta}$ reads
\begin{equation}\label{Echminus}
  \ket{C^{(-)}(n, h_-, k;\theta)}%={}\\
  {}=\left\{\!\!
    \begin{array}{ll}
      \underbrace{E^2_{\theta+n}\ldots E^2_{\theta-1}}_{-n}\cdot
      \underbrace{F^1_{\theta+n}\ldots F^1_{\theta}}_{-n+1}
      \ket{h_-,h_-  -  n(k+1),k;\theta},& n\leq0\,,\\
      \underbrace{E^1_{-\theta-n}\ldots E^1_{-\theta-1}}_{n}\cdot
      \underbrace{F^2_{1-\theta-n}\ldots F^2_{-\theta}}_{n}
      \ket{h_-,h_-  -  n(k+1),k;\theta},& n\geq1\,.
    \end{array}\right.\kern-12pt
\end{equation}
As is easy to check, this vector satisfies the highest-weight
conditions (we omit the singular vector itself)
\begin{equation}\label{hwtop1}
 E^1_{-\theta-n}\approx{}0 \qquad 
  E^2_{\theta+n}\approx{}0\qquad
  F^{12}_{1}\approx{}0
\end{equation}
together with $F^1_{\theta+n}\approx{}0$.

Similarly, whenever $h_+ +h_- = -n(k+1)$, $n\in\oZ$, the
charged singular vector in~$\mP_{h_-,h_+,k;\theta}$ is
\begin{equation}\label{Echplus}
  \ket{C^{(+)}(n, h_-, k;\theta)} %={}\\
  {}=\left\{\!\!
    \begin{array}{ll}
      \underbrace{E^2_{\theta+n}\ldots E^2_{\theta-1}}_{-n}\cdot
      \underbrace{F^1_{\theta+n+1}\ldots F^1_{\theta}}_{-n}
      \ket{h_-,-h_- - n(k+1),k;\theta},& n\leq-1,\\
      \underbrace{E^1_{-\theta-n}\ldots E^1_{-\theta-1}}_{n}
      \cdot
      \underbrace{F^2_{-\theta-n}\ldots F^2_{-\theta}}_{n+1}
      \ket{h_-,-h_- - n(k+1),k;\theta},& n\geq0.
    \end{array}\right.\kern-12pt
\end{equation}
This satisfies the highest-weight conditions~\eqref{hwtop1}
supplemented by $F^2_{-\theta-n}\approx{}0$.  It is straightforward to
check that the highest-weight conditions satisfied by
$\ket{C^{(\pm)}(n, h_-, k;\theta)}$ imply that this vector generates a
submodule in the respective~$\mP_{h_-,h_+,k;\theta}$ module.

\section{\textbf{$D(2|1;\alpha)$}}\label{app:D-algebra}
The one-parameter family of exceptional Lie superalgebras $\D$, with
$\alpha \in \oC \setminus \{0,-1,\infty \}$ are basic type II
classical simple complex Lie superalgebras in the Ka\v c
classification~\cite{[Kac77],[Cornwell]}.  At each fixed value
of~$\alpha$, $\D$ is a rank 3 superalgebra with six even roots and
eight odd roots.  Its dual Coxeter number $h^{\vee}$ is zero and its
superdimension, which is the number of bosonic generators minus the
number of fermionic generators, is $\sdim=9-8=1$.  The central charge
of the Virasoro algebra satisfied by the Sugawara energy-momentum
tensor of the affine algebra is therefore 1 for any value of the level
$\kappa$ since,
\begin{equation}
  c=\frac{\kappa\,\sdim}{\kappa + h^{\vee}}=1.
\end{equation}

\begin{figure}[bt]
  \rotatebox{270}{
    \includegraphics[height=210pt, width=210pt,
    keepaspectratio]{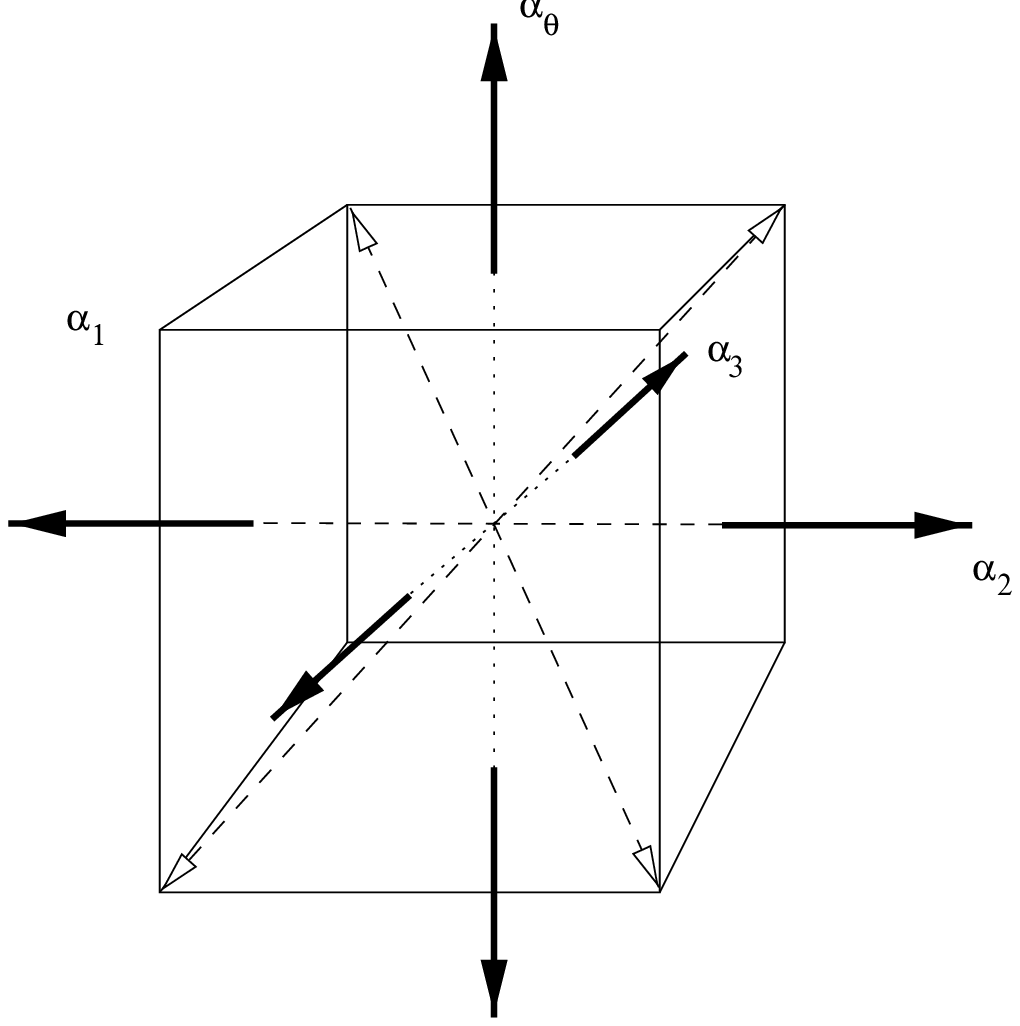}} 
  \caption[$\D$ root diagram]{\textsl{The
    $\D{}\rule[-20pt]{0pt}{20pt}$ root diagram.}}\label{fig:roots} 
\end{figure}

The bosonic part of~$\D$ is $\SL2\oplus\SL2\oplus\SL2$, and the action
of~$\D_{\zero}$ on $\D_{\one}$ is the product of 2-dimensional
representations.  The root diagram (see Fig.~\ref{fig:roots})
can be visualised in a parallelipiped in 3d space with metric
$g_{ij}=\mathrm{diag}(-1,-1,1)$.  All odd roots are at the vertices of
the parallelipiped on the light cone.  Six even roots lie on the three
lines through the centre of the faces.  The Weyl group does not act
transitively on the set of simple roots, and there are six choices of
simple root systems.

We describe in more detail the system of simple roots with one odd
root, $\alpha_1$, and two even roots, $\alpha_2$ and~$\alpha_3$. The
three regular $s\ell(2)$ subalgebras are in the directions of~$\alpha
_2, \alpha _3$ and $\alpha_{\theta}=2\alpha_1+\alpha_2+\alpha_3$.  
The relevant scalar products are summarised as
\begin{gather}
  \alpha_1^2=0\,,\qquad \alpha_2^2=-2\gamma \,,\qquad
  \alpha_3^2=-2(1-\gamma)\,,\qquad
  \alpha_{\theta}^2=2\,,\\
  \alpha_1 \cdot \alpha_2=\gamma\,,\qquad \alpha_1 \cdot
  \alpha_3=1-\gamma\,,\qquad
  \alpha_2 \cdot \alpha_3=0\,,\\
  \alpha_3 \cdot \alpha_{\theta}=0\,, \qquad\alpha_2 \cdot
  \alpha_{\theta}=0\,,\qquad \alpha_1 \cdot \alpha_{\theta}=1,
\end{gather}
where 
\begin{equation}
  \gamma=\frac{\alpha}{1+\alpha},\qquad \gamma \in \oC \setminus
  \{0,1,\infty\}.
\end{equation}
With the metric above, we have
\begin{equation}
  \alpha_1=
  \tfrac{1}{\sqrt{2}}(-\sqrt{\gamma},-\sqrt{1-\gamma},1)\,,\qquad
  \alpha_2=(\sqrt{2\gamma},0,0)\,,\qquad
  \alpha_3 = (0,\sqrt{2(1-\gamma)},0)\,.
\end{equation}

The mapping $t_1: \g \mapsto 1-\g$ interchanges the r\^oles of~$\ab$
and $\atr$, while the mapping $t_2: \g \mapsto \g^{-1}$
interchanges those of~$\ab$ and $\at$. These two transformations
generate an order-6 group defined by the relations $t_1^2=t_2^2=1$,
$t_1t_2t_1=t_2t_1t_2$, and one has the isomorphisms
\begin{equation}\label{iso}
  \DD{\tfrac{\g}{1-\g}}\simeq{}\DD{\tfrac{1-\g}{\g}}\simeq{}
  \DD{\tfrac{1}{\g-1}}
  \simeq{} \DD{-\tfrac{1}{\g}}\simeq{}\DD{\g-1}\simeq{}
  \DD{-\g}. 
\end{equation}  
If one restricts the parameter $\g$ to real values, it is sufficient
to consider the domain~$\g \in [1/2,1[$ for which $\at$ is always the
longest root.

As far as the affine superalgebra is concerned, the isomorphisms are,
\begin{multline}
  \hDD{\tfrac{\g}{1-\g}}_k\simeq{}\hDD{\tfrac{1-\g}{\g}}_k\simeq{}
  \hDD{\tfrac{1}{\g-1}}_{-\frac{k}{\g}}\\
  \simeq{}\hDD{\g-1}_{-\frac{k}{\g}}\simeq{}
  \hDD{-\tfrac{1}{\g}}_{-\frac{k}{1-\g}}\simeq{}
  \hDD{-\g}_{-\frac{k}{1-\g}}.
\end{multline}

\section{\textbf{$\protect\widehat{s\ell}(2|1)$ OPE's}}\label{app:ope}
We use a number of standard integrals, assuming whenever necessary
that they are analytically continued from the domain where they are
well-defined.  For~$\Re z>\Re w$, we have
\begin{equation}
  \int_w^z dx\,(z-x)^{\alpha}\,(x-w)^{\beta}=(z-w)^{a+b+1}\,
  \frac{\Gamma(\alpha+1)\Gamma(\beta+1)}{\Gamma(\alpha+\beta+2)}
\end{equation}
and also
\begin{multline}
  \int_z^\infty dx\,(x-z)^\alpha\,(x-w)^\beta=
  (z-w)^{\alpha+\beta+1}\,
  \frac{\Gamma(\alpha+1)\Gamma(-1-\alpha-\beta)}{\Gamma(-\beta)}={}
  \displaybreak[3]\\
  {}=
  -(z-w)^{\alpha+\beta+1}\,
  \frac{\sin\pi\beta}{\sin\pi(\alpha+\beta)}\,
  \frac{\Gamma(\alpha+1)\Gamma(\beta+1)}{\Gamma(\alpha+\beta+2)}\,.
\end{multline}

In evaluating the operator product $E^1(z)\cdot E^2(w)$, we have the
following operators in~\eqref{recipe}:
$V_1=e^{\frac{1}{\sqrt{2p}}\varphi}$,
$V_2=e^{\frac{1}{\sqrt{2p}}\varphi}$,
$V'_1=e^{\frac{1}{\sqrt{2p'}}\varphi'}$, and
$V'_2=\gamma'\,e^{\frac{1}{\sqrt{2p'}}\varphi'}$.  Using the above
integrals, we obtain
\begin{multline}
  \int_z^\infty\!\,du\,S(u)\,V_1(z)\,V_2(w)
  \int_z^\infty\! dx\,S'(x)\,V'_1(z)\,V'_2(w)={}\displaybreak[1]\\
  {}= \beta(w)\,\frac{\sin\frac{\pi}{p}}{\sin\frac{2\pi}{p}}
  (z-w)^{\frac{1}{2p}+1-\frac{2}{p}}
  \frac{\Gamma(1-\frac{1}{p})\Gamma(1-\frac{1}{p})}{
    \Gamma(2-\frac{2}{p})}\cdot (z-w)^{\frac{1}{2p'}-\frac{2}{p'}}
  \frac{\Gamma(\frac{1}{p})\Gamma(2-\frac{2}{p})}{
    \Gamma(2-\frac{1}{p})}={}\\
  {}=(z-w)^{-\half}\beta(w)\,\frac{\pi p'}{\sin\frac{2\pi}{p}}
\end{multline}
plus lower-order terms.  This is to be multiplied with the operator
product \ $e^{\frac{1}{\sqrt{2}}f(z)}\,e^{-\frac{1}{\sqrt{2}}f(w)}=
(z-w)^{-\half}+\dots$, which restores the first-order pole.  We now
recall the normalisation factors $(-1)\cdot\frac{i}{\pi
  p'}\cos\frac{\pi}{p}$ from~\eqref{E0} and the factor
$-2i\sin\tfrac{\pi}{p}$ from~\eqref{recipe}.  This gives the operator
product
\begin{equation*}
  \frac{-\beta(w)}{z-w}=\frac{E^{12}(w)}{z-w}\,,
\end{equation*}
which is in agreement with~\eqref{sl21}.  The remaining terms
in~\eqref{recipe} cancel each other.

Evaluating the operator product $E^1(z)\cdot F^1(w)$, we have
$V_1=e^{\frac{1}{\sqrt{2p}}\varphi}$,
$V_2=\gamma\,e^{\frac{1}{\sqrt{2p}}\varphi}$,
$V'_1=e^{\frac{1}{\sqrt{2p'}}\varphi'}$, and
$V'_2=\gamma'\,e^{\frac{1}{\sqrt{2p'}}\varphi'}$.
Then
\begin{multline}
  \int_z^\infty du\,S(u)\,V_1(z)V_2(w)={}\notag\displaybreak[1]\\
  {}= (z-w)^{\frac{1}{2p}} \int_z^\infty du\,(-\tfrac{1}{u-w} +
  \sqrt{\tfrac{2}{p}}\,\d\varphi(w) + \beta\gamma(w) -
  \tfrac{1}{\sqrt{2p}}\,\d\varphi(w)\,\tfrac{z-w}{u-w} + \dots)
  (u-z)^{-\frac{1}{p}}\,(u-w)^{-\frac{1}{p}}
\end{multline}
(where the dots denote higher-order terms). This equals
\begin{equation}\label{this-equals}
  {}=(z-w)^{\frac{1}{2p} - \frac{2}{p}}\,
  \frac{\Gamma(1-\frac{1}{p})\Gamma(-1+\frac{2}{p})}{
    \Gamma(\frac{1}{p})}\,\left(p-2 + (z-w)\left(\sqrt{\tfrac{p}{2}}\,
      \d\varphi(w) + \beta\gamma(w)
    \right)
  \right)\,.
\end{equation}
Next, the primed sector integral $\int_z^\infty
du\,S'(u)\,V'_1(z)V'_2(w)$ is obtained from the last expression by
simply replacing $p\mapsto p'$, $\varphi\mapsto\varphi'$,
$\beta\mapsto\beta'$, and $\gamma\mapsto\gamma'$.  Multiplying the
primed and the unprimed contributions, we thus obtain
\begin{multline}
  \int_z^\infty\!\,du\,S(u)\,V_1(z)\,V_2(w)
  \int_z^\infty\! dx\,S'(x)\,V'_1(z)\,V'_2(w)={}\\
  {}=(z-w)^{\frac{1}{2}}\,
  \frac{\Gamma(1-\frac{1}{p})\Gamma(-1+\frac{2}{p})}{
    \Gamma(\frac{1}{p})}\,\biggl(\frac{p-2}{(z-w)^2} +
  \frac{\sqrt{\frac{p}{2}}\, \d\varphi(w) + \beta\gamma(w)}{z-w}
  \biggr)\times{}\notag\displaybreak[3]\\
  {}\times\frac{\Gamma(\frac{1}{p})\Gamma(1-\frac{2}{p})}{
    \Gamma(1-\frac{1}{p})}\,\biggl(\frac{p'-2}{(z-w)^2} +
  \frac{\sqrt{\frac{p'}{2}}\, \d\varphi'(w) + \beta'\gamma'(w)}{z-w}
  \biggr)={}\\
  {}=(z-w)^{\frac{1}{2}}\,
  \frac{\pi}{\sin\frac{2\pi}{p}}\,\biggl(\frac{p'(p-2)}{(z-w)^2} +
  \frac{p'(\sqrt{\frac{p}{2}}\,\d\varphi + \beta\gamma) -
    p(\sqrt{\frac{p'}{2}}\,\d\varphi' + \beta'\gamma')}{z-w}
  \biggr)\,.
\end{multline}
This is further multiplied with %the auxiliary sector contribution
$e^{\frac{1}{\sqrt{2}}f(z)}\,e^{-\frac{1}{\sqrt{2}}f(w)}=
(z-w)^{-\half}(1 + \frac{1}{\sqrt{2}}(z-w)\d f +\dots)$; in addition,
we recall the normalisations $(-1)\cdot \frac{i}{\pi
  p'}\,\cos\frac{\pi}{p}$ in~\eqref{E0} and the factor
$-2i\sin\tfrac{\pi}{p}$ from the first term in~\eqref{recipe}.  Thus,
the first term in~\eqref{recipe} gives the contribution
\begin{multline}
  E^1(z)\,F^1(w)=
  \frac{-(p-2)}{(z-w)^2} +
  \frac{
    \frac{p}{p'}(\sqrt{\frac{p'}{2}}\,\d\varphi' + \beta'\gamma')
    -(\sqrt{\frac{p}{2}}\,\d\varphi + \beta\gamma) -
    \frac{(p-2)}{\sqrt{2}}\d f}{z-w}={}\notag\\
  {}=  \frac{-k}{(z-w)^2} +
  \frac{H^+ - H^-}{z-w}\,.
\end{multline}
The remaining terms in~\eqref{recipe} cancel each other and,
therefore, the operator product $E^1(z)\,F^1(w)$ given by the last
formula is in agreement with the respective commutator
in~\eqref{sl21}.

We will need the first regular term in the above expansion when we
calculate the~$\tSSL21$ energy-momentum tensor in
Lemma~\ref{lemma:T}, namely,
\begin{multline}
  -p'E^1\,F^1= \tfrac{p'}{\sqrt{2}}\beta \gamma \d f -
  \tfrac{\p}{\sqrt{2}}\beta' \gamma' \d f -
  \sqrt{\tfrac{\p}{2}}\,\beta' \gamma' \d\varphi +
  \sqrt{\tfrac{p'}{2}}\beta\gamma\d\varphi' - \sqrt{\tfrac{\p(\p -
      1)}{2}} \beta' \gamma' \d\varphi'\\
  {}- \tfrac{\sqrt{\p}}{\sqrt{2} (1 - \p)}\beta\gamma\d\varphi
  {}+ \tfrac{(\p - 2)\sqrt{\p}}{2 \sqrt{2} (\p - 1)} \d^2\varphi +
  \tfrac{\p + 2}{4(\p - 1)} \d\varphi \d\varphi + \tfrac{(\p -
    2)}{2\sqrt{2}}p'\d^2f - \tfrac{\p - 2}{2 \sqrt{\p - 1}} \d\varphi
  \d\varphi'\\
  {}+ \tfrac{1}{4}(2 - 3 \p) \d\varphi' \d\varphi'
  {}+ \tfrac{\p - 2}{2}\sqrt{\tfrac{p'}{2}}\d^2\varphi' +
  \tfrac{p'}{2}\sqrt{\p}\,\d\varphi \d f -
  \tfrac{\p}{2}\sqrt{p'}\d\varphi' \d f + \tfrac{p'}{4}\,(\p - 2) \d f
  \d f + \mathfrak{X}
\end{multline}
where $\mathfrak{X}$ is a contribution that cancels against similar
terms coming from $E^2\,F^2$.

As regards $E^2(z)\cdot F^2(w)$, we again use~\eqref{recipe}, where
now $V_1=e^{\frac{1}{\sqrt{2p}}\varphi}$,
$V_2=\gamma\,e^{\frac{1}{\sqrt{2p}}\varphi}$,
$V'_1=\gamma'\,e^{\frac{1}{\sqrt{2p'}}\varphi'}$, and
$V'_2=e^{\frac{1}{\sqrt{2p'}}\varphi'}$.  We already know
from~\eqref{this-equals} the unprimed integral $\int_z^\infty
du\,S(u)\,V_1(z)\,V_2(w)$; the primed sector contributes is evaluated
similarly.  We only quote the first regular term
\begin{multline}
  -p'E^2\,F^2= - \p\beta' \d\gamma' - p'\d\beta \gamma +
  \sqrt{\tfrac{p'}{2}} \beta \gamma \d\varphi' -
  \sqrt{\tfrac{\p}{2}}\beta' \gamma' \d\varphi +
  \tfrac{p'}{\sqrt{2}}\beta \gamma \d f +
  \tfrac{\p}{\sqrt{2}}\beta' \gamma' \d f \\
  {}- \tfrac{\sqrt{\p}}{\sqrt{2} (1 - \p)}\beta \gamma \d\varphi -
  \sqrt{\tfrac{\p(\p - 1)}{2}}\beta' \gamma' \d\varphi' + \tfrac{(\p -
    2)p'}{2 \sqrt{2}} \d^2f -
  \tfrac{p'}{2}\sqrt{\tfrac{\p}{2}}\d^2\varphi -
  \tfrac{\p}{2}\sqrt{\tfrac{p'}{2}}\d^2\varphi'
  - \tfrac{p'}{4}(\p - 2)\d f\d f  \\
  {}+ \tfrac{p'}{2}\sqrt{\p}\,\d\varphi \d f +
  \tfrac{\p}{2}\sqrt{p'}\d\varphi' \d f - \tfrac{\p - 2}{4 (\p - 1)}
  \d\varphi \d\varphi + \tfrac{2 - \p}{2 \sqrt{\p - 1}} \d\varphi
  \d\varphi' + \tfrac{1}{4}(2 - \p) \d\varphi' \d\varphi' +
  \mathfrak{X}\,,
\end{multline}
where $\mathfrak{X}$ is a contribution that cancels between $E^1\,F^1$
and $E^2\,F^2$.

The operator product $F^1(z)\cdot F^2(w)$ is evaluated similarly; we
have $V_1=\gamma\,e^{\frac{1}{\sqrt{2p}}\varphi}$,
$V_2=\gamma\,e^{\frac{1}{\sqrt{2p}}\varphi}$,
$V'_1=\gamma'\,e^{\frac{1}{\sqrt{2p'}}\varphi'}$, and
$V'_2=e^{\frac{1}{\sqrt{2p'}}\varphi'}$.  One then multiplies the 
contributions of the primed and unprimed sectors,
given by
\begin{equation*}
  \int_z^\infty du\,S'(u)\,V'_1(z)\,V'_2(w)=
  -(z-w)^{\frac{1}{2p'} - \frac{2}{p'}}\,
  \frac{\Gamma(-\frac{1}{p'})\Gamma(\frac{2}{p'})}{
    \Gamma(\frac{1}{p'})}\,,
\end{equation*}
and
\begin{multline}
  \int_z^\infty du\,S(u)\,V_1(z)\,V_2(w)={}\notag\displaybreak[0]\\
  {}=\int_z^\infty du\,\Bigl( \beta\gamma^2 - \frac{\gamma(z)}{u-w} -
  \frac{\gamma(w)}{u-z}\Bigr) \Bigl(1 -
  \sqrt{\tfrac{2}{p}}\,(u-w)\,\d\varphi +
  \tfrac{1}{\sqrt{2p}}\,(z-w)\,\d\varphi\Bigr)
  (u-z)^{-\frac{1}{p}}\,(u-w)^{-\frac{1}{p}}\displaybreak[1]\\{}=
  (z-w)^{\frac{1}{2p} 
    - \frac{2}{p} + 1}\,
  \frac{\Gamma(1-\frac{1}{p})\Gamma(-1+\frac{1}{p})}{
    \Gamma(\frac{1}{p})}\, (\beta\gamma^2 + \sqrt{2p}\,\gamma\d\varphi
  + (p-2)\d\gamma)\,,
\end{multline}
where we see the $J^-$ current from~\eqref{Wakimoto}.

\end{document}